%% file: MFV.tex
\documentclass[12pt]{article}
\usepackage{amsmath,amssymb,amsfonts,color,graphicx,cite,color}
\usepackage{latexsym}
\usepackage{epsfig,psfrag,rotating,soul}
\usepackage{rotfloat}
\input paperdef

\evensidemargin \oddsidemargin
\allowdisplaybreaks

\begin{document}
\thispagestyle{empty}

\def\thefootnote{\fnsymbol{footnote}}

\begin{flushright}
\mbox{}
\end{flushright}

\vspace{0.5cm}

\begin{center}

\begin{large}
\textbf{Effects of Sfermion Mixing induced by RGE Running}
\\[2ex]
\textbf{in the Minimal Flavor Violating CMSSM}
\end{large}

\vspace{1cm}

{\sc
M.E.~G{\'o}mez$^{1}$%
\footnote{email: mario.gomez@dfa.uhu.es}%
, S.~Heinemeyer$^{2}$%
\footnote{email: Sven.Heinemeyer@cern.ch}%
~and M.~Rehman$^{2}$%
\footnote{email: rehman@ifca.unican.es}%
\footnote{MulitDark Scholar}
 
}

\vspace*{.7cm}

{\sl
$^1$ Department of Applied Physics, University of Huelva, 21071 Huelva, Spain

\vspace*{0.1cm}

\vspace*{0.1cm}
$^2$Instituto de F\'isica de Cantabria (CSIC-UC),  39005 Santander, Spain

}

\end{center}

\vspace*{0.1cm}

\begin{abstract}
\noindent

Within the Constrained Minimal Supersymmetric Standard Model (CMSSM)
with Minimal Flavor Violation (MFV) for scalar quarks
we study
the effects of intergenerational squark mixing on $B$-physics
observables, electroweak precision observables (EWPO) and the Higgs boson
mass predictions. Squark mixing is generated through the
Renormalization Group Equations (RGE) running from the GUT scale to the 
electroweak scale due to presence of non diagonal Yukawa matrices in the
RGE's, e.g.\ due to the CKM matrix.
We find that the $B$-Physics observables as well as the Higgs mass
predictions do not receive sizable corrections. On the other hand, 
the EWPO such as
the $W$~boson mass can receive corrections by far exceeding the current
experimental precision. These contributions can place new upper
bounds on the CMSSM parameter space.
We extend our analysis to the CMSSM extended with a
mechanism to explain neutrino masses (\CMSSMI), which induces
flavor violation in the scalar lepton sector.
Effects from slepton mixing on the analyzed observables are in general smaller
than from squark mixing, but can reach the level of the current experimenal
uncertainty for the EWPO.

\end{abstract}

\def\thefootnote{\arabic{footnote}}
\setcounter{page}{0}
\setcounter{footnote}{0}

\newpage


\input{intro}

\input{MFV_Sfermion}

\input{Comp_Setup}

\input{Results}
\input{Conclusions}
\vspace{-0.5em}
\subsection*{Acknowledgments}

The work of S.H.\ and M.R.\ was partially supported by CICYT (grant FPA
2013-40715-P). 
M.G., S.H.\ and M.R.\ were supported by 
the Spanish MICINN's Consolider-Ingenio 2010 Programme under grant
MultiDark CSD2009-00064. 
M.E.G.\  acknowledges further support from the
MICINN project FPA2011-23781



\include{bibliography}
\end{document}

%% file: intro.tex

\section{Introduction}

Supersymmetric (SUSY) extensions of the Standard
Model (SM) are broadly considered as the most motivated
and promising New Physics (NP) theories beyond
the SM. The solution of the hierarchy problem, the
gauge coupling unification and the possibility of having a
natural cold dark matter candidate, constitute the most
convincing arguments in favor of SUSY.

Within the Minimal Supersymmetric Standard Model
(MSSM)~\cite{mssm}, flavor mixing can occur in both scalar quark and scalar
lepton sector. Here the possible presence of soft 
SUSY-breaking parameters in the squark and slepton sector, which are
off-diagonal in flavor space (mass parameters as well as trilinear
couplings) are the most general way to introduce flavor mixing within the
MSSM. This, however, yields many new sources of flavor and
$\cp$-violation, which potentially lead to large non-standard effects in
flavor processes in conflict with experimental bounds.

The SM has been very successfully tested by
low-energy flavor observables both from the kaon and $B_d$
sectors. In particular, the two $B$~factories have established that
$B_d$ flavor and $\cp$-violating processes are well described
by the SM up to an accuracy of the $\sim 10\%$
level~\cite{HFAgroup}.
This immediately implies a tension between the solution
of the hierarchy problem, calling for a NP scale at or below
the TeV scale, and the explanation of the Flavor Physics data
requiring a multi-TeV NP scale if the new flavor-violating
couplings are of generic size.

An elegant way to simultaneously solve the above problems
is provided by the Minimal Flavor Violation (MFV)
hypothesis~\cite{MFV1,MFV2}, where flavor and $\cp$-violation in quark sector
are assumed to be entirely described by the CKM matrix. Even
in theories beyond the SM. For example in MSSM,
the off-diagonality in the sfermion mass matrix reflects the   
misalignment (in flavor space) between fermions and sfermions mass
matrices, that cannot be diagonalized simultaneously. 
This misalignment can be produced from various
origins. For instance, off-diagonal sfermion
mass matrix entries can be generated by Renormalization Group Equations
(RGE) running. Going from a high energy scale, where no flavor violation is
assumed, down to the electroweak (EW) scale can generate such entries due to
presence of non diagonal Yukawa matrices in RGE's. For example, in the
Constrained Minimal  
Supersymmetric Standard Model (CMSSM, see \citere{AbdusSalam:2011fc} and
references therein), the left-handed scalar-quark soft
SUSY-breaking parameter RGE's
have a general form:
\begin{align}
\frac{d}{dt}(m_{\tilde{Q}}^{2})_{ij}\propto a \id_{ij}+b(Y_{U}^{\dag}Y_{U})_{ij}
\label{eq:dmQ2}
\end{align}
where $a$ and $b$ are some constants, the up-quark Yukawa matrix,
$Y_U$, is non-diagonal, and $t= \log \frac{\mu}{\mu_{0}}$ with 
$\mu\,(\mu_{0})$  is the running (fixed) scale. 
 
It is convenient to work in the basis in which the Yukawa couplings are
given by
\begin{align}
Y_{D} = {\rm diag}(y_{d},y_{s},y_{b}), \quad
Y_{U} = V_{\rm CKM}^{\dag} {\rm diag}(y_{u},y_{c},y_{t})
\end{align}
and hence all flavor violation in the quark and squark sector is
controlled by the CKM matrix. 

The situation is somewhat different in the slepton sector where
neutrinos are strictly massless (in the SM and the MSSM)
Consequently, there is no slepton mixing, which would induce 
Lepton Flavor Violation (LFV) in the 
charged sector, allowing not yet observed processes like $l_i \to l_j \ga$ 
($i > j$; $l_{3,2,1} = \tau, \mu, e$)~\cite{bm}. 
However in the neutral sector, we have strong experimental
evidence that shows that the neutrinos are massive and mix among
themselves~\cite{Neutrino-Osc}. 
In order to incorporate this one needs to go beyond the MSSM to
introduce a mechanism that generates neutrino masses. 
The simplest way would be to introduce Dirac masses, however, leaving
the extreme smallness of the neutrino masses unexplained. To overcome
this problem, typically a see-saw mechanism is
used to generate neutrino masses, and the PMNS matrix plays the role of
the CKM matrix in the lepton sector. Extending the MFV hypothesis for
leptons~\cite{MFVinlepton} we can assume that the flavor mixing in the
lepton and slepton sector is induced and controlled by the see-saw mechanism. 

Consequently, in this paper we will investigate two models (more
detailed definitions are given in the next section):
\begin{itemize}
\item[(i)] 
the CMSSM, where only flavor violation in the squark sector is present.
\item[(ii)]
the CMSSM augmented by the seesaw type~I mechanism\cite{seesaw:I}, 
called ``\CMSSMI'' below.
\end{itemize}

\medskip
In many analyses of the CMSSM, or extensions such as the NUHM1 or NUHM2
(see \citere{AbdusSalam:2011fc} and references therein), 
the hypothesis of MFV has been used, and it has been assumed
that the contributions coming
from MFV are negligible for other observables as well, see, e.g.,
\citere{CMSSM-NUHM}. 
In this paper we will analyze whether this assumption is justified, and
whether including these MFV effects could lead to additional constraints
on the CMSSM parameter space. In this respect we evaluate in the CMSSM
and in the \CMSSMI\ the following set of observables: $B$~physics
observables (BPO), in particular \bsg, \bmm\ and \dmbs, 
electroweak precision observables (EWPO), in particular $\MW$ and the
effective weak leptonic mixing angle, $\sweff$, as well as the masses of
the neutral and charged Higgs bosons in the MSSM.

In order to perform our calculations, we used SPheno~\cite{Porod:2003um}
to generate the CMSSM (containing also the type~I seesaw) particle
spectrum by running RGE from the GUT down to the EW scale. 
The particle spectrum was handed over in the form of an SLHA
file~\cite{SLHA} to 
\fh~\cite{feynhiggs,mhiggslong,mhiggsAEC,mhcMSSMlong,Mh-logresum} to
calculate  EWPO and Higgs boson masses. The $B$-Physics observables were
calculated by the {\tt BPHYSICS} subroutine included in the SuFla
code~\cite{sufla} (see also \citeres{arana,arana-NMFV2} for the improved
version used here).

The paper is organized as follows: First we review the main features of
the MSSM with sfermion flavor mixing in MFV in \refse{sec:MFV_Sfermions}. 
The computational setup is given in \refse{sec:CompSetup}. The numerical
results are presented in \refse{sec:NResults}, where first we discuss the
effect of squarks mixing in the CMSSM. In a second step we analyze
effects of slepton mixing i.e.\ the \CMSSMI.
Our conclusions can be found in \refse{sec:conclusions}. 


%% file: MFV_Sfermion.tex
\section{Model set-up}
\label{sec:MFV_Sfermions}

In this section we will first review the CMSSM and the concept of
MFV. Subsequently, we will discuss the MSSM, its seesaw extension and
parameterization of sfermion mixing at low energy. 


\subsection{The CMSSM and MFV}
 
The  MSSM is the simplest Supersymmetric structure we can build from the SM 
particle content. The general set-up for the soft SUSY-breaking
parameters is given by~\cite{mssm}
\begin{eqnarray}
\label{softbreaking}
-\cL_{\rm soft}&=&(m_{\tilde Q}^2)_i^j {\tilde q}_{L}^{\dagger i}
{\tilde q}_{Lj}
+(m_{\tilde u}^2)^i_j {\tilde u}_{Ri}^* {\tilde u}_{R}^j
+(m_{\tilde d}^2)^i_j {\tilde d}_{Ri}^* {\tilde d}_{R}^j
\nonumber \\
& &+(m_{\tilde L}^2)_i^j {\tilde l}_{L}^{\dagger i}{\tilde l}_{Lj}
+(m_{\tilde e}^2)^i_j {\tilde e}_{Ri}^* {\tilde e}_{R}^j
\nonumber \\
& &+{\tilde m}^2_{1}h_1^{\dagger} h_1
+{\tilde m}^2_{2}h_2^{\dagger} h_2
+(B \mu h_1 h_2
+ {\rm h.c.})
\nonumber \\
& &+ ( A_d^{ij}h_1 {\tilde d}_{Ri}^*{\tilde q}_{Lj}
+A_u^{ij}h_2 {\tilde u}_{Ri}^*{\tilde q}_{Lj}
+A_l^{ij}h_1 {\tilde e}_{Ri}^*{\tilde l}_{Lj}
\nonumber \\
& & +\frac{1}{2}M_1 {\tilde B}_L^0 {\tilde B}_L^0
+\frac{1}{2}M_2 {\tilde W}_L^a {\tilde W}_L^a
+\frac{1}{2}M_3 {\tilde G}^a {\tilde G}^a + {\rm h.c.}).
\end{eqnarray}
Here $m_{\tilde Q}^2$ and $m_{\tilde L}^2$ are $3 \times 3$
matrices in family space (with $i,j$ being the
generation indeces) for the soft masses of the
left handed squark ${\tilde q}_{L}$ and slepton ${\tilde l}_{L}$
$SU(2)$ doublets, respectively. $m_{\tilde u}^2$, $m_{\tilde d}^2$ and
$m_{\tilde e}^2$ contain the soft masses for right handed up-type squark
${\tilde u}_{R}$,  down-type squarks ${\tilde d}_{R}$ and charged
slepton ${\tilde e}_{R}$ $SU(2)$ singlets, respectively. $A_u$, $A_d$
and $A_l$ are the $3 \times 3$ matrices for the trilinear
couplings for up-type squarks, down-type 
squarks and charged slepton, respectively  
${\tilde m}_1$ and ${\tilde m}_2$ are the soft
masses of the higgs sector. In the last line $M_1$, $M_2$ and $M_3$
defines the bino, wino  and gluino mass terms, respectively.

Within the Constrained MSSM the soft SUSY-breaking
parameters are assumed to be universal at the Grand Unification scale
$M_{\rm GUT} \sim 2 \times 10^{16} \gev$,
\begin{eqnarray}
\label{soft}
& (m_Q^2)_{i j} = (m_U^2)_{i j} = (m_D^2)_{i j} = (m_L^2)_{i j}
 = (m_E^2)_{i j} = m_0^2\  \delta_{i j}, & \nonumber \\
& m_{H_1}^2 = m_{H_2}^2 = m_0^2, &\\
& m_{\tilde{g}}\ =\ m_{\tilde{W}}\ =\ m_{\tilde{B}}\ =\ m_{1/2}, &  
\nonumber \\
& (A_U)_{i j}= A_0 e^{i \phi_A} (Y_U)_{i j},\ \ \ (A_D)_{i j}= A_0 e^{i \phi_A}
(Y_D)_{i j},\ \ \ 
(A_E)_{i j}= A_0e^{i \phi_A} (Y_E)_{i j}. & \nonumber 
\end{eqnarray}
There is a common mass for all the scalars, $m_0^2$, a single gaugino 
mass, $m_{1/2}$, and all the trilinear soft-breaking terms are directly 
proportional to the corresponding Yukawa couplings in the superpotential 
with a proportionality constant $A_0 e^{i \phi_A}$, containing a
potential non-trivial complex phase.

With the use of the Renormalization Group Equations (RGE) of the MSSM,
one can obtain the SUSY spectrum at the EW scale.
All the SUSY masses and mixings are then given as 
a function of $m_0^2$, $m_{1/2}$, $A_0$, and 
$\tb = v_2/v_1$, the ratio of the two vacuum expectation values (see
below). We require radiative symmetry breaking to fix $|\mu|$ and 
$|B \mu|$ \cite{rge,bertolini} with the tree--level Higgs potential.

By definition, this model fulfills the MFV hypothesis, since the only
flavor violating terms stem from the CKM matrix. 
The important point is that, even in a model with universal soft
SUSY-breaking terms at some high energy scale as the CMSSM, some
off-diagonality in the squark mass matrices appears at the EW scale. 
Working in the basis where the squarks are rotated parallel to the
quarks, the so-called  Super CKM (SCKM) basis, the squark mass
matrices are not flavor diagonal at the EW scale.
This is due to the fact that at $M_{\rm GUT}$ there exist two non-trivial 
flavor structures, namely the two Yukawa matrices for the up and down quarks, 
which are not simultaneously diagonalizable. This implies that 
through RGE evolution some flavor mixing leaks into the sfermion mass matrices.
In a general SUSY model the presence of new flavor structures
in the soft SUSY-breaking terms would generate large flavor mixing in
the sfermion mass matrices. However, in the CMSSM, which we are
investigating here, the two Yukawa matrices are the only 
source of flavor change. As always in the SCKM basis, any off-diagonal entry 
in the sfermion mass matrices at the EW scale will be necessarily
proportional to a product of Yukawa couplings, see \refeq{eq:dmQ2}.
The RGE's for the soft SUSY-breaking terms
are sets of linear equations, and thus, to match the correct chirality of the 
coupling, Yukawa couplings or tri-linear soft terms must enter the RGE in 
pairs. (The same holds for the \CMSSMI, see below.)


\subsection{MSSM and its seesaw extension}
\label{sec:cmssmI}

One can write the most general $SU(3)_{C}\times SU(2)_{L}\times
U(1)_{Y}$ gauge invariant and renormalizable superpotential as
\begin{eqnarray}
\label{superpotential}
W_{\rm MSSM}&=&Y_e^{ij}\epsilon_{\alpha \beta} H_1^{\alpha} E_i^c  L_j^{\beta}
+ Y_{d}^{ij} \epsilon_{\alpha \beta} H_1^{\alpha} D_i^c  Q_j^{\beta}
+ Y_{u}^{ij} \epsilon_{\alpha \beta} H_2^{\alpha} U_i^c Q_j^{\beta}
\nonumber \\
&&+ \mu \epsilon_{\alpha \beta} H_1^{\alpha} H_2^{\beta}
\end{eqnarray}
where $L_i$ represents the chiral multiplet of a $SU(2)_L$ doublet
lepton, $E_i^c$ a $SU(2)_L$ singlet charged lepton, $H_1$ and $H_2$ two Higgs doublets with opposite hypercharge.
Similarly $Q$, $U$ and $D$ represent chiral multiplets of quarks of a
$SU(2)_L$ doublet and two singlets with different $U(1)_Y$ charges.
Three generations of leptons and quarks are assumed and thus the
subscripts $i$ and $j$ run over 1 to 3. The symbol $\epsilon_{\alpha
\beta}$ is an anti-symmetric tensor with $\epsilon_{12}=1$.  

In order to provide an explanation for the (small) neutrino masses, the
MSSM can be extended by the type-I seesaw mechanism~\cite{seesaw:I}.
The superpotential for \CMSSMI\ can be written as

\begin{eqnarray}
\label{superpotentialSeesaw1}
W&=&W_{\rm MSSM}+ Y_{\nu}^{ij}\epsilon_{\alpha \beta} H_2^{\alpha} N_i^c L_j^{\beta}
+ \frac{1}{2} M_{N}^{ij} N_i^c N_j^c,
\end{eqnarray}

Where $W_{\rm MSSM}$ is given in \refeq{superpotential} and $N_i^c$
is the additional superfield that contains the three right-handed neutrinos,
$\nu_{Ri}$, and 
their scalar partners, $\tilde \nu_{Ri}$. $M_N^{ij}$ denotes the $3\times3$ Majorana mass matrix for heavy right handed neutrino.
The full set of soft SUSY-breaking terms is given by,
\begin{eqnarray}
\label{softbreakingSeesaw1}
-\cL_{\rm soft,SI} &=& - \cL_{\rm soft}
+(m_{\tilde \nu}^2)^i_j {\tilde \nu}_{Ri}^* {\tilde \nu}_{R}^j
+ (\frac{1}{2}B_{\nu}^{ij} M_{N}^{ij} {\tilde \nu}_{Ri}^* {\tilde \nu}_{Rj}^*
+A_{\nu}^{ij}h_2 {\tilde \nu}_{Ri}^* {\tilde l}_{Lj}+ {\rm h.c.})~,
\end{eqnarray}
with $\cL_{\rm soft}$ given by \refeq{softbreaking}, 
$(m_{\tilde \nu}^2)^i_j$,  $A_{\nu}^{ij}$ and $B_{\nu}^{ij}$
are the new soft breaking parameters.

By the seesaw mechanism three of the neutral fields acquire heavy masses and
decouple at high energy scale that we will denote as $M_N$, below this scale the effective theory
contains the MSSM plus an operator that provides masses to the neutrinos.

\begin{equation}
W=W_{\rm MSSM}+ \frac{1}{2}(Y_{\nu} L  H_2)^{T}  M_{N}^{-1} (Y_{\nu} L  H_2).
\end{equation}

This framework naturally explains neutrino oscillations in agreement with
experimental data~\cite{Neutrino-Osc}. At the electroweak scale an
effective Majorana mass matrix for light neutrinos, 
\begin{equation}
\label{meff}
m_{\rm eff}=-\frac{1}{2}v_u^2 Y_{\nu}\cdot M_{N}^{-1}\cdot Y^{ T}_{\nu}, 
\end{equation}
arises from Dirac neutrino Yukawa $Y_{\nu}$ (that can be assumed of
the same order as the charged-lepton and quark Yukawas), and heavy Majorana
masses $M_N$.  The smallness of the neutrino masses implies that the scale
$M_N$ is very high, \order{10^{14} \gev}. 

From \refeqs{superpotentialSeesaw1} and (\ref{softbreakingSeesaw1}) 
we can observe that 
one can choose a basis such that the Yukawa coupling matrix,
$Y_l^{ij}$, and the mass matrix of the right-handed neutrinos, $M_N^{ij}$, are
diagonalized as $Y_l^\delta$ and $M_R^\delta$, respectively. In this case
the neutrino Yukawa couplings $Y_{\nu}^{ij}$ are not generally diagonal,
giving rise to LFV. Here it is important to note that the lepton-flavor
conservation is not a consequence of the SM gauge symmetry, even in the absence
of the right-handed neutrinos. 
Consequently, slepton mass terms can violate
the lepton-flavor conservation in a manner consistent with the gauge
symmetry.  Thus the scale of LFV can be identified with the
EW scale, much lower than the right-handed neutrino scale $M_N$, leading
to potentially observable rates.

In the SM augmented by 
right-handed neutrinos, the flavor violating processes such as
$\mu \to e \gamma$, $\tau \to \mu \gamma$ etc., 
whose rates are proportional to inverse powers of $M_R^\delta$, would be
highly suppressed with such a large $M_N$ scale, and hence are far
beyond current experimental bounds. However, in SUSY theories, the
neutrino Dirac couplings $Y_\nu$ 
enter in the RGE's of
the soft SUSY-breaking sneutrino and slepton masses, generating LFV. In
the basis where the charged-lepton masses $Y_{\ell}$ is diagonal, the soft
slepton-mass matrix acquires corrections that contain off-diagonal
contributieons from the RGE running from $M_{\rm GUT}$ down to the 
Majorana mass scale $M_N$,
of the following form (in the leading-log approximation)~\cite{LFVhisano}: 
\begin{align}
(m_{\tilde L}^2)_{ij} &\sim \frac 1{16\pi^2} (6m^2_0 + 2A^2_0)
\left({Y_{\nu}}^{\dagger} Y_{\nu}\right)_{ij}  
\log \KL \frac{M_{\rm GUT}}{M_N} \KR \, \nonumber\\
(m_{\tilde e}^2)_{ij} &\sim 0  \, \nonumber\\
(A_l)_{ij} &\sim  \frac 3{8\pi^2} {A_0 Y_{l}}_i
\left({Y_{\nu}}^{\dagger} Y_{\nu}\right)_{ij}  
\log \KL \frac{M_{\rm GUT}}{M_N} \KR \,
\label{offdiagonal}
\end{align}
Consequently, even if the soft scalar masses were universal  at the
unification scale, quantum corrections between the GUT scale and 
the see-saw scale $M_N$
would modify this structure via renormalization-group 
running, which generates off-diagonal 
contributions~\cite{Cannoni:2013gq,gllv,Mismatch,Antusch,EGL,casas-ibarra} at 
$M_N$ in a basis such that $Y_{\ell}$ is diagonal. Below this  
scale, the off-diagonal contributions remain almost unchanged. 
   
Therefore the see-saw mechanism induces non trivial values for slepton
$\deFABij$ resulting in a prediction for LFV decays  
$l_i \to l_j \ga$, $(i >  j)$ that can be much larger than
the non-SUSY case. These rates depend on the structure of $Y_\nu$ at a see-saw
scale $M_N$ in a basis where $Y_l$ and $M_N$ are diagonal.  
By using the approach of \citere{casas-ibarra} a general form of $Y_\nu$
containing all neutrino experimental information can be wtritten as: 
\begin{equation}
Y_\nu = \frac{\sqrt{2}} {v_u} \sqrt{M_R^\delta} R  \sqrt{m_\nu^\delta} U^\dagger~,
\label{eq:casas} 
\end{equation}
where $R$ is a general orthogonal matrix and $m_\nu^\de$ denotes the
diagonalized neutrino mass matrix. In this basis the matrix~$U$ can be
identified with the $U_{\rm PMNS}$ matrix obtained as: 
\begin{equation}
m_\nu^\delta=U^T m_{\rm eff} U~.
\end{equation}

In order to find values for the slepton generation mixing
parameters we need a
specific form of the product $Y_\nu^\dagger Y_\nu$ as shown in
\refeq{offdiagonal}. The 
simple consideration of direct hierarchical neutrinos with a common
scale for right handed neutrinos provides a representative reference
value. In this case using \refeq{eq:casas} we find 
\begin{equation}
Y_\nu^\dagger Y_\nu= \frac{2}{v_u^2}M_R U m_\nu^\delta U^\dagger~.
\label{eq:ynu2}
\end{equation}
Here $M_R$ is the common mass assigned to the $\nu_R$'s. In the conditions
considered here, LFV effects are independent of the matrix $R$. 

For the numerical analysis the values of the Yukawa couplings
etc.\ have to be set to yield values in agreement with the
experimental data for neutrino masses and mixings.
In our computation, by considering a normal  hierarchy among the neutrino 
masses, we fix 
$m_{\nu_3} \sim \sqrt{\Delta m^2_{\text{atm}}} \sim 0.05 \ev$ and
require $m_{\nu_2}/m_{\nu_3}=0.17$, 
$m_{\nu_2} \sim  100 \cdot m_{\nu_1}$ consistent with the measured values of 
$\Delta m^2_{\text{sol}}$ and $\Delta m^2_{\text{atm}}$~\cite{neu-fits}. 
The matrix $U$ is identified with $U_{\rm PMNS}$ with the $\cp$-phases set to
zero and neutrino mixing angles set to the center of their
experimental values.  

One can observe that $m_{\rm eff}$ remains unchanged by consistent
changes on the scales of $M_N$ and $Y_\nu$. This is no longer correct
for the off-diagonal entries in the slepton mass matrices
(parameterized by slepton $\deFABij$, 
see the next subsection). These quantities have a quadratic dependence
on $Y_\nu$ and a logarithmic in $M_N$, see \refeq{offdiagonal}. 
Therefore larger values of $M_N$ imply larger LFV effects. 
By setting $M_N=10^{14} \gev$, the largest values of
$Y_\nu$ are of about 0.29, this
implies an important restriction on the parameters space arising from the
$\br(\mu\to e \ga)$ as will be discussed in \refses{sec:CompSetup}
and \ref{sec:NResults}. An  
example of models  with almost degenerate $\nu_R$ can be found in
\cite{Cannoni:2013gq}. For our numerical analysis we tested several scenarios
and we found that the one defined here is the simplest and also the
one with larger LFV prediction.




\subsection{Scalar fermion sector with flavor mixing}
\label{sec:sfermions}

In this section we give a brief description about how we parameterize
flavor mixing at the EW scale. We are using the same
notation as in \citeres{drhoLFV,arana-LFV,arana,arana-NMFV2}. 
However, while in this section we give a general description, in our
analysis below, contrary to our previous analyses~\cite{drhoLFV}, 
this time we concentrate 
on the origin of the flavor mixing as discussed in the previous sections. 

\medskip
The most general hypothesis for flavor mixing assumes a mass matrix that
is not diagonal in flavor space, both for squarks and sleptons. 
In the squarks sector and charged slepton sector we have $6 \times 6$
mass matrices, based on the corresponding six electroweak interaction
eigenstates,  
${\tilde U}_{L,R}$ with $U = u, c, t$ for up-type squarks, 
${\tilde D}_{L,R}$ with $D = d, s, b$ for down-type squarks and 
${\tilde L}_{L,R}$ with $L=e, \mu, \tau$ for charged sleptons.
For the sneutrinos 
we have a $3 \times 3$ mass matrix, since within the MSSM even with type~I
seesaw 
(right handed neutrinos decouple below their respective mass scale) 
we have only three electroweak interaction eigenstates, ${\tilde \nu}_{L}$
with $\nu=\nu_e, \nu_\mu, \nu_\tau$. 

The non-diagonal entries in this $6 \times 6$ general matrix for sfermions
can be described in terms of a set of 
dimensionless parameters $\deFABij$ ($F=Q,U,D,L,E; A,B=L,R$; $i,j=1,2,3$, 
$i \neq j$) where  $F$ identifies the sfermion type, $L,R$ refer to the 
``left-'' and ``right-handed'' SUSY partners of the corresponding
fermionic degrees of freedom, and $i,j$
indexes run over the three generations. 
(Non-zero values for the $\deFABij$ are generated via the processes
discussed in the previous subsections.)

One usually writes the $6\times 6$ non-diagonal mass matrices,  
${\mathcal M}_{\tilde u}^2$ and ${\mathcal M}_{\tilde d}^2$, referred to
the Super-CKM basis, being ordered respectively as $(\SupL, \SchaL,
\StopL, \SupR, \SchaR, \StopR)$,  $(\SdownL, \SstrL, \SbotL, \SdownR,
\SstrR, \SbotR)$ and ${\mathcal M}_{\tilde l}^2$ referred to the
Super-PMNS basis, being ordered as $(\SelL, \SmuL, \StauL, \SelR, \SmuR,
\StauR)$, and write them in terms of left- and right-handed blocks
$M^2_{\tilde q \, AB}$, $M^2_{\tilde l \, AB}$ ($q=u,d$, $A,B=L,R$),
which are non-diagonal $3\times 3$ matrices, 
\begin{equation}
\cM_{\tilde q}^2 =\left( \begin{array}{cc}
M^2_{\tilde q \, LL} & M^2_{\tilde q \, LR} \\[.3em] 
M_{\tilde q \, LR}^{2 \, \dagger} & M^2_{\tilde q \,RR}
\end{array} \right), \qquad \tilde q= \tilde u, \tilde d~,
\label{eq:blocks-matrix}
\end{equation} 
 where:
 \begin{alignat}{5}
 M_{\tilde u \, LL \, ij}^2 
  = &  m_{\tilde U_L \, ij}^2 + \left( m_{u_i}^2
     + (T_3^u-Q_u\sw^2 ) M_Z^2 \cos 2\beta \right) \delta_{ij},  \notag\\
 M^2_{\tilde u \, RR \, ij}
  = &  m_{\tilde U_R \, ij}^2 + \left( m_{u_i}^2
     + Q_u\sw^2 M_Z^2 \cos 2\beta \right) \delta_{ij} \notag, \\
  M^2_{\tilde u \, LR \, ij}
  = &  \left< \cH_2^0 \right> {\cal A}_{ij}^u- m_{u_{i}} \mu \cot \beta \, \delta_{ij},
 \notag, \\
 M_{\tilde d \, LL \, ij}^2 
  = &  m_{\tilde D_L \, ij}^2 + \left( m_{d_i}^2
     + (T_3^d-Q_d \sw^2 ) M_Z^2 \cos 2\beta \right) \delta_{ij},  \notag\\
 M^2_{\tilde d \, RR \, ij}
  = &  m_{\tilde D_R \, ij}^2 + \left( m_{d_i}^2
     + Q_d\sw^2 M_Z^2 \cos 2\beta \right) \delta_{ij} \notag, \\
  M^2_{\tilde d \, LR \, ij}
  = &  \left< \cH_1^0 \right> {\cal A}_{ij}^d- m_{d_{i}} \mu \tb \, \delta_{ij}~,
\label{eq:SCKM-entries}
\end{alignat}
and
\begin{equation}
{\mathcal M}_{\tilde l}^2 =\left( \begin{array}{cc}
M^2_{\tilde l \, LL} & M^2_{\tilde l \, LR} \\[.3em]
M_{\tilde l \, LR}^{2 \, \dagger} & M^2_{\tilde l \,RR}
\end{array} \right),
\label{eq:slep-6x6}
\end{equation} 
 where:
 \begin{alignat}{5}
M_{\tilde l \, LL \, ij}^2 
  = &  m_{\tilde L \, ij}^2 + \left( m_{l_i}^2
     + (-\edz + \sw^2 ) \MZ^2 \cos 2\beta \right) \delta_{ij},  \notag\\
 M^2_{\tilde l \, RR \, ij}
  = &  m_{\tilde E \, ij}^2 + \left( m_{l_i}^2
     -\sw^2 \MZ^2 \cos 2\beta \right) \delta_{ij} \notag, \\
  M^2_{\tilde l \, LR \, ij}
  = &  \left< \cH_1^0 \right> {\cal A}_{ij}^l- m_{l_{i}} \mu \tb \, \delta_{ij},
\label{eq:slep-matrix}
\end{alignat}
with, $i,j=1,2,3$, $Q_u=2/3$, $Q_d=-1/3$, $T_3^u=1/2$ and
$T_3^d=-1/2$. $M_{Z,W}$ denote the $Z$~and $W$~boson masses, with
$\sw^2 = 1 - \MW^2/\MZ^2$, 
and $(m_{u_1},m_{u_2}, m_{u_3})=(m_u,m_c,m_t)$, $(m_{d_1},m_{d_2},
m_{d_3})=(m_d,m_s,m_b)$ are the quark masses and $(m_{l_1},m_{l_2},
m_{l_3})=(m_e,m_\mu,m_\tau)$ are the lepton masses. $\mu$ is the
Higgsino mass term and $\tb = v_2/v_1$
with  $v_1=\left< \cH_1^0 \right>$ and $v_2=\left< \cH_2^0
\right>$ being the two vacuum expectation values of the corresponding
neutral Higgs boson in the Higgs $SU(2)_L$ doublets, 
$\cH_1= (\cH^0_1\,\,\, \cH^-_1)$ and $\cH_2= (\cH^+_2 \,\,\,\cH^0_2)$.

It should be noted that the non-diagonality in flavor comes
exclusively from the soft SUSY-breaking parameters, that could be
non-vanishing for $i \neq j$, namely: the masses $m_{\tilde Q \, ij}$
and $m_{\tilde L \, ij}$ 
for the sfermion $SU(2)$ doublets, the masses $m_{\tilde U_L \, ij}^2$,
$m_{\tilde U_R \, ij}^2$, $m_{\tilde D_L \, ij}^2$, 
$m_{\tilde D_R \, ij}^2$,  $m_{\tilde E \, ij}$ for the sfermion $SU(2)$ 
singlets and the trilinear couplings, ${\cal A}_{ij}^f$.   

In the sneutrino sector there is, correspondingly, a one-block $3\times
3$ mass matrix, that is referred to the $(\tinu_{eL}, \tinu_{\mu L},
\tinu_{\tau L})$ electroweak interaction basis: 
\begin{equation}
{\mathcal M}_{\tilde \nu}^2 =\left( \begin{array}{c}
M^2_{\tilde \nu \, LL}   
\end{array} \right),
\label{eq:sneu-3x3}
\end{equation} 
 where:
\begin{equation} 
  M_{\tilde \nu \, LL \, ij}^2 
  =   m_{\tilde L \, ij}^2 + \left( 
      \frac{1}{2} \MZ^2 \cos 2\beta \right) \delta_{ij},   
\label{eq:sneu-matrix}
\end{equation} 
 
It is important to note that due to $SU(2)_L$ gauge invariance
the same soft masses $m_{\tilde Q \, ij}$ enter in both up-type and
down-type squarks mass matrices similarly $m_{\tilde L \, ij}$ enter in
both the slepton and sneutrino $LL$ mass matrices. 
The soft SUSY-breaking parameters for the up-type squarks differ from
corresponding ones for down-type squarks by a rotation with CKM
matrix. The same would hold for sleptons i.e.\ the soft SUSY-breaking
parameters of the sneutrinos would differ from the corresponding ones
for charged sleptons by a rotation with the PMNS matrix. However, taking
the neutrino masses and oscillations 
into account in the SM leads to LFV effects that are extremely small. 
(For instance, in $\mu \to e \gamma$  they are of \order{10^{-47}} in case
of Dirac neutrinos with mass around 1~eV and maximal
mixing~\cite{Kuno:1999jp,DiracNu,MajoranaNu}, and of \order{10^{-40}} in case
of Majorana neutrinos~\cite{Kuno:1999jp,MajoranaNu}.) Consequently we do not
expect large effects from the inclusion of neutrino mass effects here
and neglect a rotation with the PMNS matrix. 
The sfermion mass matrices in terms of the $\deFABij$ are given as
\begin{equation}  
m^2_{\tilde U_L}= \left(\begin{array}{ccc}
 m^2_{\tilde Q_{1}} & \de_{12}^{QLL} m_{\tilde Q_{1}}m_{\tilde Q_{2}} & 
 \de_{13}^{QLL} m_{\tilde Q_{1}}m_{\tilde Q_{3}} \\
 \de_{21}^{QLL} m_{\tilde Q_{2}}m_{\tilde Q_{1}} & m^2_{\tilde Q_{2}}  & 
 \de_{23}^{QLL} m_{\tilde Q_{2}}m_{\tilde Q_{3}}\\
 \de_{31}^{QLL} m_{\tilde Q_{3}}m_{\tilde Q_{1}} & 
 \de_{32}^{QLL} m_{\tilde Q_{3}}m_{\tilde Q_{2}}& m^2_{\tilde Q_{3}} 
\end{array}\right)~,
\label{mUL}
\end{equation}
 
\noindent
\begin{equation}
m^2_{\tilde D_L}= V_{\rm CKM}^\dagger \, m^2_{\tilde U_L} \, V_{\rm CKM}~,
\label{mDL}
\end{equation}
 
\noindent 
\begin{equation}  
m^2_{\tilde U_R}= \left(\begin{array}{ccc}
 m^2_{\tilde U_{1}} & \de_{12}^{URR} m_{\tilde U_{1}}m_{\tilde U_{2}} & 
 \de_{13}^{URR} m_{\tilde U_{1}}m_{\tilde U_{3}}\\
 \de_{{21}}^{URR} m_{\tilde U_{2}}m_{\tilde U_{1}} & m^2_{\tilde U_{2}}  & 
 \de_{23}^{URR} m_{\tilde U_{2}}m_{\tilde U_{3}}\\
 \de_{{31}}^{URR}  m_{\tilde U_{3}} m_{\tilde U_{1}}& 
 \de_{{32}}^{URR} m_{\tilde U_{3}}m_{\tilde U_{2}}& m^2_{\tilde U_{3}} 
\end{array}\right)~,
\end{equation}

\noindent 
\begin{equation}  
m^2_{\tilde D_R}= \left(\begin{array}{ccc}
 m^2_{\tilde D_{1}} & \de_{12}^{DRR} m_{\tilde D_{1}}m_{\tilde D_{2}} & 
 \de_{13}^{DRR} m_{\tilde D_{1}}m_{\tilde D_{3}}\\
 \de_{{21}}^{DRR} m_{\tilde D_{2}}m_{\tilde D_{1}} & m^2_{\tilde D_{2}}  & 
 \de_{23}^{DRR} m_{\tilde D_{2}}m_{\tilde D_{3}}\\
 \de_{{31}}^{DRR}  m_{\tilde D_{3}} m_{\tilde D_{1}}& 
 \de_{{32}}^{DRR} m_{\tilde D_{3}}m_{\tilde D_{2}}& m^2_{\tilde D_{3}} 
\end{array}\right)~,
\end{equation}

\noindent 
\begin{equation}
v_2 {\cal A}^u  =\left(\begin{array}{ccc}
 m_u A_u & \de_{12}^{ULR} m_{\tilde Q_{1}}m_{\tilde U_{2}} & 
 \de_{13}^{ULR} m_{\tilde Q_{1}}m_{\tilde U_{3}}\\
 \de_{{21}}^{ULR}  m_{\tilde Q_{2}}m_{\tilde U_{1}} & 
 m_c A_c & \de_{23}^{ULR} m_{\tilde Q_{2}}m_{\tilde U_{3}}\\
 \de_{{31}}^{ULR}  m_{\tilde Q_{3}}m_{\tilde U_{1}} & 
 \de_{{32}}^{ULR} m_{\tilde Q_{3}} m_{\tilde U_{2}}& m_t A_t 
\end{array}\right)~,
\label{v2Au}
\end{equation}

\noindent 
\begin{equation}
v_1 {\cal A}^d  =\left(\begin{array}{ccc}
 m_d A_d & \de_{12}^{DLR} m_{\tilde Q_{1}}m_{\tilde D_{2}} & 
 \de_{13}^{DLR} m_{\tilde Q_{1}}m_{\tilde D_{3}}\\
 \de_{{21}}^{DLR}  m_{\tilde Q_{2}}m_{\tilde D_{1}} & m_s A_s & 
 \de_{23}^{DLR} m_{\tilde Q_{2}}m_{\tilde D_{3}}\\
 \de_{{31}}^{DLR}  m_{\tilde Q_{3}}m_{\tilde D_{1}} & 
 \de_{{32}}^{DLR} m_{\tilde Q_{3}} m_{\tilde D_{2}}& m_b A_b 
\end{array}\right)~.
\label{v1Ad}
\end{equation}

\noindent \begin{equation}  
m^2_{\tilde L}= \left(\begin{array}{ccc}
 m^2_{\tilde L_{1}} & \delta_{12}^{LLL} m_{\tilde L_{1}}m_{\tilde L_{2}} & \delta_{13}^{LLL} m_{\tilde L_{1}}m_{\tilde L_{3}} \\
 \delta_{21}^{LLL} m_{\tilde L_{2}}m_{\tilde L_{1}} & m^2_{\tilde L_{2}}  & \delta_{23}^{LLL} m_{\tilde L_{2}}m_{\tilde L_{3}}\\
\delta_{31}^{LLL} m_{\tilde L_{3}}m_{\tilde L_{1}} & \delta_{32}^{LLL} m_{\tilde L_{3}}m_{\tilde L_{2}}& m^2_{\tilde L_{3}} \end{array}\right)\end{equation}

\noindent \begin{equation}
v_1 {\cal A}^l  =\left(\begin{array}{ccc}
m_e A_e & \delta_{12}^{ELR} m_{\tilde L_{1}}m_{\tilde E_{2}} & \delta_{13}^{ELR} m_{\tilde L_{1}}m_{\tilde E_{3}}\\
\delta_{21}^{ELR}  m_{\tilde L_{2}}m_{\tilde E_{1}} & m_\mu A_\mu & \delta_{23}^{ELR} m_{\tilde L_{2}}m_{\tilde E_{3}}\\
\delta_{31}^{ELR}  m_{\tilde L_{3}}m_{\tilde E_{1}} & \delta_{32}^{ELR}  m_{\tilde L_{3}} m_{\tilde E_{2}}& m_{\tau}A_{\tau}\end{array}\right)\label{v1Al}\end{equation}

\noindent \begin{equation}  
m^2_{\tilde E}= \left(\begin{array}{ccc}
 m^2_{\tilde E_{1}} & \delta_{12}^{ERR} m_{\tilde E_{1}}m_{\tilde E_{2}} & \delta_{13}^{ERR} m_{\tilde E_{1}}m_{\tilde E_{3}}\\
 \delta_{21}^{ERR} m_{\tilde E_{2}}m_{\tilde E_{1}} & m^2_{\tilde E_{2}}  & \delta_{23}^{ERR} m_{\tilde E_{2}}m_{\tilde E_{3}}\\
\delta_{31}^{ERR}  m_{\tilde E_{3}} m_{\tilde E_{1}}& \delta_{32}^{ERR} m_{\tilde E_{3}}m_{\tilde E_{2}}& m^2_{\tilde E_{3}} \end{array}\right)\end{equation}

In all this work, for simplicity, we are assuming that all $\deFABij$
parameters are real, therefore, hermiticity of 
${\mathcal M}_{\tilde Q}^2$, ${\mathcal M}_{\tilde l}^2$ and 
${\mathcal M}_{\tilde \nu}^2$ implies $\delta_{ij}^{FAB}= \delta_{ji}^{FBA}$.

The next step is to rotate the squark states from the Super-CKM basis, 
${\tilde q}_{L,R}$, to the physical basis. 
If we set the order in the Super-CKM basis as above, 
$(\SupL, \SchaL, \StopL, \SupR, \SchaR, \StopR)$ and  
$(\SdownL, \SstrL, \SbotL, \SdownR, \SstrR, \SbotR)$, 
and in the physical basis as
${\tilde u}_{1,..6}$ and ${\tilde d}_{1,..6}$, respectively, these last
rotations are given by two $6 \times 6$ matrices, $R^{\tilde u}$ and
$R^{\tilde d}$,  
\BE
\VL  \tiu_{1} \\ \tiu_{2}  \\ \tiu_{3} \\
                                    \tiu_{4}   \\ \tiu_{5}  \\\tiu_{6}   \VR
  \; = \; R^{\tiu}  \VL \SupL \\ \SchaL \\\StopL \\ 
  \SupR \\ \SchaR \\ \StopR \VR ~,~~~~
\VL  \tid_{1} \\ \tid_{2}  \\  \tid_{3} \\
                                   \tid_{4}     \\ \tid_{5} \\ \tid_{6}  \VR             \; = \; R^{\tid}  \VL \SdownL \\ \SstrL \\ \SbotL \\
                                      \SdownR \\ \SstrR \\ \SbotR \VR ~,
\label{newsquarks}
\end{equation} 
yielding the diagonal mass-squared matrices for squarks as follows,
\BEA
{\rm diag}\{m_{\tiu_1}^2, m_{\tiu_2}^2, 
          m_{\tiu_3}^2, m_{\tiu_4}^2, m_{\tiu_5}^2, m_{\tiu_6}^2 
           \}  & = &
R^{\tiu}  \;  {\cal M}_{\tiu}^2   \; 
 R^{\tiu \dagger}    ~,\\
{\rm diag}\{m_{\tid_1}^2, m_{\tid_2}^2, 
          m_{\tid_3}^2, m_{\tid_4}^2, m_{\tid_5}^2, m_{\tid_6}^2 
          \}  & = &
R^{\tid}  \;   {\cal M}_{\tid}^2   \; 
 R^{\tid \dagger}    ~.
\EEA 

Similarly we need to rotate the sleptons and sneutrinos from the electroweak interaction basis to the physical mass eigenstate basis, 
\BE
\VL  \til_{1} \\ \til_{2}  \\ \til_{3} \\
                                    \til_{4}   \\ \til_{5}  \\\til_{6}   \VR
  \; = \; R^{\til}  \VL \SelL \\ \SmuL \\\StauL \\ 
  \SelR \\ \SmuR \\ \StauR \VR ~,~~~~
\VL  \tinu_{1} \\ \tinu_{2}  \\  \tinu_{3}  \VR             \; = \; R^{\tinu}  \VL \tinu_{eL} \\ \tinu_{\mu L}  \\  \tinu_{\tau L}   \VR ~,
\label{rotsquarks}
\end{equation} 
with $R^{\til}$ and $R^{\tinu}$ being the respective $6\times 6$ and
$3\times 3$ unitary rotating matrices that yield the diagonal
mass-squared matrices as follows, 
\BEA
{\rm diag}\{m_{\til_1}^2, m_{\til_2}^2, 
          m_{\til_3}^2, m_{\til_4}^2, m_{\til_5}^2, m_{\til_6}^2 
           \}  & = &
R^{\til}  \;  {\cal M}_{\til}^2   \; 
 R^{\til \dagger}    ~,\\
{\rm diag}\{m_{\tinu_1}^2, m_{\tinu_2}^2, 
          m_{\tinu_3}^2  
          \}  & = &
R^{\tinu}  \;   {\cal M}_{\tinu}^2   \; 
 R^{\tinu \dagger}    ~.
\EEA

%% file: Comp_Setup.tex
\section{Computational setup}
\label{sec:CompSetup}

Here we briefly describe our numerical set-up. We first give some
details on the running from the GUT to the EW scale, and subsequently
describe the calculations of the observables evaluated at the EW scale.

\subsection{From the GUT scale to the EW scale}
\label{sec:GUTEW}

The SUSY spectra have been generated with the code 
{\tt SPheno 3.2.4}~\cite{Porod:2003um} (for the CMSSM and the \CMSSMI). 
We defined the SLHA~\cite{SLHA} file at the GUT scale. 
In a first step within {\tt SPheno}, gauge and
Yukawa couplings at $\MZ$ scale are calculated using tree-level
formulas. Fermion masses, the $Z$~boson pole mass, the fine structure constant
$\alpha$, the Fermi constant $G_F$ and the strong coupling constant
$\alpha_s(\MZ)$ are used as input parameters. The gauge and Yukawa
couplings, calculated at $\MZ$, are then used as 
input for the one-loop
RGE's to obtain the corresponding values at the GUT scale
which is calculated from the requirement that $g_1 = g_2$
(where $g_{1,2}$ denote the gauge couplings of the $U(1)$ and
$SU(2)$, respectively). The CMSSM boundary
conditions are then applied to the complete set of
two-loop RGE's and are evolved to the EW scale.  
At this point the SM and SUSY radiative
corrections are applied to the gauge and Yukawa couplings, and the 
two-loop RGE's are again evolved to GUT scale. 
After applying the CMSSM  boundary conditions again
the two-loop RGE's are run down to EW scale to get SUSY spectrum. This
procedure is iterated until the required precision is achieved. 
The output is then written in the form of an SLHA, file which is used as 
input to calculate low energy observables discussed below. 

For the \CMSSMI\ a similar procedure is applied, where the neutrino
related input parameters are
included in the respective SLHA input blocks (see \citere{SLHA} for
details), (the relevant numerical values are given in
\refse{sec:cmssmI}). 
For our scans of the \CMSSMI\ parameter space we use 
{\tt SPheno 3.2.4}~\cite{Porod:2003um} with
the model ``see-saw type-I''. The value for $Y_\nu$ is  
implemented as explained in \refse{sec:cmssmI}, adjusting the matrix
elements such that neutrino experimental parameters achieve the
desired results after RGE's. The predictions for $\br(l_i \to l_j \gamma)$ 
are also obtained with {\tt SPheno 3.2.4} , 
see the discussion in \refse{sec:Sl}.  
We checked that the use of this code
produces results similar to the ones obtained by our private codes used
in \citere{Cannoni:2013gq}.


\subsection{Calculations at the EW scale}
\label{sec:EWcalc}
 
Here we briefly review the various observables that we compute at the EW
scale, either taking the non-zero $\deFABij$ into account, or setting
them to zero.


\subsubsection{The MSSM Higgs sector}
\label{sec:higgs}

The MSSM Higgs sector consist of two Higgs
doublets and predicts five physical
Higgs bosons, the light and heavy $\cp$-even $h$ and $H$, the $\cp$-odd $A$,
and the charged Higgs boson, $H^\pm$. At tree-level the Higgs sector is
described with the help of two parameters: the mass of the $A$~boson, $\MA$,
and $\tb := v_2/v_1$, the ratio of the two vacuum expectation values.
The tree-level relations receive large higher-order corrections, see,
e.g., \citere{MHreviews} and references therein.

The lightest MSSM Higgs boson, with mass $\Mh$, can be interpreted as
the new state discovered at the LHC around $\sim 125 \gev$. 
The present experimental uncertainty at the LHC for $\Mh$, 
is about~\cite{Aad:2014aba,CMS:2014ega},
\begin{align}
\de\Mh^{\rm exp,today} \sim 200 \mev~.
\end{align}
This can possibly be reduced below the level of 
\begin{align}
\de\Mh^{\rm exp,future} \lsim 50 \mev
\end{align}
at the ILC~\cite{dbd}.  Similarly, for the masses of the heavy neutral Higgs 
$\MH$ and charged Higgs boson $\MHp$, an uncertainty at the $1\%$ level 
could be expected at the LHC~\cite{cmsHiggs}. 

Effects of sfermion mixing in the MSSM Higgs sector has already been calculated
in a model independent way in the scalar quark
sector~\cite{delrhoNMFV,arana,arana-NMFV2} and in the scalar lepton sector
~\cite{drhoLFV} and there are sizable corrections to higgs boson masses
specially to the charged higgs  boson mass $\MHp$, assuming general NMFV
in the squark and slepton sector.
 
\medskip
In the Feynman diagrammatic approach that we are following here, the
higher-order corrected  $\cp$-even Higgs boson masses are derived by finding the
poles of the $(h,H)$-propagator 
matrix. The inverse of this matrix is given by
\BE
\left(\Delta_{\rm Higgs}\right)^{-1}
= - i \ML p^2 -  \mHtree^2 + \hSi_{HH}(p^2) &  \hSi_{hH}(p^2) \\
     \hSi_{hH}(p^2) & p^2 -  \mhtree^2 + \hSi_{hh}(p^2) \MR~.
\label{higgsmassmatrixnondiag}
\end{equation}
Determining the poles of the matrix $\Delta_{\rm Higgs}$ in
\refeq{higgsmassmatrixnondiag} is equivalent to solving
the equation
\begin{equation}
\left[p^2 - \mhtree^2 + \hSi_{hh}(p^2) \right]
\left[p^2 - \mHtree^2 + \hSi_{HH}(p^2) \right] -
\left[\hSi_{hH}(p^2)\right]^2 = 0\,.
\label{eq:proppole}
\end{equation}

Similarly, in the case of the charged Higgs sector, the corrected Higgs
mass is derived  by the position of the pole in the charged Higgs
propagator, which is defined by:  
\noindent \begin{equation}
p^{2}-m^{2}_{H^{\pm},{\rm tree}} +
\hat{\Sigma}_{H^{-}H^{+}}\left(p^{2}\right)=0.
\label{eq:proppolech}
\end{equation}

The flavor violating parameters enter into the one-loop prediction of
the various (renormalized) Higgs-boson self-energies, where details can
be found in \citeres{arana,arana-NMFV2,drhoLFV}. Numerically the results
have been obtained using the code 
\fh~\cite{feynhiggs,mhiggslong,mhiggsAEC,mhcMSSMlong,Mh-logresum}, 
which contains the complete set of one-loop corrections from (flavor
violating) squark and slepton contributions (based on
\citeres{delrhoNMFV,arana,drhoLFV}). Those are supplemented with leading
and sub-leading two-loop corrections as well as a resummation of leading
and sub-leading logarithmic contributions from the $t/\Stop$ sector, all
evaluated in the flavor conserving MSSM.


\subsubsection{Electroweak precision observables}
\label{sec:ewpo}

EWPO that are known with an accuracy at the per-mille level or better 
have the potential to allow a discrimination between quantum effects of 
the SM and SUSY models, see \citere{PomssmRep} for a review.  Examples 
are the $W$-boson mass $\MW$ and the $Z$-boson observables, such as the 
effective leptonic weak mixing angle $\sweff$, whose present 
experimental uncertainties are \cite{LEPEWWG}
\begin{equation}
\label{eq:EWPO-today}
\de\MW^{\text{exp,today}} \sim 15 \mev, \quad
\de\sweff^{\text{exp,today}} \sim 15 \times 10^{-5}\,, 
\end{equation}
The experimental uncertanity will further be reduced~\cite{Baak:2013fwa} 
to
\begin{equation}
\label{eq:EWPO-future}
\de\MW^{\text{exp,future}} \sim 4\mev, \quad
\de\sweff^{\text{exp,future}} \sim 1.3 \times 10^{-5}\,,
\end{equation}
at the ILC and at the GigaZ option of the ILC, respectively. 
Even higher precision could be expected from the FCC-ee, see, e.g.,
\citere{fcc-ee-paris}.

The $W$-boson mass can be evaluated from
\begin{equation}
\MW^2 \KL 1 - \frac{\MW^2}{\MZ^2} \KR = 
\frac{\pi \al}{\wz \Gmu} (1 + \De r)
\end{equation}
where $\al$ is the fine-structure constant and $\Gmu$ the Fermi 
constant.  This relation arises from comparing the prediction for muon 
decay with the experimentally precisely known Fermi constant.  
The one-loop contributions to $\De r$ can be written as
\begin{equation}
\De r = \De\al - \frac{\cw^2}{\sw^2}\De\rho + (\De r)_{\text{rem}},
\end{equation}
where $\De\al$ is the shift in the fine-structure constant due to the 
light fermions of the SM, $\De\al \propto \log(\MZ/m_f)$, and $\De\rho$ 
is the leading contribution to the $\rho$ parameter~\cite{rho} from 
(certain) fermion and sfermion loops (see below).  The remainder part 
$(\De r)_{\text{rem}}$ contains in particular the contributions from the 
Higgs sector.

\medskip
The effective leptonic weak mixing angle at the $Z$-boson resonance, 
$\sweff$, is defined through the vector and axial-vector couplings 
($g_{\text{V}}^\ell$ and $g_{\text{A}}^\ell$) of leptons ($\ell$) to the 
$Z$~boson, measured at the $Z$-boson pole.  If this vertex is written as 
$i\bar \ell\ga^\mu (g_{\text{V}}^\ell - g_{\text{A}}^\ell \ga_5) \ell 
Z_\mu$ then
\begin{equation}
\sweff = \frac 14 \KL 1 -
  \re\frac{g_{\text{V}}^\ell}{g_{\text{A}}^\ell}\KR\,.
\end{equation}
Loop corrections enter through higher-order contributions to 
$g_{\text{V}}^\ell$ and $g_{\text{A}}^\ell$.

\medskip
Both of these (pseudo-)observables are affected by shifts in the 
quantity $\De\rho$ according to
\begin{equation}
\label{eq:precobs}
\De\MW \approx \frac{\MW}{2}\frac{\cw^2}{\cw^2 - \sw^2} \De\rho\,, \quad
\De\sweff \approx - \frac{\cw^2 \sw^2}{\cw^2 - \sw^2} \De\rho\,.
\end{equation}
The quantity $\De\rho$ is defined by the relation
\begin{equation}
\De\rho = \frac{\Si_Z^{\text{T}}(0)}{\MZ^2} -
          \frac{\Si_W^{\text{T}}(0)}{\MW^2}
\end{equation} 
with the unrenormalized transverse parts of the $Z$- and $W$-boson 
self-energies at zero momentum, $\Si_{Z,W}^{\text{T}}(0)$.  It 
represents the leading universal corrections to the electroweak 
precision observables induced by mass splitting between partners in 
isospin doublets~\cite{rho}. Consequently, it is sensitive to the 
mass-splitting effects induced by flavor mixing.
The effects from flavor violation in the squark and slepton sector,
entering via $\De\rho$ have been evaluated in
\citeres{delrhoNMFV,drhoLFV} and included in \fh. 
In particular, in \citere{delrhoNMFV} it has been shown that for the
squark contributions $\De\rho$ constitutes an excellent approximation to
$\De r$. We use \fh\ for our numerical evaluation.


\subsubsection{\boldmath{$B$}-physics observables}
\label{sec:bpo}

 We also calculate several $B$-physics observables (BPO):
\bsg, \bmm\ and \dmbs. 
Concerning \bsg\ included in the calculation are the most 
relevant loop contributions to the Wilson coefficients:
(i)~loops with Higgs bosons (including the resummation of large $\tb$
effects~\cite{Isidori:2002qe}),  
(ii)~loops with charginos and 
(iii)~loops with gluinos. 
For \bmm\ there are three types of relevant one-loop corrections
contributing to the relevant Wilson coefficients:
(i)~Box diagrams, 
(ii)~$Z$-penguin diagrams and 
(iii)~neutral Higgs boson $\phi$-penguin diagrams, where $\phi$ denotes the
three neutral MSSM Higgs bosons, $\phi = h, H, A$ (again large resummed
$\tb$ effects have been taken into account).
In our numerical evaluation there are included
what are known to be the dominant contributions to
these three types of diagrams \cite{Chankowski:2000ng}: chargino
contributions to box and $Z$-penguin diagrams and chargino and gluino
contributions to $\phi$-penguin diagrams.   
Concerning \dmbs, in the MSSM 
there are in general three types of one-loop diagrams that contribute:
(i)~Box diagrams, 
(ii)~$Z$-penguin diagrams and 
(iii)~double Higgs-penguin diagrams (again including the resummation of
large $\tb$ enhanced effects).
In our numerical evaluation there are included again what are known to be the
dominant contributions to these three types of diagrams in scenarios
with non-minimal flavor violation (for a review see, for instance,
\cite{Foster:2005wb}): gluino contributions to box 
diagrams, chargino contributions to box and $Z$-penguin diagrams, and 
chargino and gluino contributions to double $\phi$-penguin diagrams. 
More details about the calculations employed can be
found in \citeres{arana,arana-NMFV2}.
We perform our numerical calculation with the {\tt BPHYSICS} subroutine
taken from the {\tt SuFla} code~\cite{sufla} (with some additions and
improvements as detailed in \citeres{arana,arana-NMFV2}), which has
been implemented as a subroutine into (a private version of) \fh.
The Present experimental status and SM prediction of these observables
is given in the 
\refta{tab:ExpStatus-BPO}~\cite{hfag:rad,Misiak:2009nr,Chatrchyan:2013bka,Aaij:2013aka,Buras:2012ru,hfag:pdg,Buras:1990fn,Golowich:2011cx}.

\begin{table}[htb!]
\BC
\renewcommand{\arraystretch}{1.5}
\begin{tabular}{|c|c|c|}
\hline
Observable & Experimental Value & SM Prediction  \\\hline
\bsg & $3.43\pm 0.22 \times 10^{-4}$ & $3.15\pm 0.23 \times 10^{-4}$ \\
\bmm & $(3.0)^{+1.0}_{-0.9} \times 10^{-9}$ & $3.23\pm 0.27 \times 10^{-9}$ \\
\dmbs & $116.4\pm 0.5 \times 10^{-10} \mev $ 
      & $(117.1)^{+17.2}_{-16.4} \times 10^{-10} \mev$   \\
\hline
\end{tabular}
\caption{Present experimental status of $B$-physics observables with
their SM prediction.}  
\label{tab:ExpStatus-BPO}
\renewcommand{\arraystretch}{1.0}
\EC
\end{table}

%% file: Results.tex
\section{Numerical Results}
\label{sec:NResults}

\subsection{Effects of squark mixing in the CMSSM}
\label{sec:Sq}

In this section we analyze the effects from RGE induced flavor violating
mixing in the scalar quark sector in the CMSSM (i.e.\ with no mixing in
the slepton sector). The RGE running from the GUT scale to the EW has
been performed as described in \refse{sec:GUTEW}, with the subsequent
evaluation of the low-energy observables as discussed in
\refse{sec:EWcalc}. 
In order to get an overview about the size of the effects in the CMSSM
parameter space, the relevant parameters $m_0$, $m_{1/2}$ have been
scanned as, or in case of $A_0$ and
$\tb$ have been set to all combinations of 
\begin{align}
m_0 &\eq 500 \gev \ldots 5000 \gev~, \\
m_{1/2} &\eq 1000 \gev \ldots 3000 \gev~, \\
A_0 &\eq -3000, -2000, -1000, 0 \gev~, \\
\tb &\eq 10, 20, 35, 45~,
\end{align}
with $\mu > 0$.  
Primarily we are not interested in the absolute values for all these observables but the
effects that comes from flavor violation within the MFV framework,
i.e.\ the effect from the off-diagonal entries in the sfermion mass matrices.
We first calculate the low-energy observables by setting all $\deFABij=0$
by hand. In a second step we evaluate the observables with the values of 
$\deFABij$ obtained through RGE running. We then evaluate the ``pure
MFV effects'', 
\begin{align}
\Dbsg &\eq \br(B \to X_s \gamma) - \br^{\rm MSSM})(B \to X_s \gamma)~, \\
\Dbmm &\eq \br(B_{s} \to \mu^+ \mu^-) 
         - \br^{\rm MSSM}(B_{s} \to \mu^+ \mu^-)~, \\
\Ddmbs &\eq \De M_{B_s} - \De M_{B_s}^{\rm MSSM}~,
\end{align}
where $\br^{\rm MSSM}(B \to X_s \gamma)$, 
$\br^{\rm MSSM}(B_s \to \mu^+ \mu^-)$ and $\De M_{B_{S}}^{\rm MSSM}$
corresponds to the  values of relevant observables with all $\deFABij =
0$. Furthermore we use 
\begin{align}
\DMh &\eq \Mh - \Mh^{\rm MSSM} \\
\DMH &\eq \MH - \MH^{\rm MSSM} \\
\DMHp &\eq \MHp - \MHp^{\rm MSSM}
\end{align}
where $\Mh^{\rm MSSM}$, $\MH^{\rm MSSM}$ and 
$\MHp^{\rm MSSM}$ corresponds to the 
Higgs masses with all $\deFABij = 0$.  Similarly we use for the EWPO
\begin{align}
\Drho &\eq \De\rho-\De\rho^{\rm MSSM} \\
\DMW &\eq \MW-\MW^{\rm MSSM} \\
\Dsweff &\eq \sweff-\sweff^{\rm MSSM}
\end{align}
where $\De\rho^{\rm MSSM}$, $\MW^{\rm MSSM}$ and $\sweff^{\rm MSSM}$ are the
values of the relavant observables with all $\deFABij = 0$. 

\begin{figure}[ht!]
\begin{center}
\vspace{3.0cm}
\psfig{file=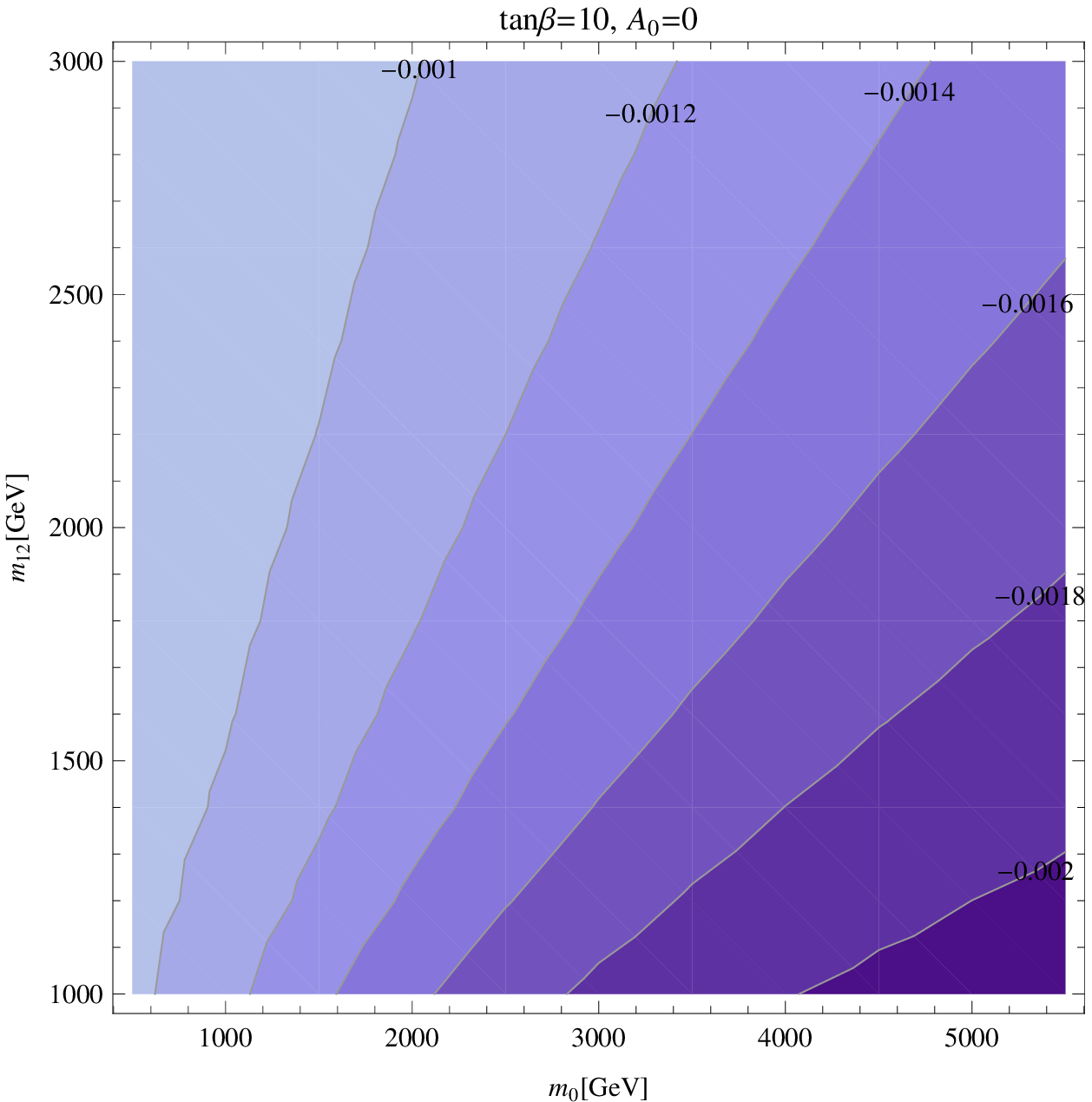  ,scale=0.57,angle=0,clip=}
\psfig{file=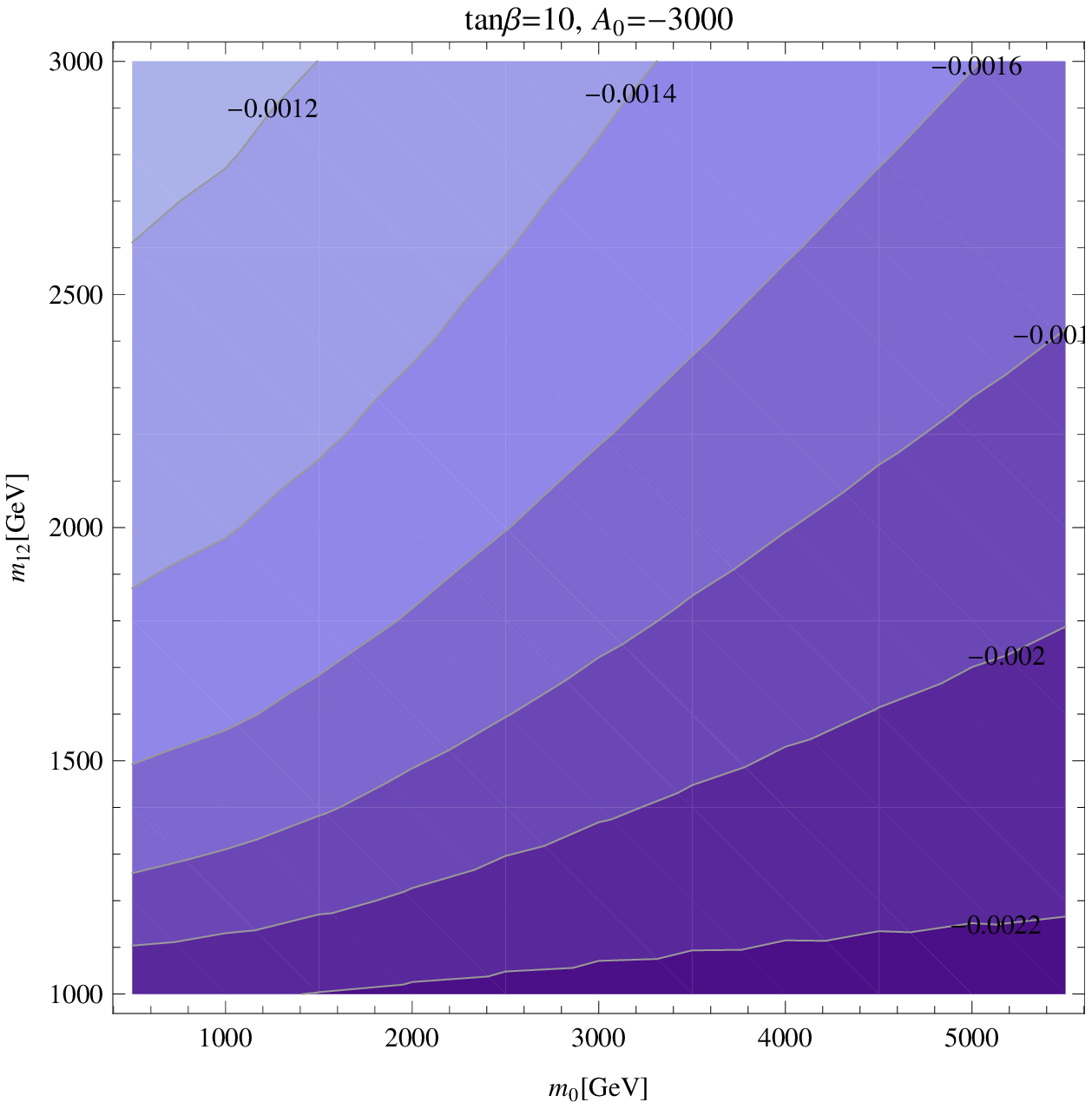  ,scale=0.57,angle=0,clip=}\\
\vspace{2.0cm}
\psfig{file=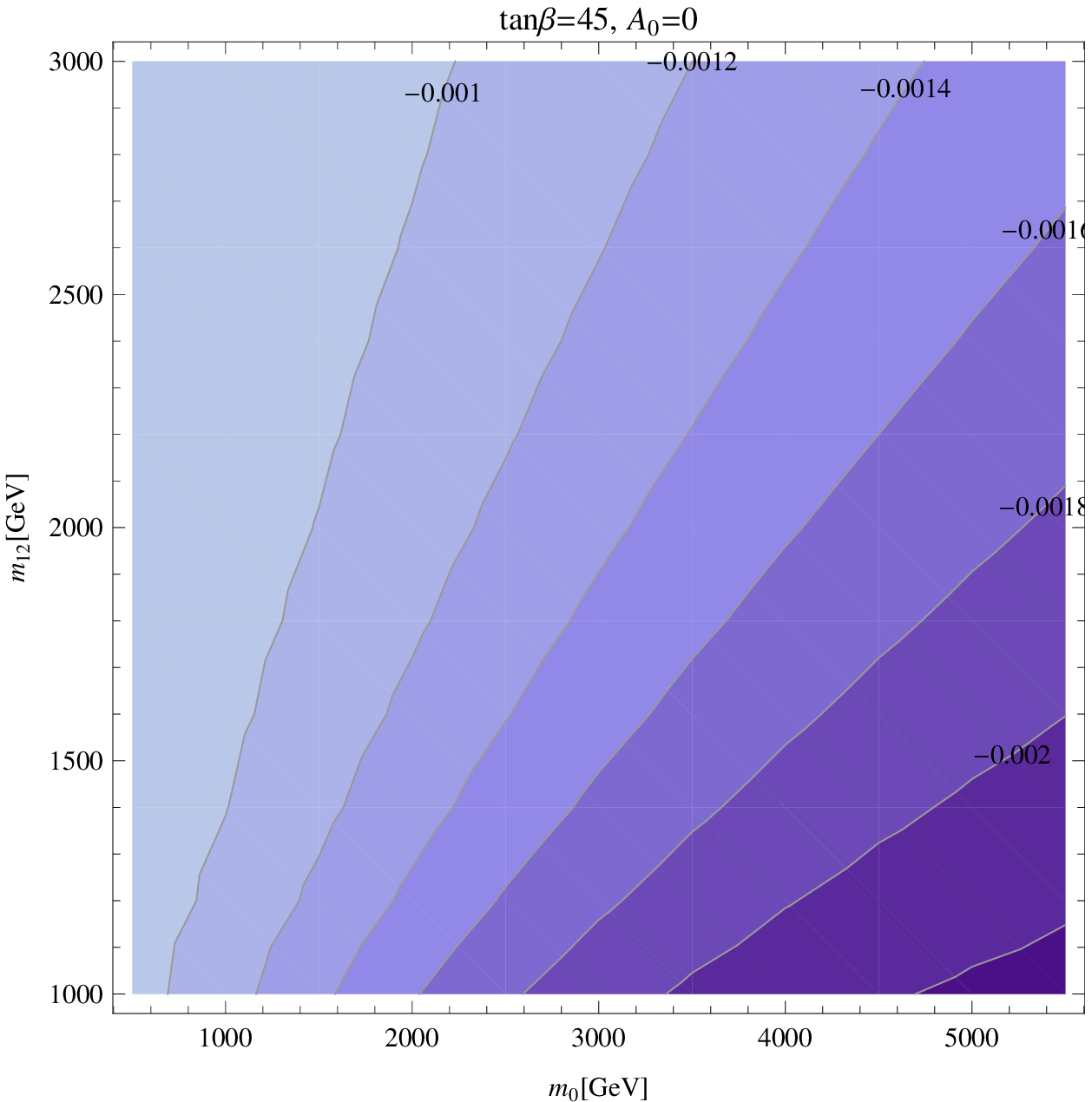 ,scale=0.56,angle=0,clip=}
\psfig{file=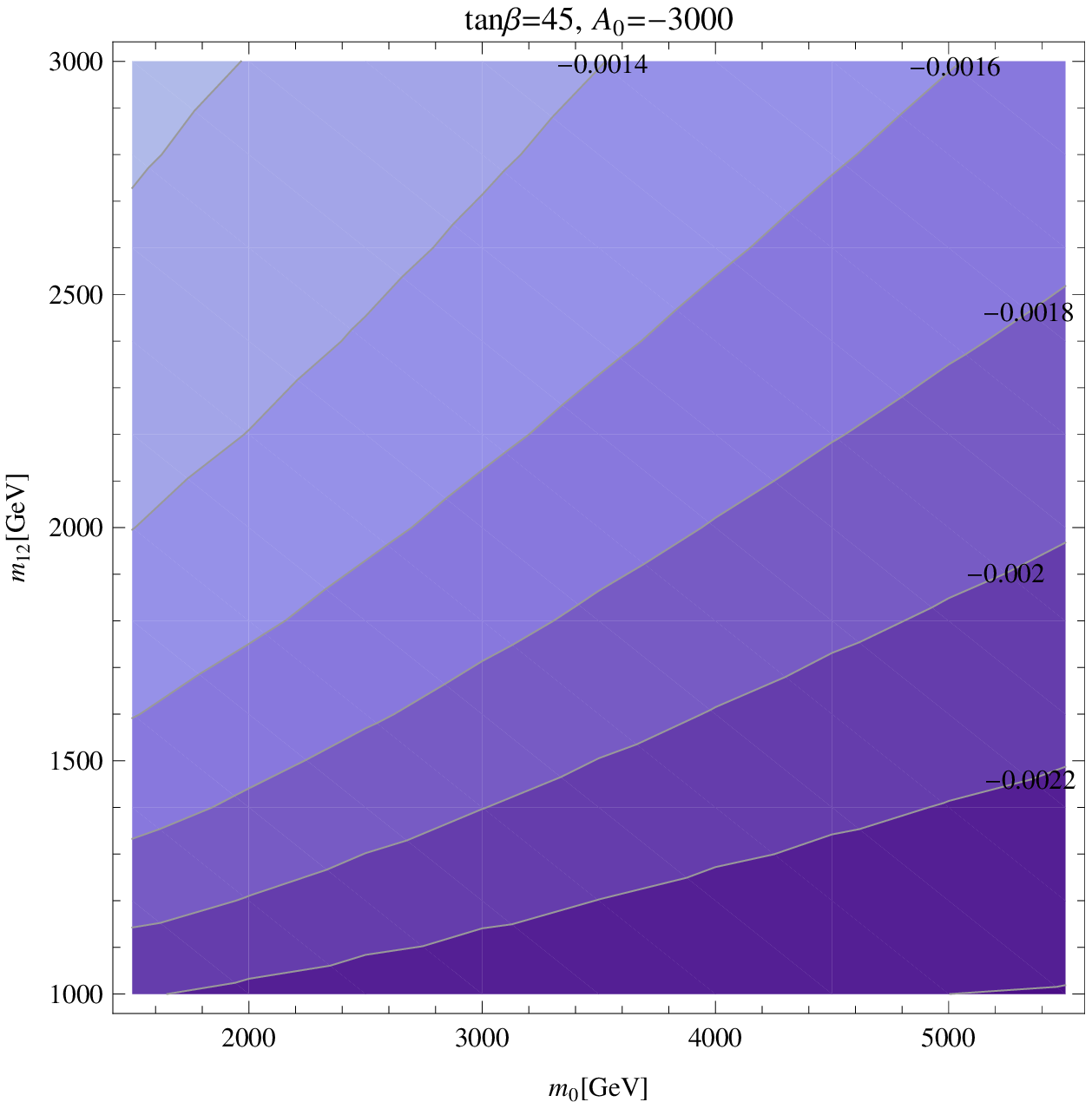   ,scale=0.56,angle=0,clip=}\\
\vspace{0.2cm}
\end{center}
\caption{Contours of $\delta^{QLL}_{13}$  in the
  $m_0$--$m_{1/2}$ plane for different values of $\tb$ and $A_0$ in
  the CMSSM.}   
\label{fig:DelQLL13}
\end{figure}

\begin{figure}[ht!]
\begin{center}
\vspace{3.0cm}
\psfig{file=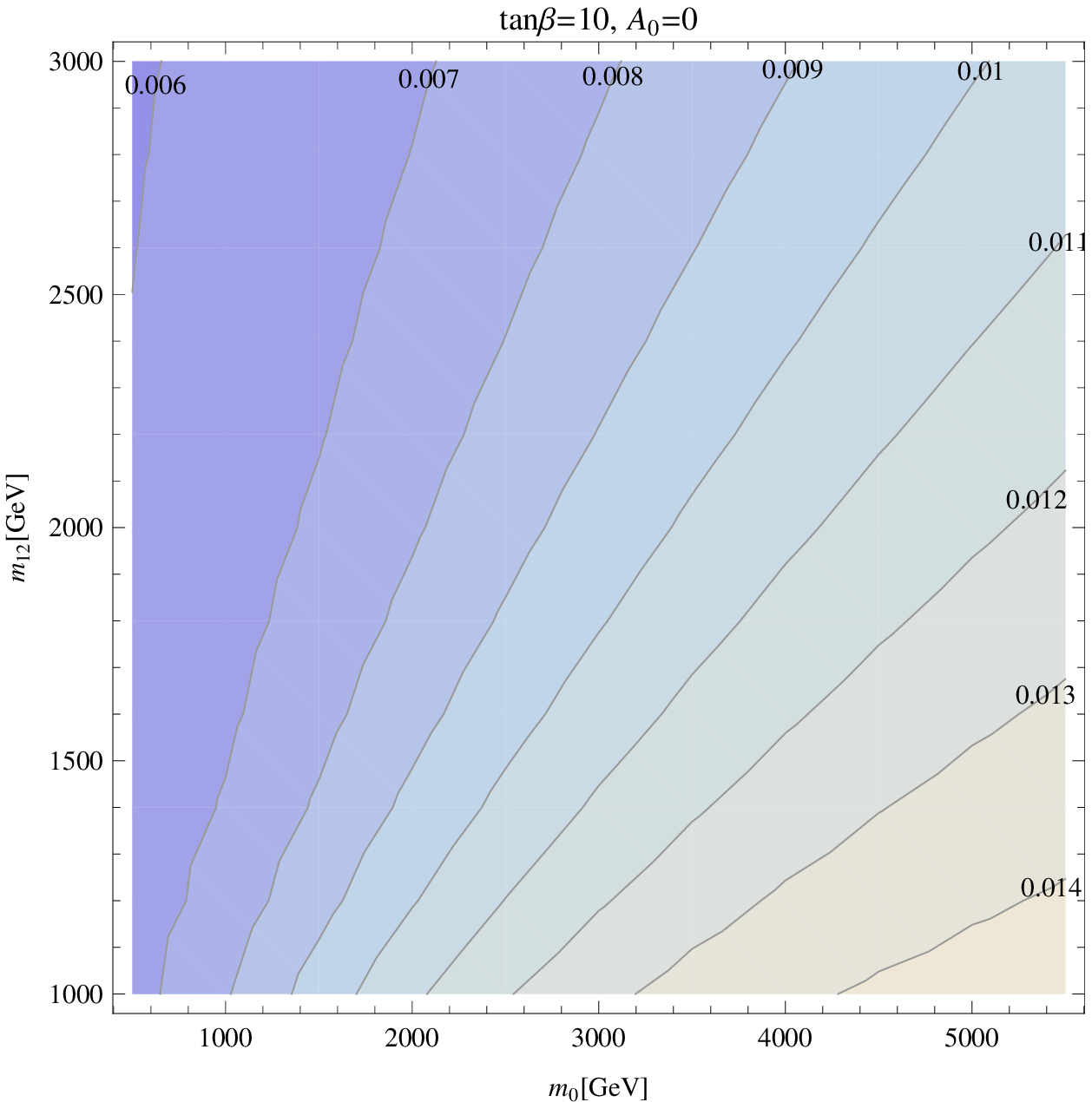  ,scale=0.57,angle=0,clip=}
\psfig{file=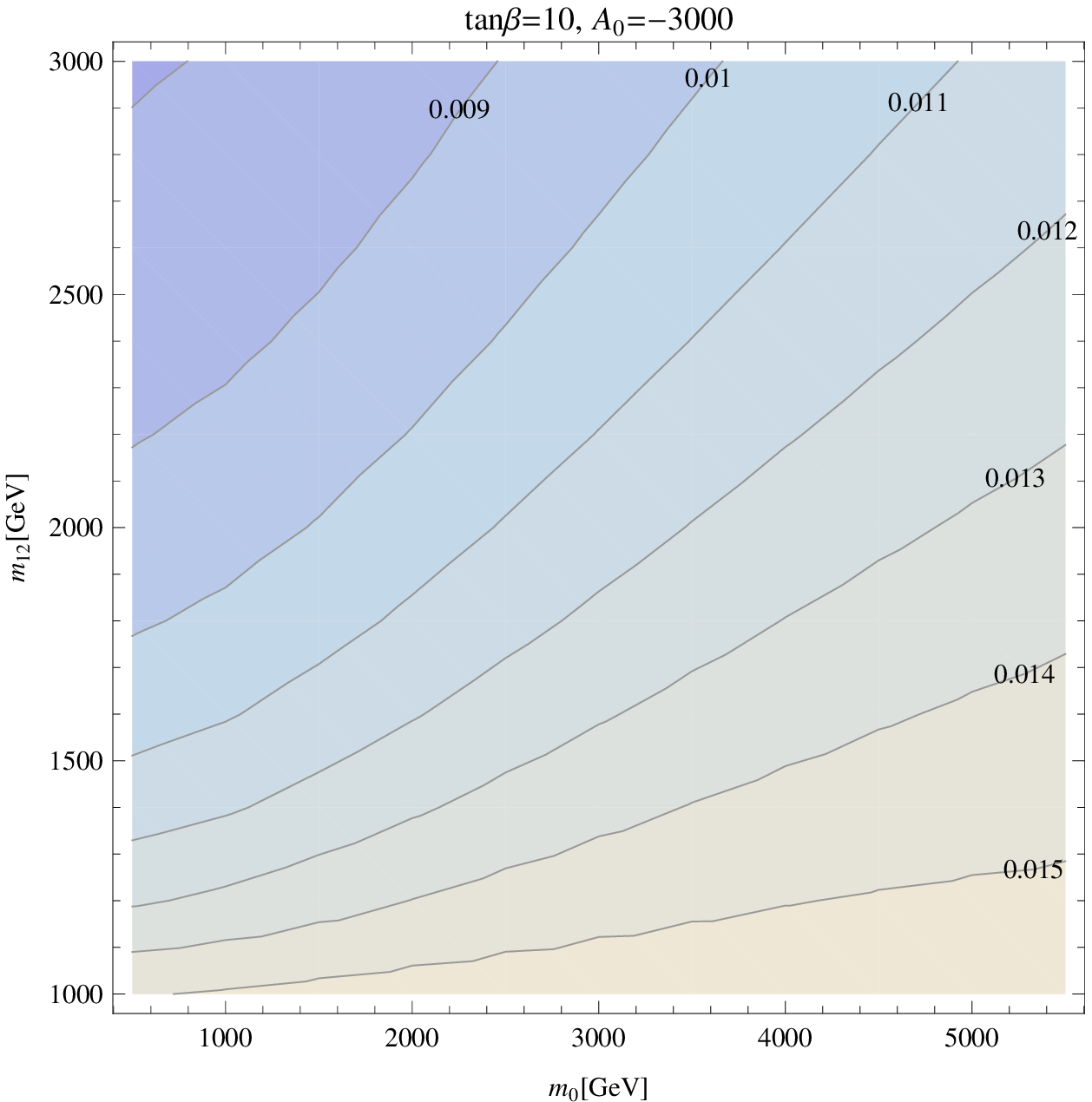  ,scale=0.57,angle=0,clip=}\\
\vspace{2.cm}
\psfig{file=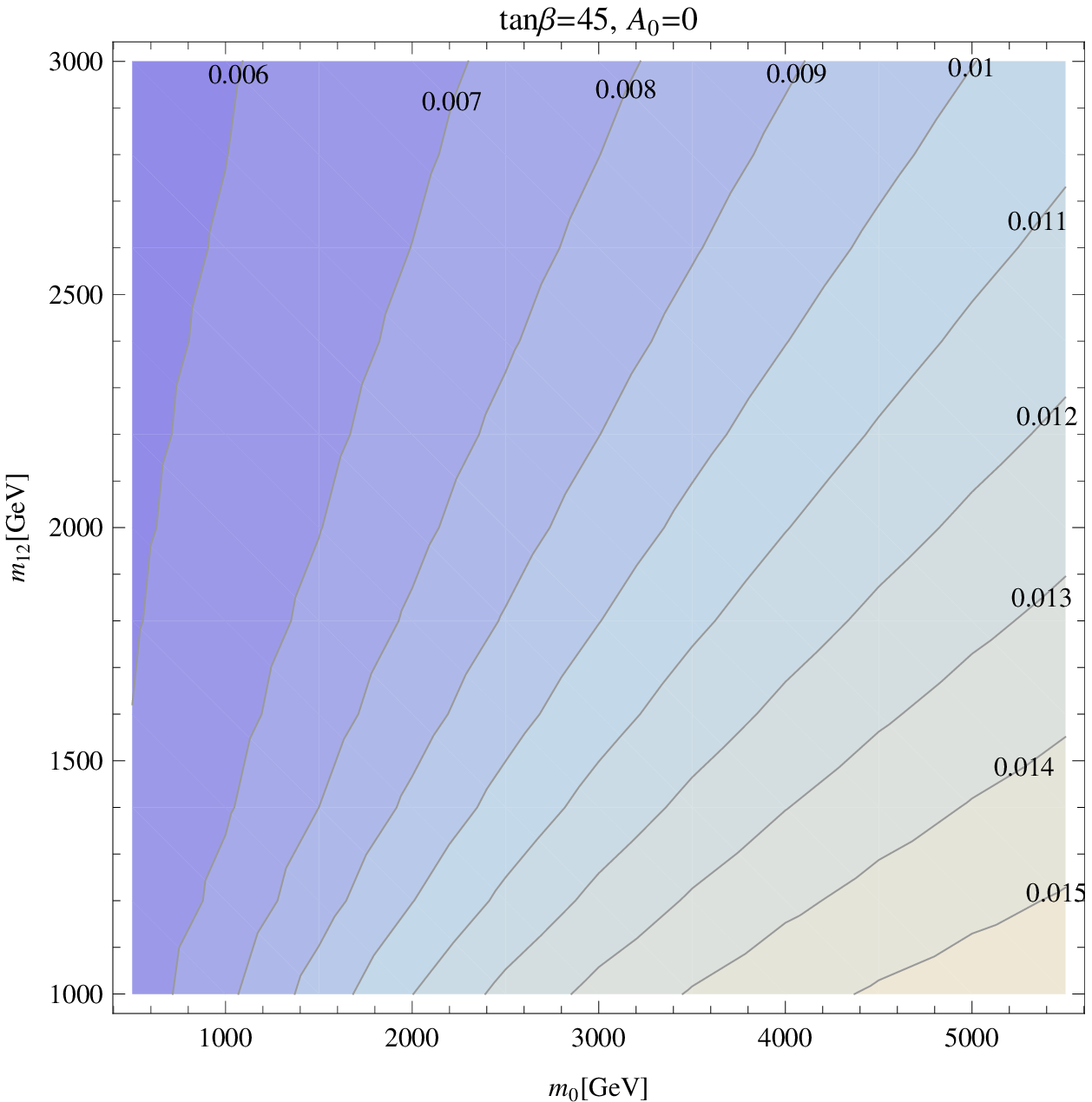 ,scale=0.56,angle=0,clip=}
\psfig{file=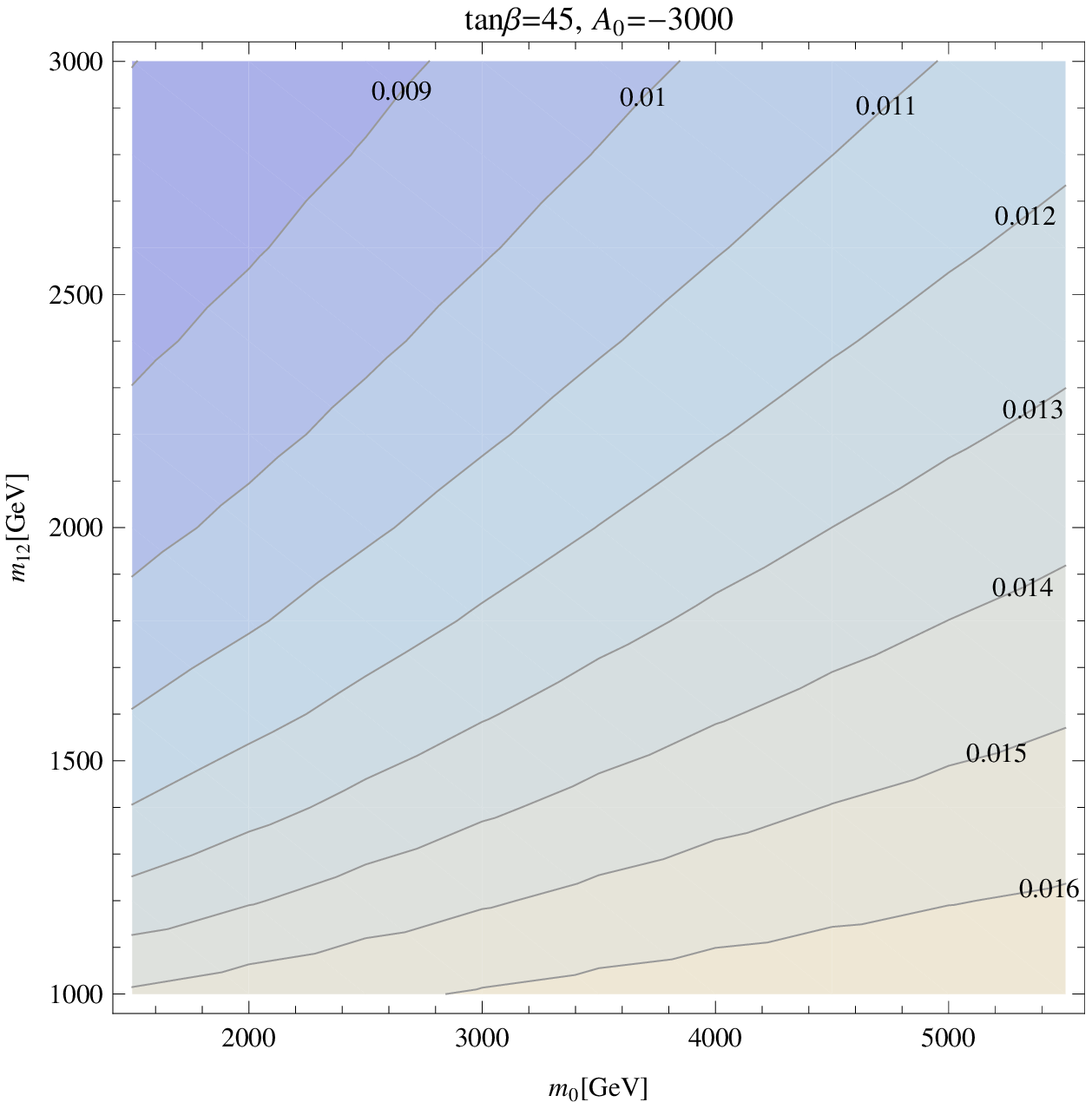   ,scale=0.56,angle=0,clip=}\\
\vspace{0.2cm}
\end{center}
\caption{Contours of $\delta^{QLL}_{23}$ in the
  $m_0$--$m_{1/2}$ plane for different values of $\tb$ and    
$A_0$ in the CMSSM.}  
\label{fig:DelQLL23}
\end{figure}

\begin{figure}[ht!]
\begin{center}
\vspace{3.0cm}
\psfig{file=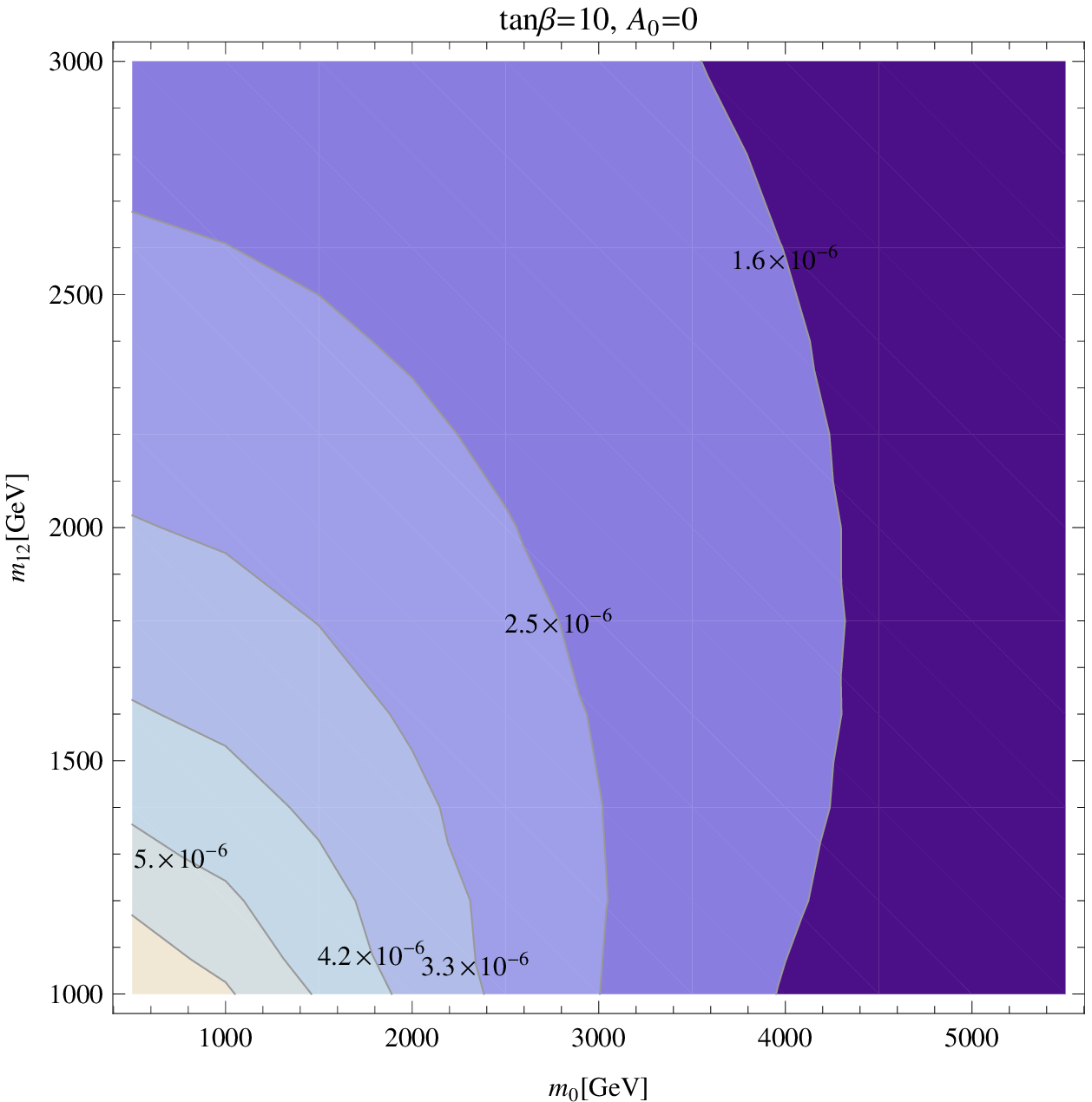  ,scale=0.57,angle=0,clip=}
\psfig{file=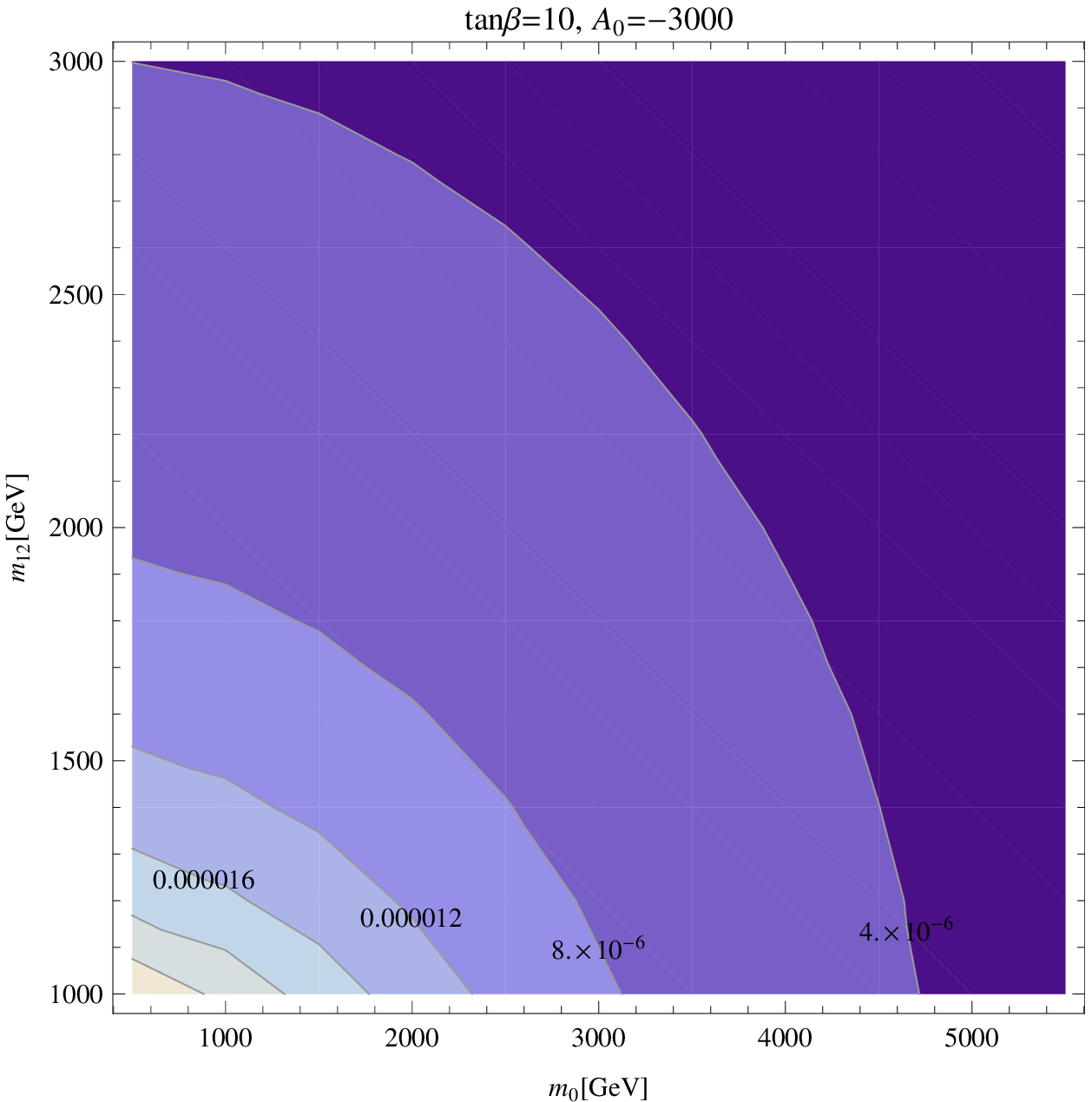  ,scale=0.57,angle=0,clip=}\\
\vspace{2.0cm}
\psfig{file=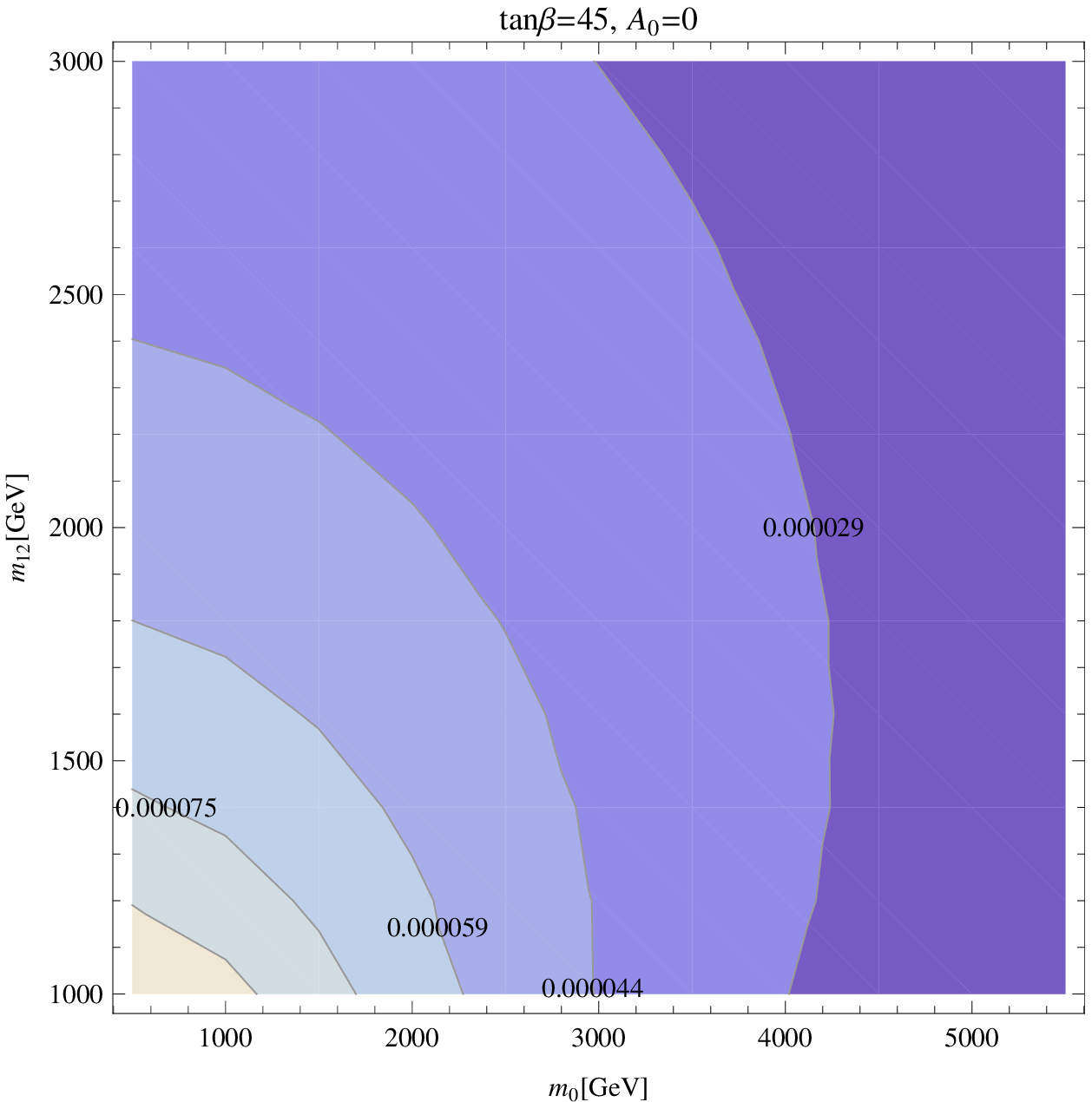 ,scale=0.56,angle=0,clip=}
\psfig{file=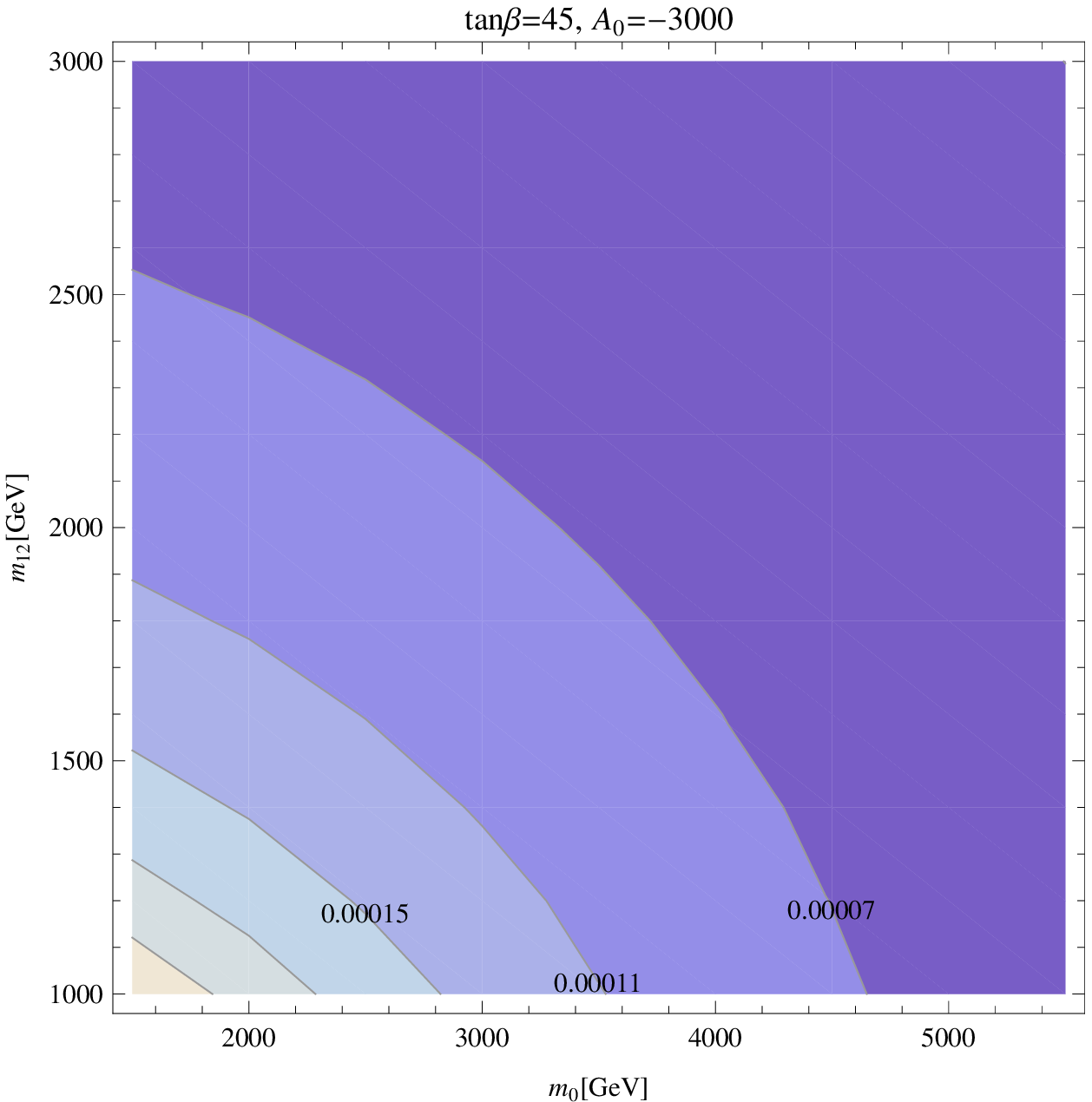   ,scale=0.56,angle=0,clip=}\\
\vspace{0.2cm}
\end{center}
\caption{Contours of $\delta^{ULR}_{23}$ in the
  $m_0$--$m_{1/2}$ plane for different values of $\tb$ and    
$A_0$ in the CMSSM.}  
\label{fig:DelULR23}
\end{figure}

\begin{figure}[ht!]
\begin{center}
\vspace{3.0cm}
\psfig{file=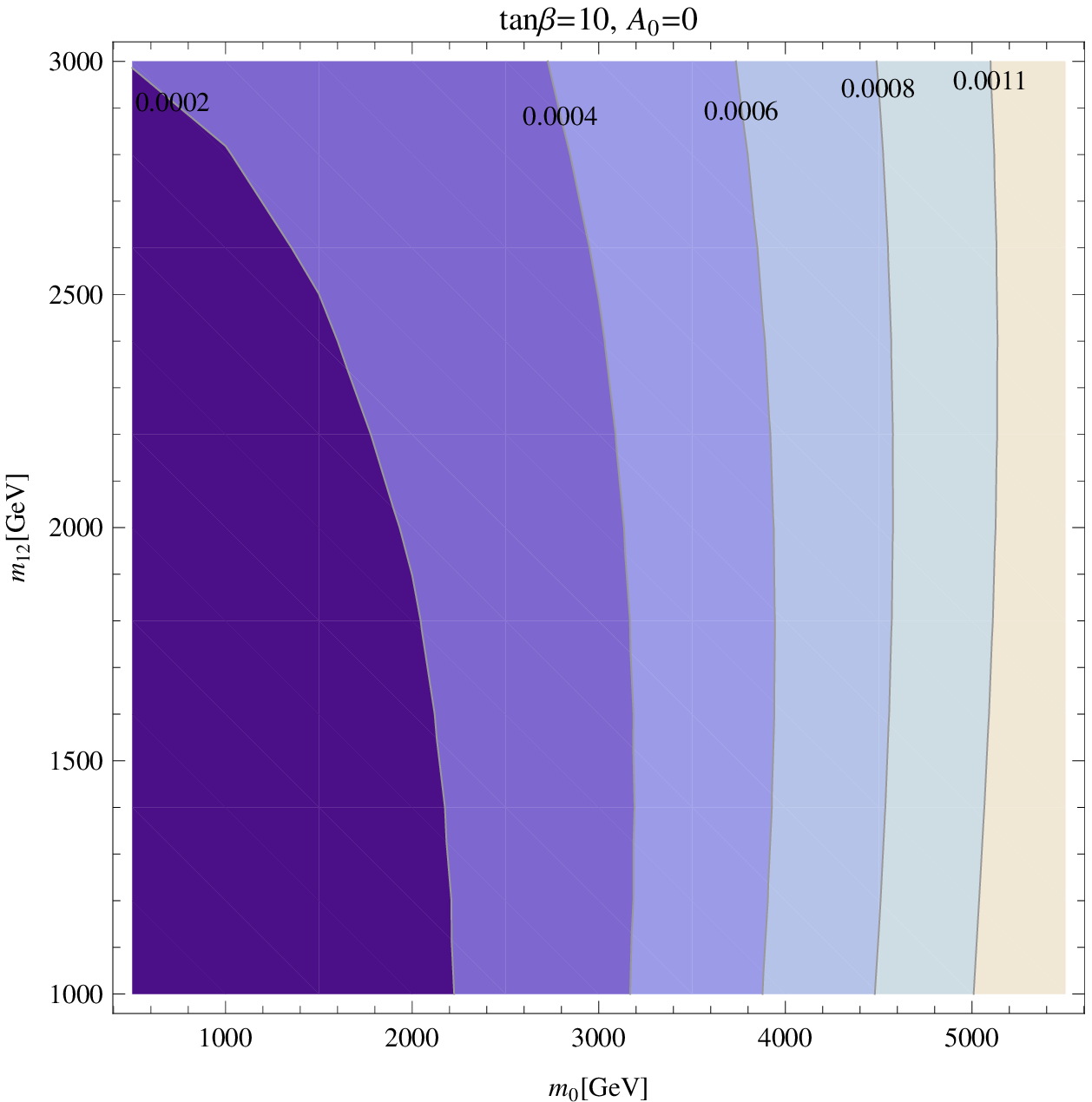  ,scale=0.57,angle=0,clip=}
\psfig{file=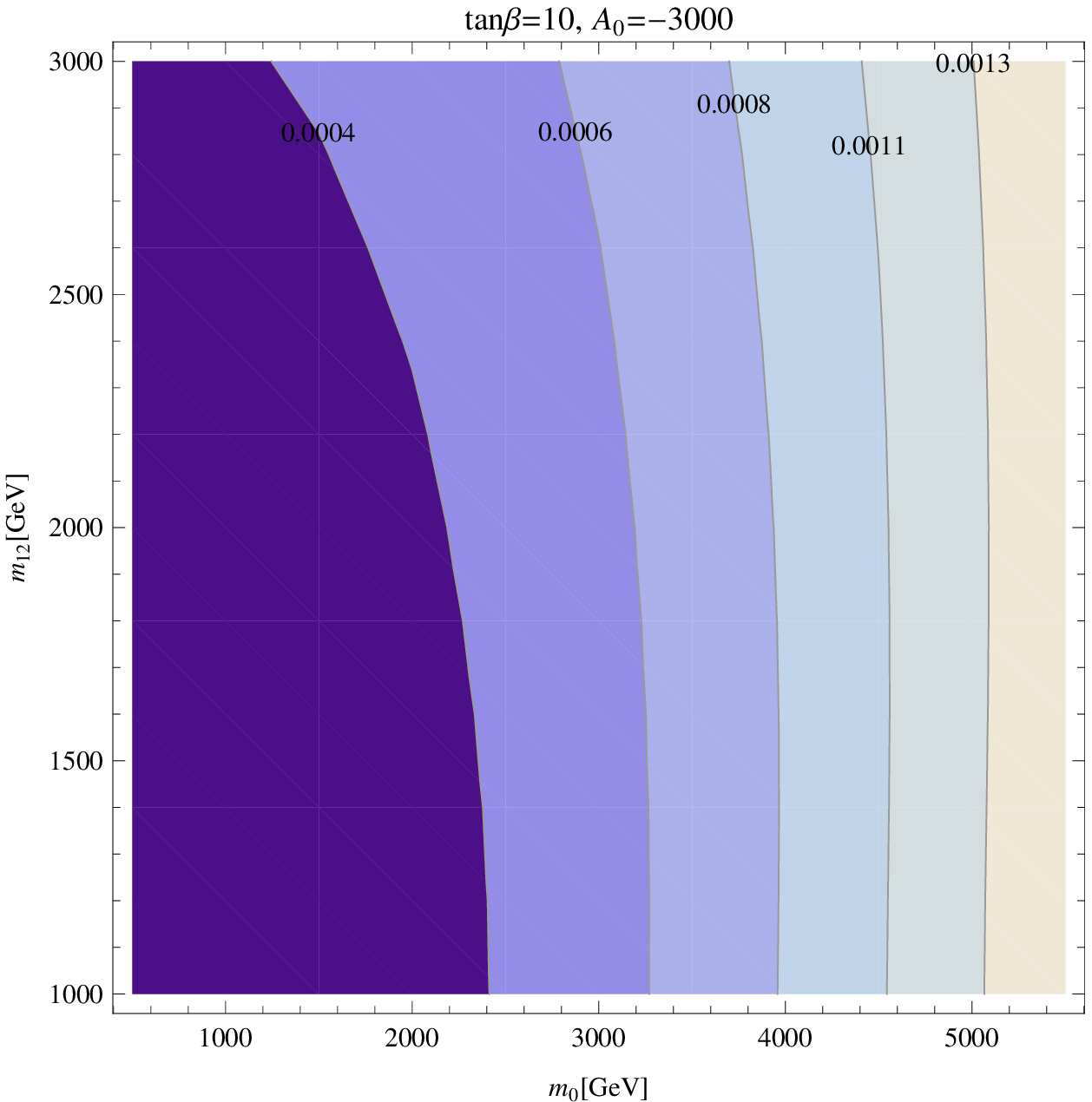  ,scale=0.57,angle=0,clip=}\\
\vspace{2.0cm}
\psfig{file=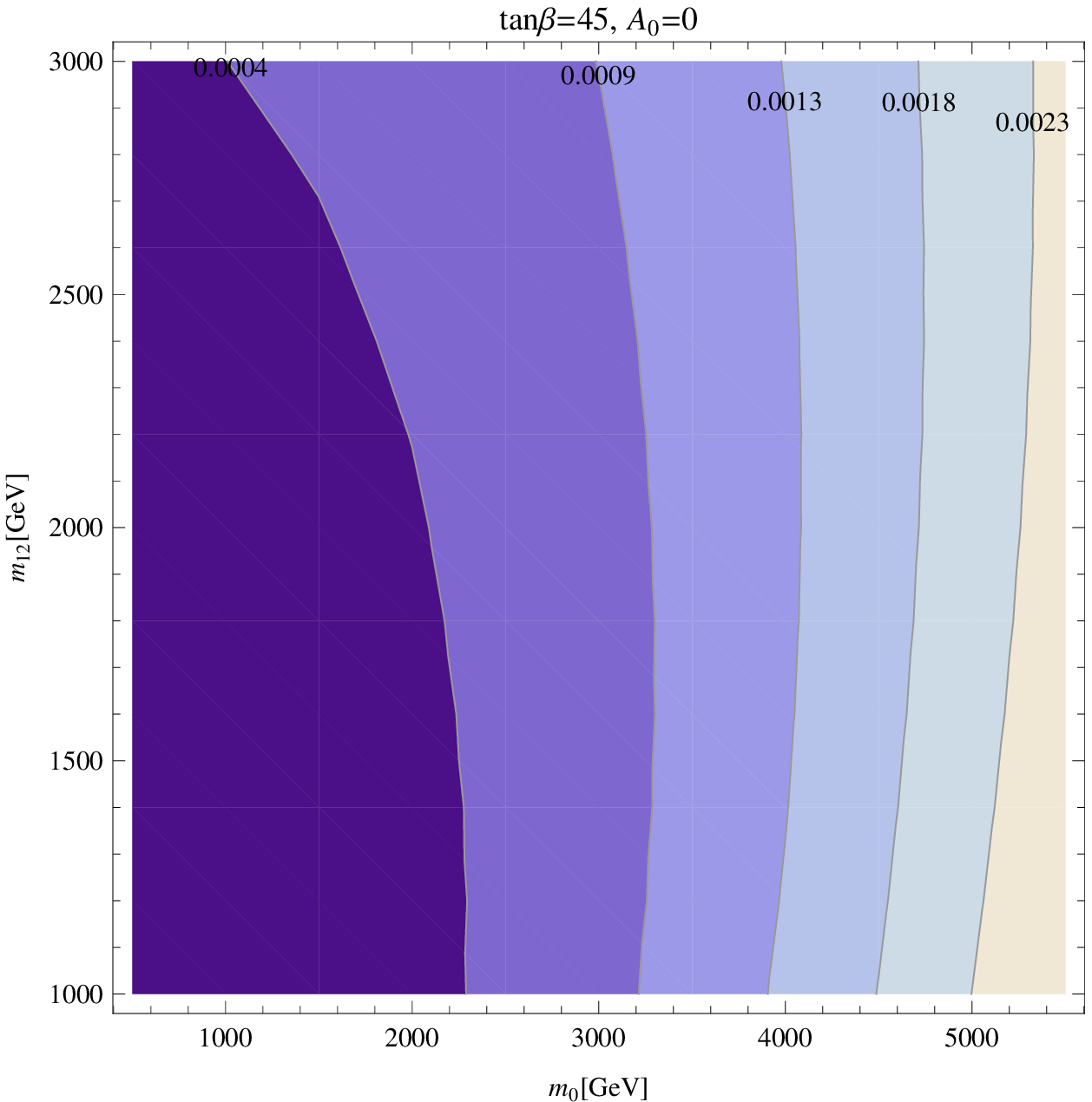 ,scale=0.56,angle=0,clip=}
\psfig{file=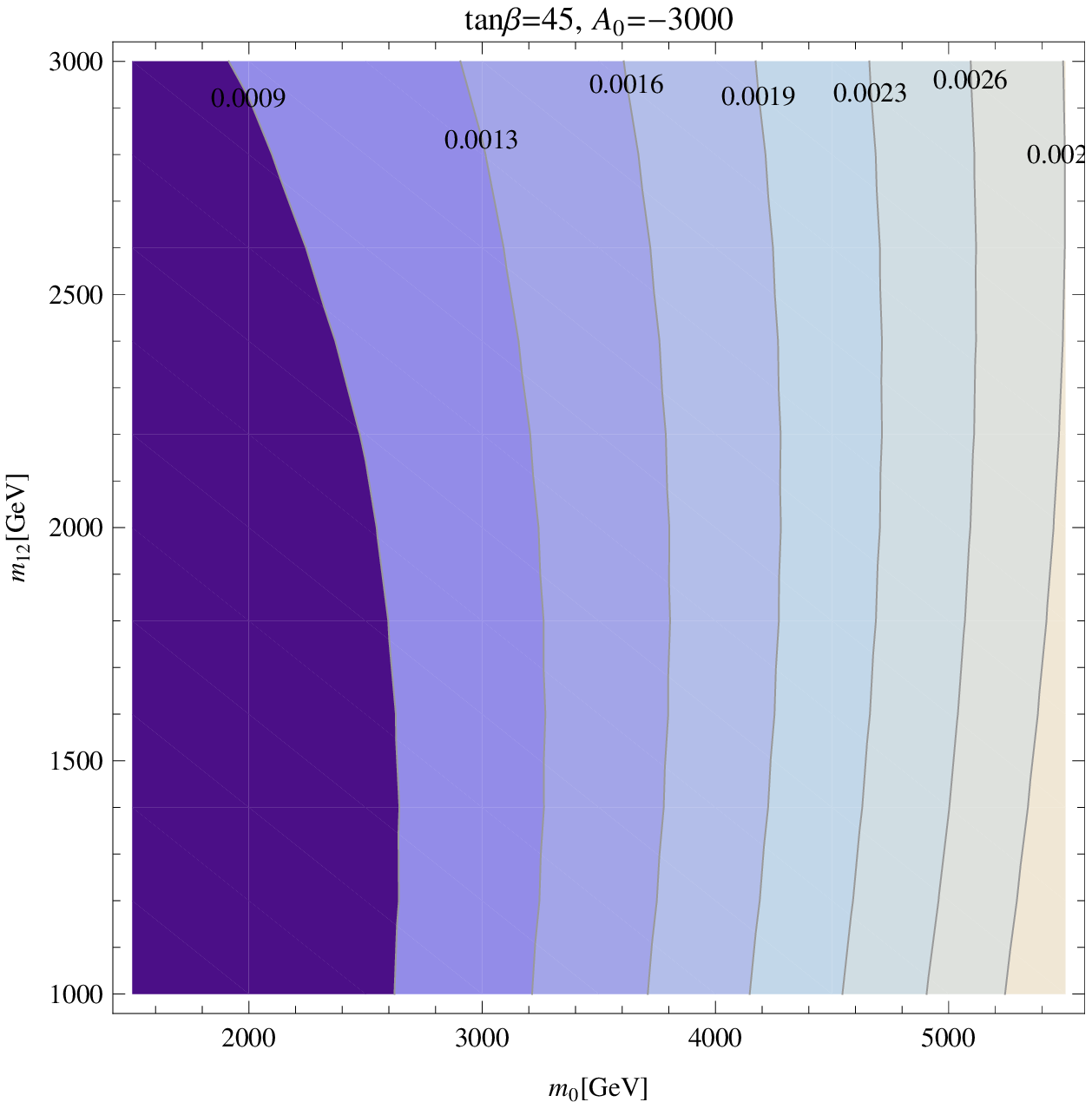   ,scale=0.56,angle=0,clip=}\\
\vspace{0.2cm}
\end{center}
\caption{Contours of \Drho\ in the $m_0$--$m_{1/2}$ plane for different
  values of $\tb$ and $A_0$ in the CMSSM.}  
\label{fig:SQ-delrho}
\end{figure}

\begin{figure}[ht!]
\begin{center}
\vspace{3.0cm}
\psfig{file=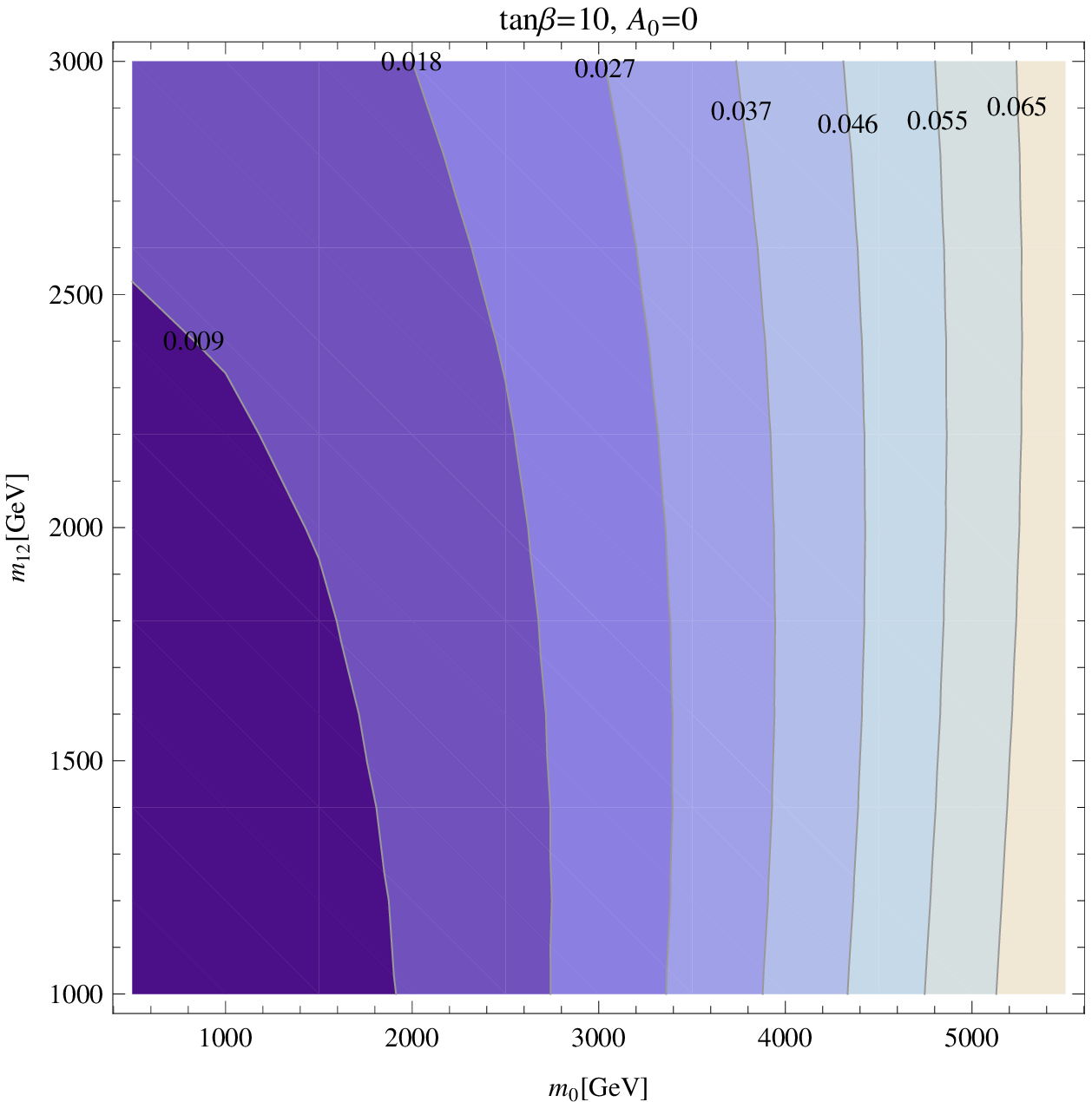  ,scale=0.57,angle=0,clip=}
\psfig{file=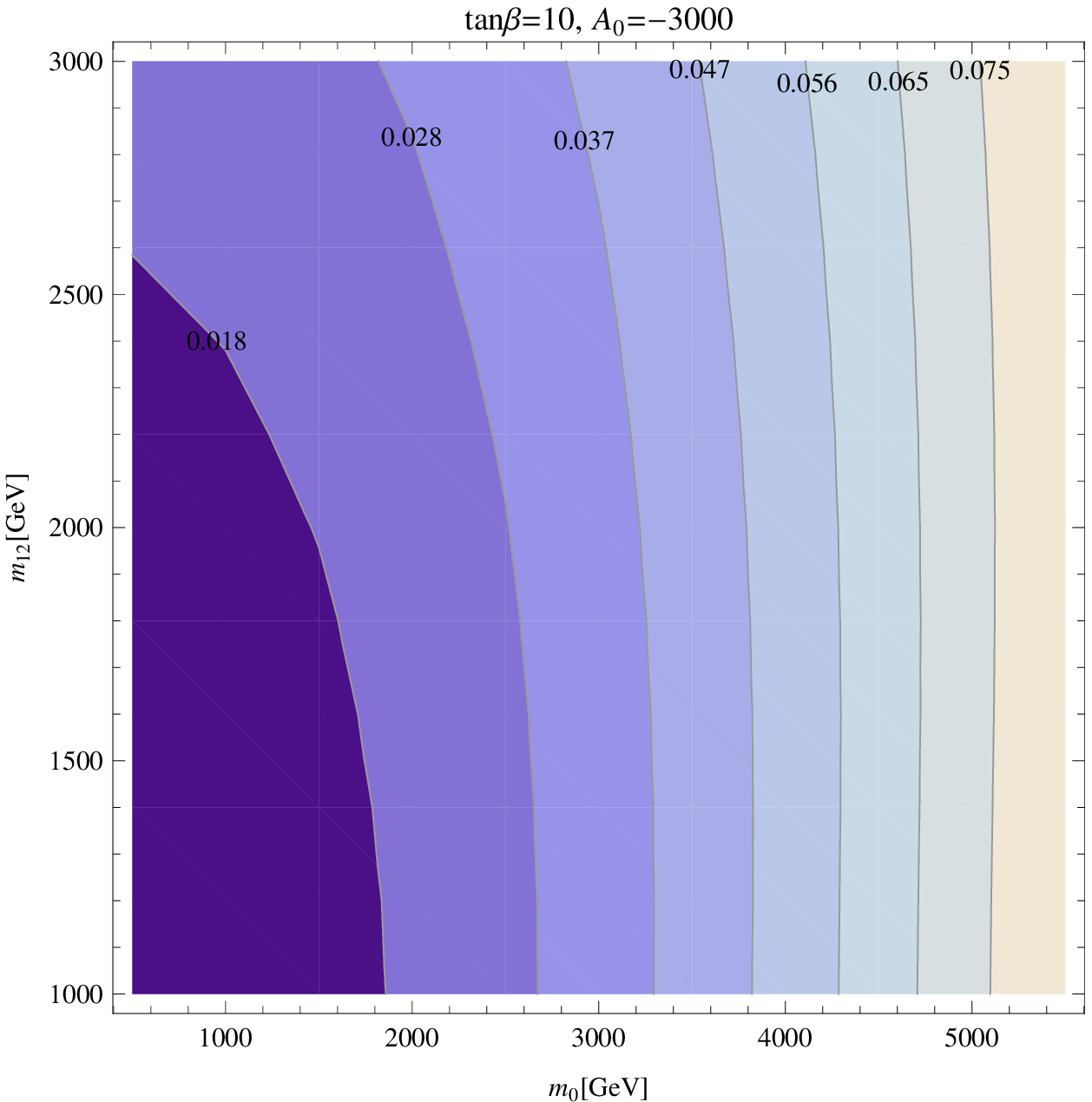  ,scale=0.57,angle=0,clip=}\\
\vspace{2.0cm}
\psfig{file=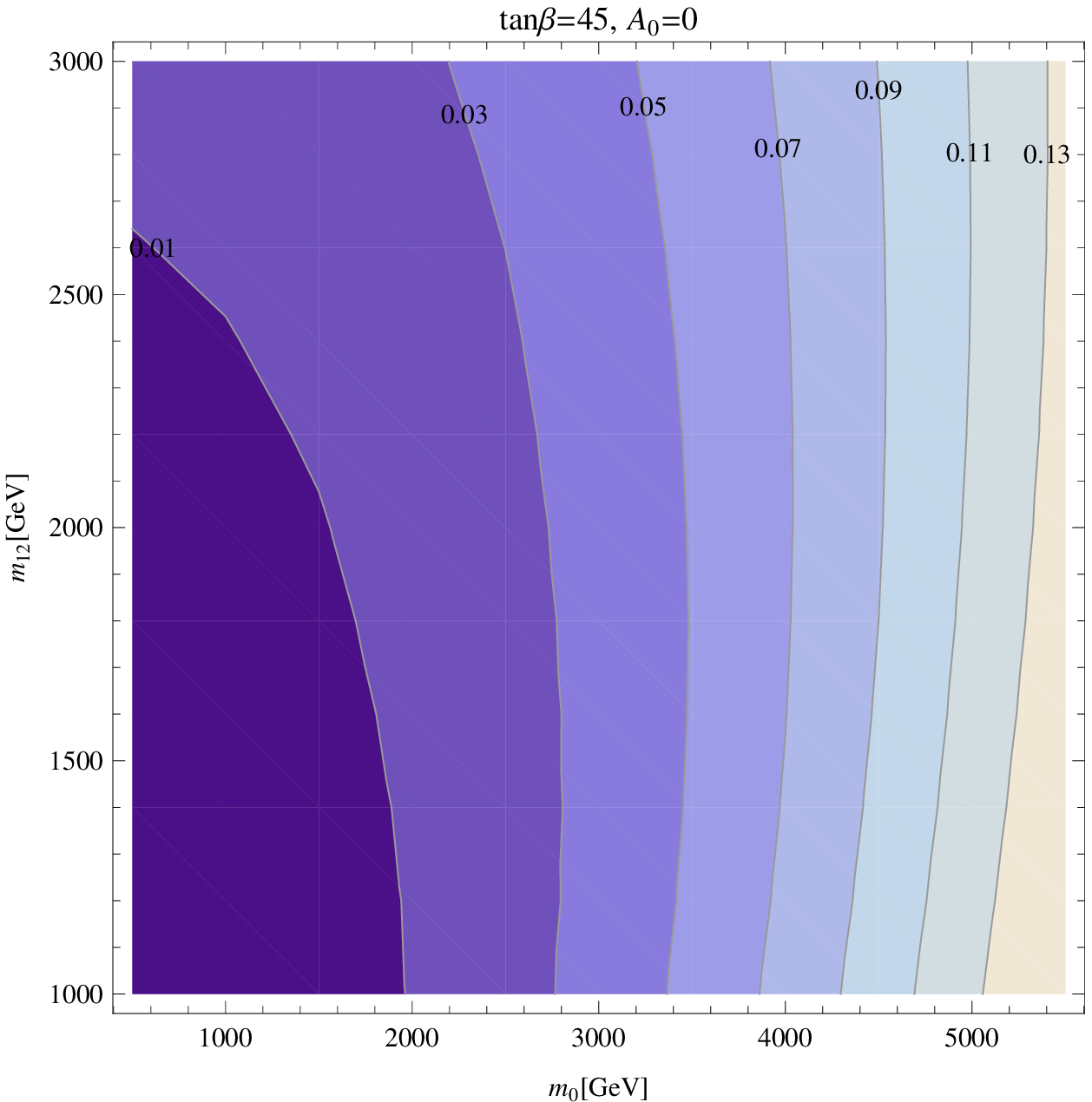 ,scale=0.56,angle=0,clip=}
\psfig{file=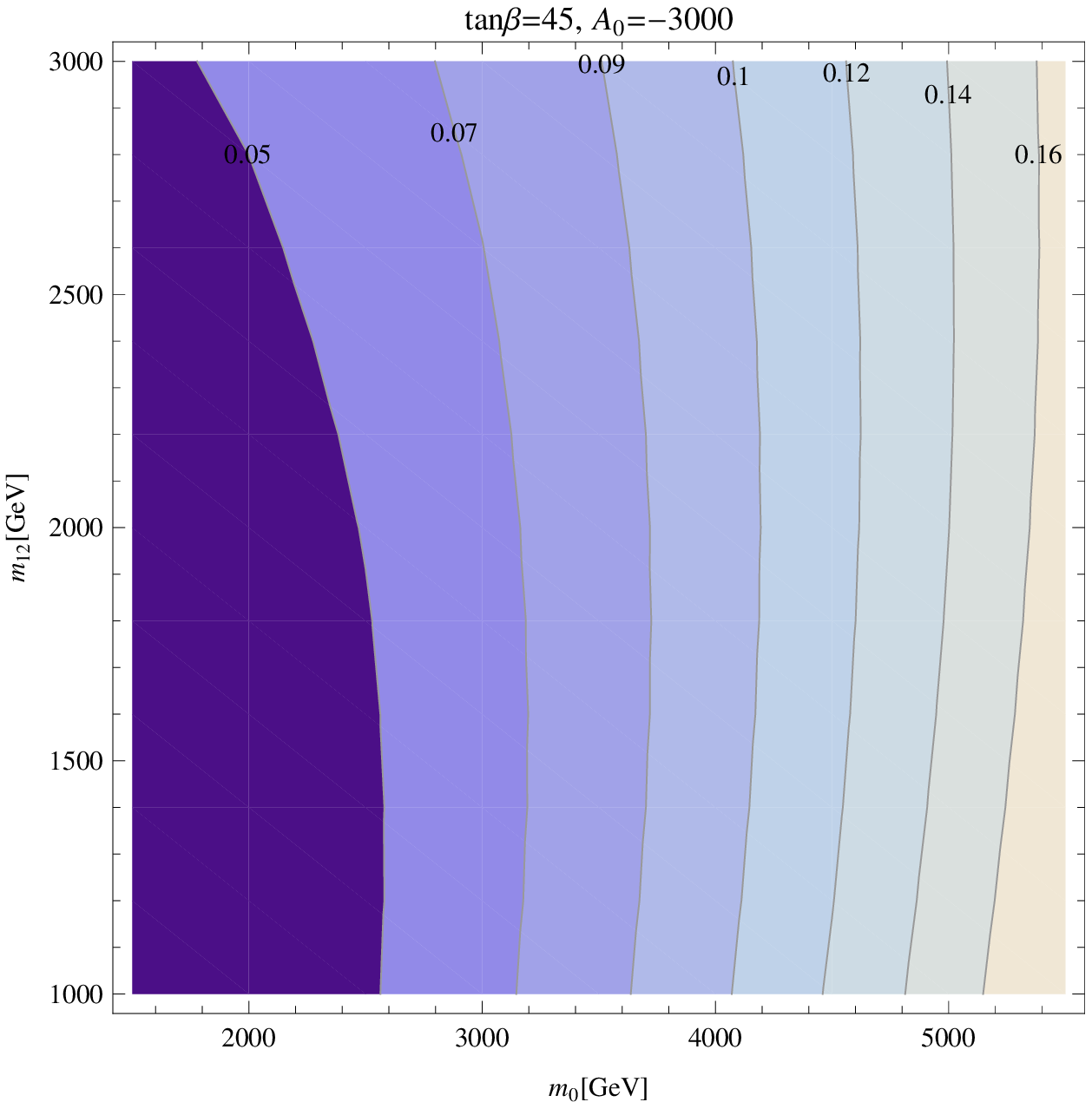   ,scale=0.56,angle=0,clip=}\\
\vspace{0.2cm}
\end{center}
\caption{Contours of \DMW\ in GeV in the
  $m_0$--$m_{1/2}$ plane for different values of $\tb$ and    
$A_0$ in the CMSSM.}  
\label{fig:SQ-delMW}
\end{figure}

\begin{figure}[ht!]
\begin{center}
\vspace{3.0cm}
\psfig{file=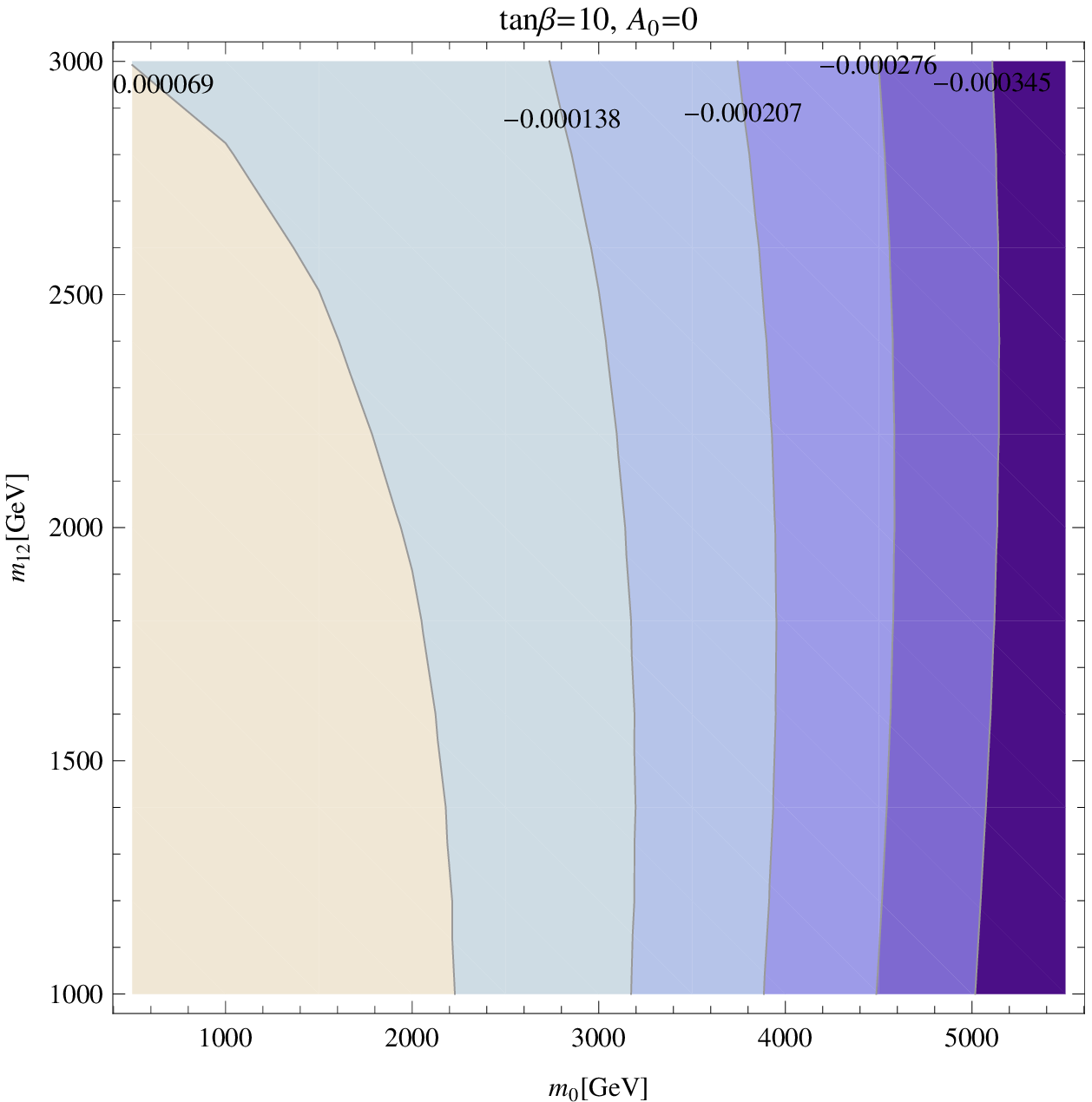  ,scale=0.57,angle=0,clip=}
\psfig{file=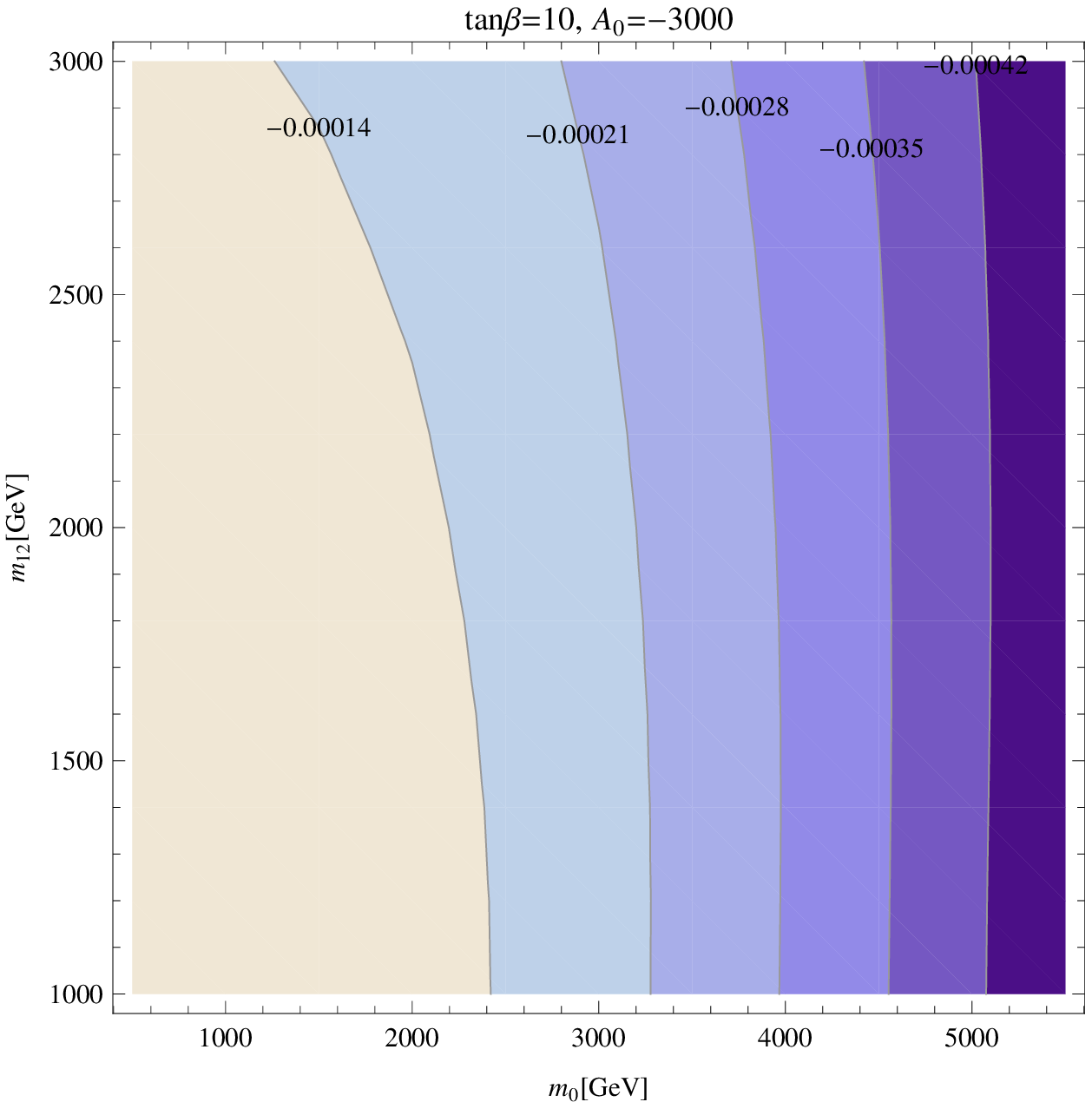  ,scale=0.57,angle=0,clip=}\\
\vspace{2.0cm}
\psfig{file=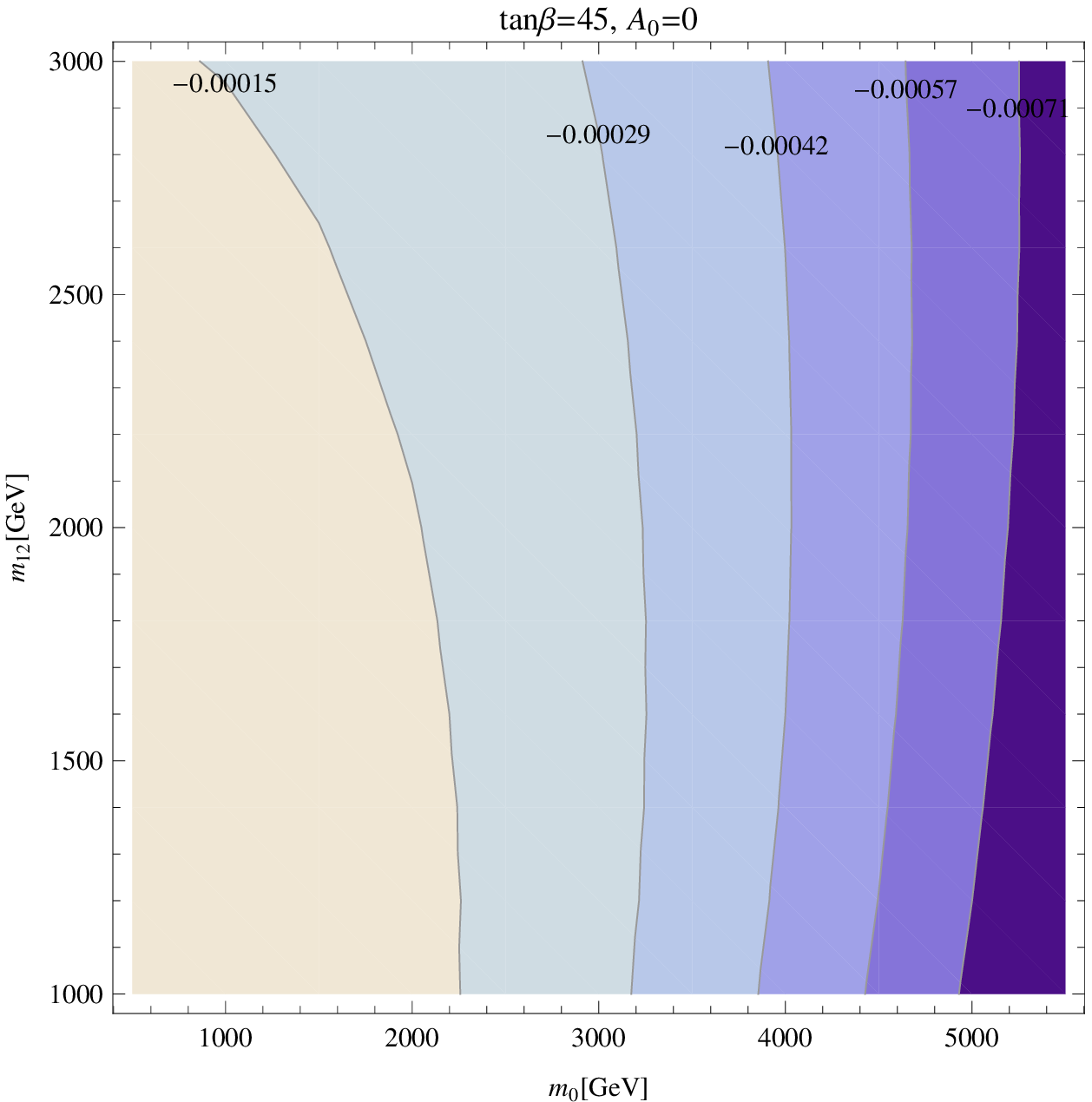 ,scale=0.56,angle=0,clip=}
\psfig{file=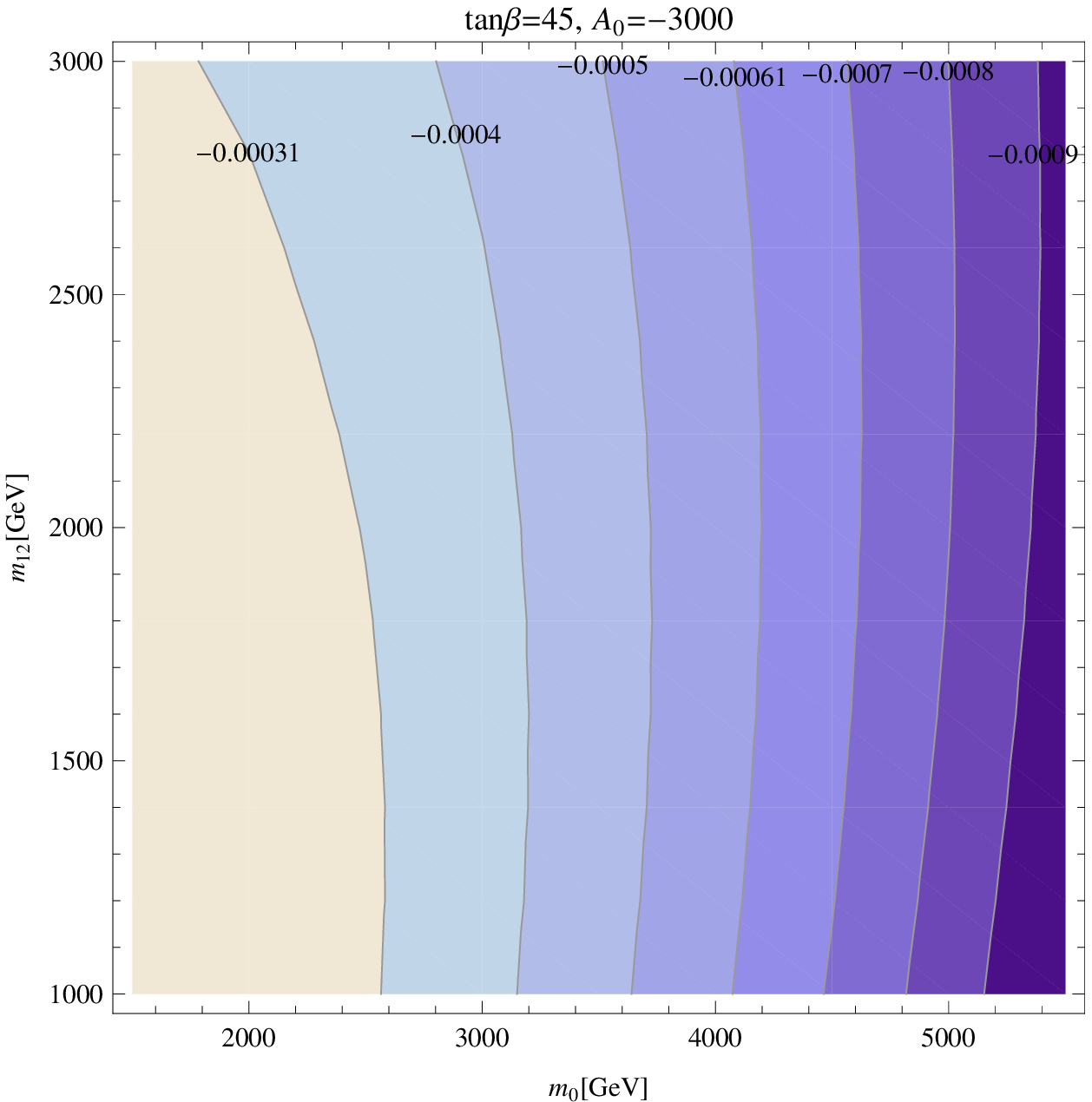   ,scale=0.56,angle=0,clip=}\\
\vspace{0.2cm}
\end{center}
\caption{Contours of \Dsweff\ in the
  $m_0$--$m_{1/2}$ plane for different values of $\tb$ and    
$A_0$ in the CMSSM.}  
\label{fig:SQ-delSW2}
\end{figure}

\begin{figure}[ht!]
\begin{center}
\psfig{file=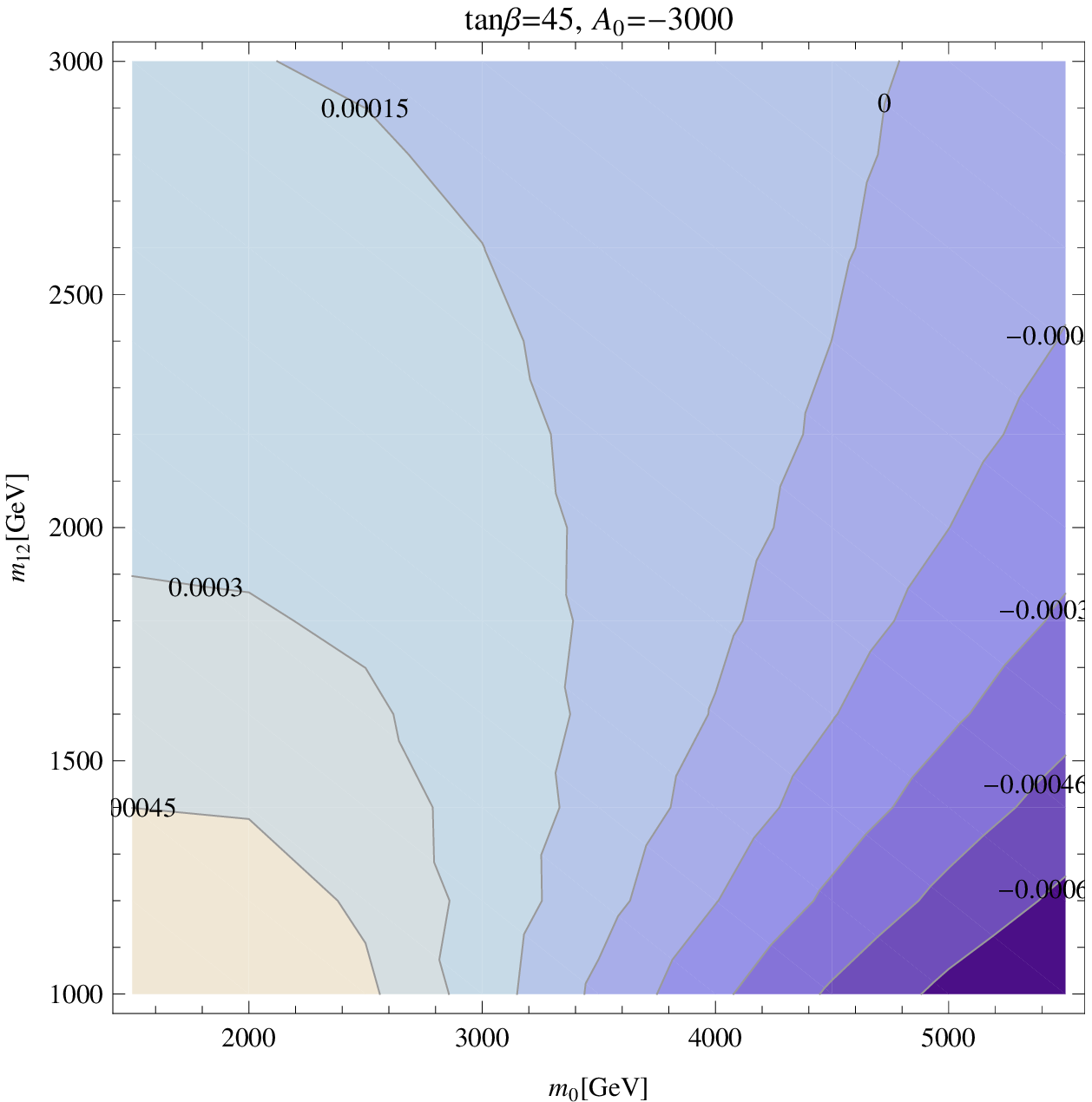,scale=0.50,angle=0,clip=}
\psfig{file=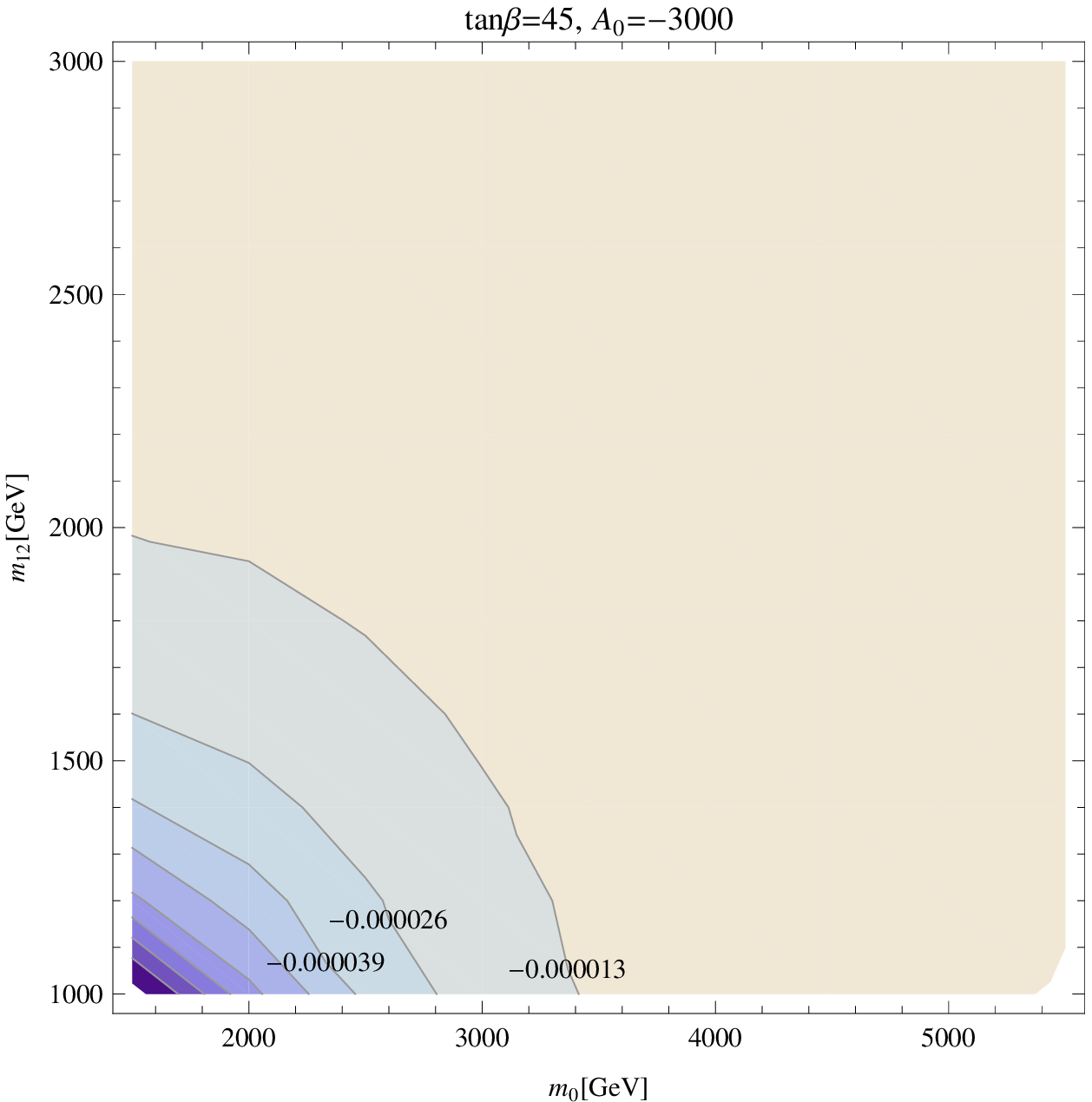,scale=0.50,angle=0,clip=}\\
\vspace{0.2cm}
\psfig{file=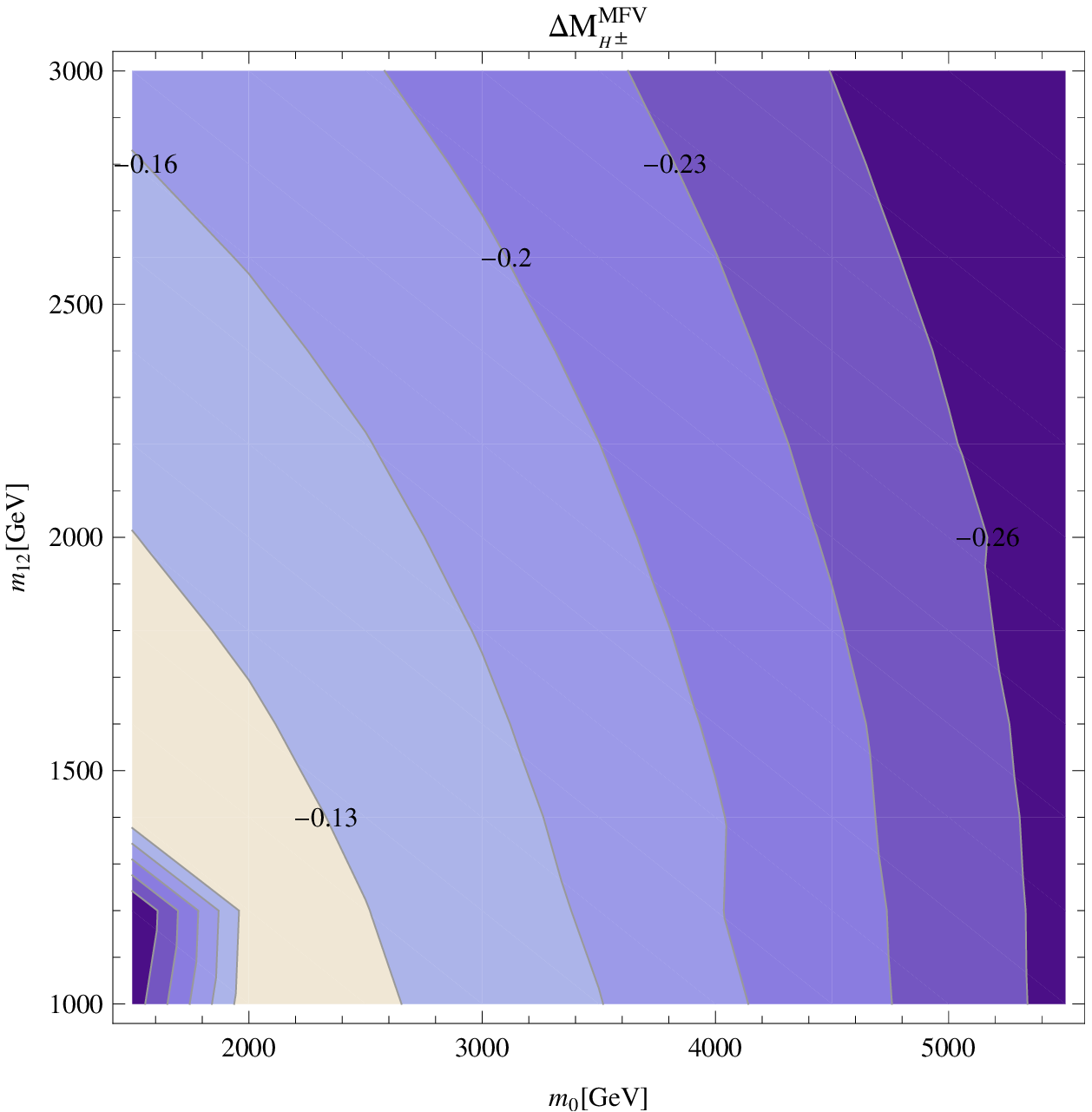,scale=0.50,angle=0,clip=}
\psfig{file=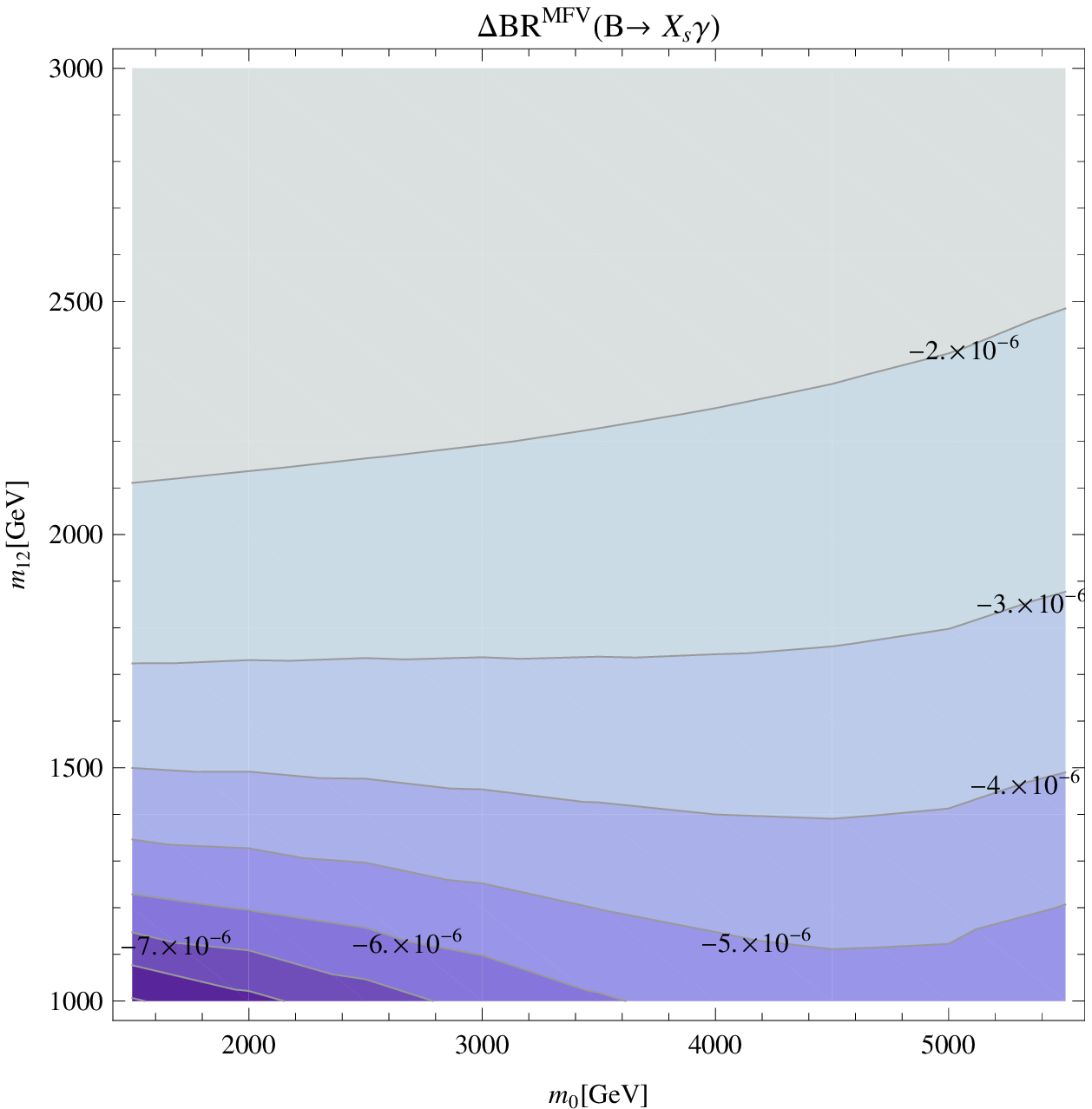,scale=0.50,angle=0,clip=}\\
\vspace{0.5cm}
\psfig{file=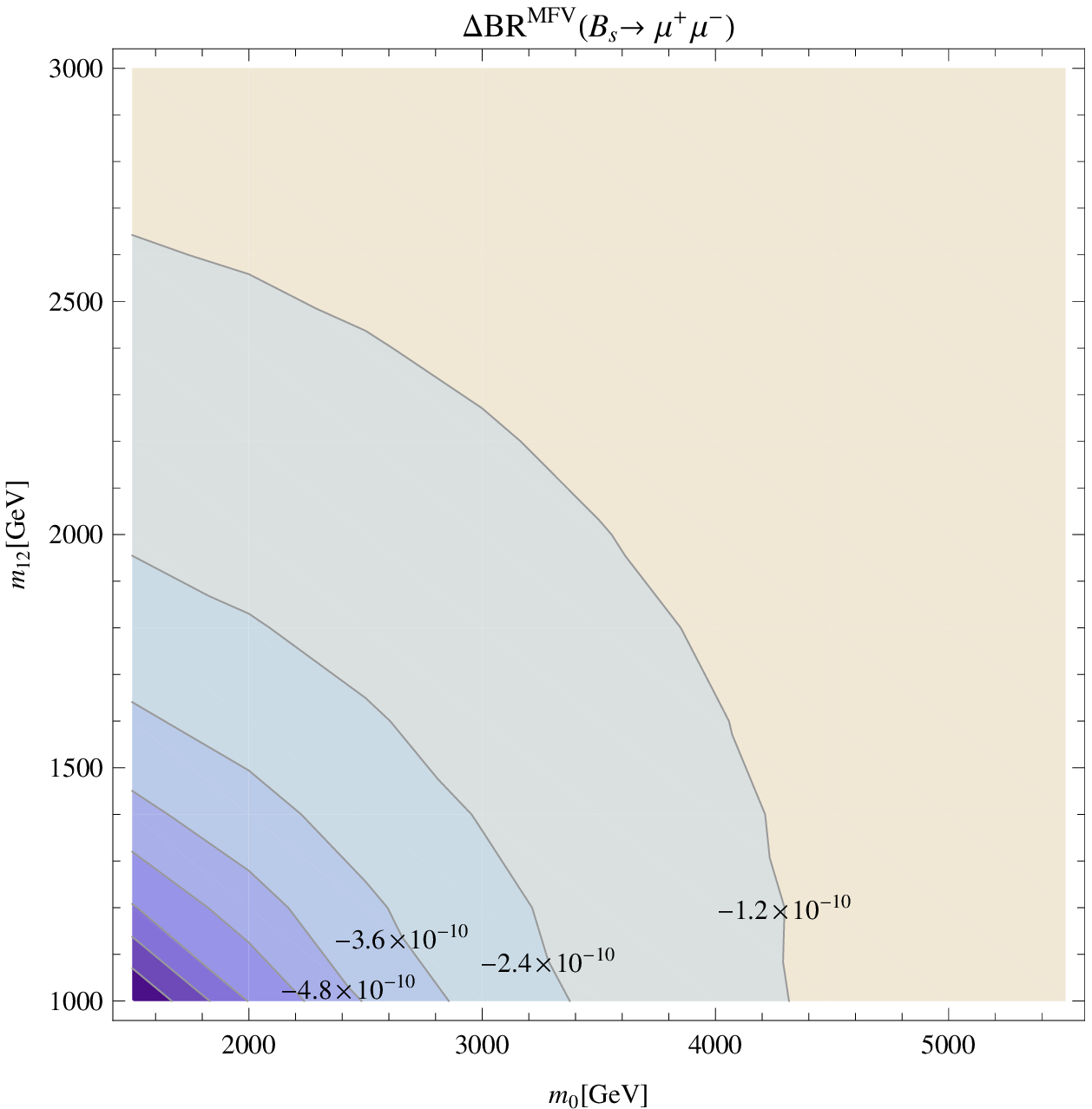,scale=0.50,angle=0,clip=}
\psfig{file=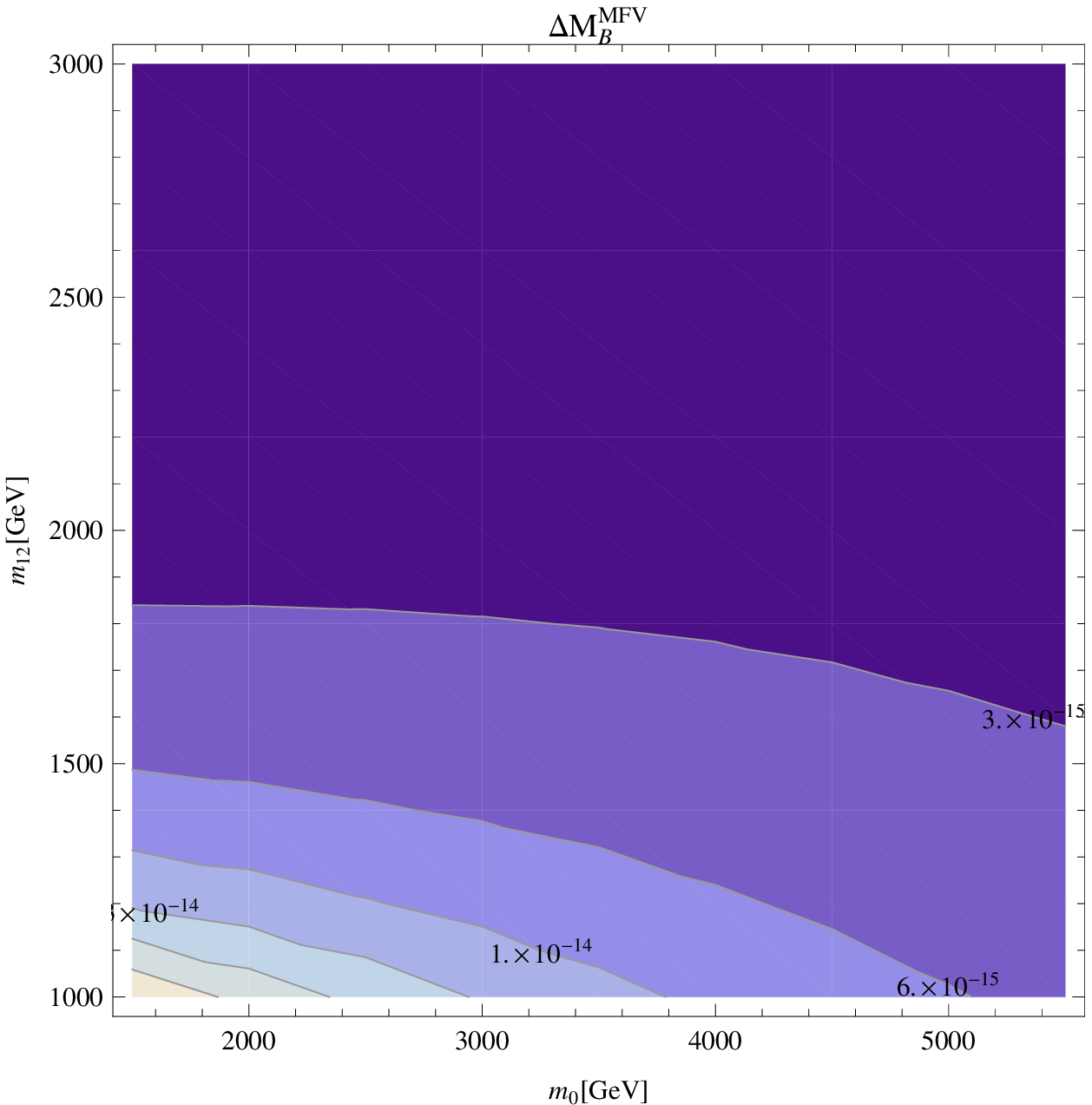,scale=0.50,angle=0,clip=}\\
\vspace{0.5cm}
\end{center}
\caption{Contours of Higgs mass corrections (\DMh,
\DMH\ and \DMHp\ in GeV) and BPO (\Dbsg, \Dbmm\ and \Ddmbs) 
in the $m_0$--$m_{1/2}$ plane for $\tb = 45$ and \
$A_0 = -3000 \gev$ in the CMSSM.}
\label{fig:Sq-MH-BPO} 
\vspace{-3em}
\end{figure}

\bigskip
In \reffis{fig:DelQLL13}-\ref{fig:Sq-MH-BPO} we show the results of
our CMSSM analysis in the $m_0$--$m_{1/2}$ plane for four
different combinations of $\tb = 10, 45$ (left and right column) 
and $A_0 = 0, -3000 \gev$ (upper and lower row).
This set represents four ``extreme'' cases of the parameter space and give
an overview about the possible sizes of the effects and their
dependences on $\tb$ and $A_0$ (which we verified with other, not
shown, combinations).
We start with the three most relevant $\deFABij$'s. In
\reffis{fig:DelQLL13}-\ref{fig:DelULR23} we show the results for
$\del{QLL}{13}$, $\del{QLL}{23}$ and $\del{ULR}{23}$, respectively,
which are expected to yield the largest results.
The values show the expected pattern of their size with 
$\del{QLL}{23} \sim \order{10^{-2}}$ being the largest one, and
$\del{QLL}{13}$ and $\del{ULR}{23}$ about one or two orders of magnitude
smaller. All other $\deFABij$ which are not shown reach only values of
\order{10^{-5}}. 
One can observe an interesting pattern in these figures: the
values of $\deFABij$ increase with larger values of either $\tb$ or
$A_0$. 
As discussed above, these $\deFABij \neq 0$ are often neglected
in phenomenological analyses of the CMSSM (see, e.g.,
\citere{CMSSM-NUHM}).

In \reffis{fig:SQ-delrho}-\ref{fig:SQ-delSW2} we analyze the effects of
the non-zero $\deFABij$ on the EWPO \Drho, \DMW\ and \Dsweff,
respectively. 
Here the same pattern is reflected for the EWPO, i.e.\ by increasing
the value of $\tb$ or $A_0$, we find larger contributions to the
EWPO. In particular one can observe a non-decoupling effect for
large values of $m_0$. Larger soft SUSY-breaking
parameters with the non-zero values in particular of $\del{QLL}{23}$,
see above, lead to an enhanced splitting in masses belonging to an
$SU(2)$ doublet, and thus to an enhanced contribution to the
$\rho$-parameter. The corresponding effects on $\MW$ and $\sweff$, 
for $m_0 \gsim 3 \tev$, exhibit corrections that are several times
larger than the current experimental accuracy (whereas the SUSY
corrections with all $\deFABij = 0$ decouple and go to
zero). Consequently, including the non-zero values of the $\deFABij$
and correctly taking these corrections into account, would yield an 
{\em upper} limit on $m_0$, which in the known analyses so far
is unconstrained from above~\cite{CMSSM-NUHM}. 
A more detailed analysis within the CMSSM will be needed to determine
the real upper bound on $m_0$, which, however, is beyond the scope of
this paper.

In \reffi{fig:Sq-MH-BPO} we show the results of
our CMSSM analysis with the effects of
the non-zero $\deFABij$ on the Higgs mass calculations and on the 
BPO in the $m_0$--$m_{1/2}$ plane for $\tb = 45$ and
$A_0 = -3000$. 
We only show this ``extreme'' case, where smaller values of
$\tb$ and $A_0$ would lead to smaller effects.
In the upper left, upper right and middle left plot we show \DMh,
\DMH\ and \DMHp, respectively. It can be seen that the effects on the
neutral Higgs boson masses are negligible w.r.t.\ the experimental
accuracy. The effects on $\MHp$ can reach \order{100 \mev}, 
where largest effects are found for both very small values
of $m_0$ and $m_{1/2}$ (dominated by $\del{ULR}{23}$) or very large values of 
$m_0$ and $m_{1/2}$ (dominated by $\del{QLL}{13,23}$). Corrections of up to
$-300 \mev$ are found, but still remaining
below the foreseeable future precision. Consequently,
also in the Higgs mass evaluation not taking into account the non-zero
values of the $\deFABij$ is a good approximation.

In the middle right, lower left and lower right plot of
\reffi{fig:Sq-MH-BPO} we show the results for the BPO
\Dbsg, \Dbmm\ and \Ddmbs,
respectively. The effects in \Dbsg\ are of \order{-10^{-5}} and thus one
order of magnitude smaller than the experimenal accuracay. Similarly, we
find $\Dbmm \sim \order{10^{-10}}$ and 
$\Ddmbs \sim \order{10^{-15} \gev}$, i.e.\ one or several orders of
magnitude below the experimental precision. This shows that for the BPO
neglecting the effects of non-zero $\deFABij$ in the CMSSM is a good
approximation.

\newpage
\subsection{Effects of slepton mixing.}
\label{sec:Sl}

In this section we analyze the effects of non-zero $\deFABij$ values in
the \CMSSMI.
In order to investigate the effects induced just by the mixings
in the slepton sector, such that we can compare their contribution from
the one produced by the mixings in the squak sector (and to discriminate
it from effects from mixings in the squark sector) we present here the
results with only $\deFABij$ in 
the slepton sector non-zero, i.e.\ after the RGE running with both CKM
and see-saw parameters non-zero, the $\deFABij$ from the squark
sector are set to zero by hand at the EW scale. The effects of the 
squark mixing in the \CMSSMI\ are nearly
indistinguishable from the ones analyzed in the previous subsection.

As mentioned in \refse{sec:cmssmI}, the calculations in this section are
done by using the values of $Y_\nu$ constructed from \refeq{eq:casas} with
degenerate $M_R$'s. The matrix $R$ is set to the identity since it does not 
enter in \refeq{eq:ynu2} and therefore the slepton $\deFABij$'s do not depend
on it. The matrix $m_\nu^\delta$ is a diagonal mass matrix adjusted to
reproduce neutrino masses at low energy compatible with the experimental
observations and with hierarchical neutrino masses. We performed our
computation by using the seesaw scale $M_N=10^{14} \gev$. With this
choice the bound $BR(\mu \to e \gamma) < 5.7 \times 10^{-13}$ ~\cite{Adam:2013mnn}
imposes severe restrictions on the $m_0$--$m_{1/2}$ plane, excluding 
values of $m_0$  below 2--3~TeV (depending on $\tb$ and $A_0$). The values of
the slepton $\deFABij$ will increase as the scale $M_N$  increases but also
does the parameter space excluded by the $\br(\mu \to e \gamma)$
bound. For example, by increasing $M_N$ by an order of magnitude, the largest
entries in the matrix $Y_\nu$ will become of \order{1} and the bound on
$\br(\mu \to e \gamma)$ will only be satisfied if $m_0\approx 5 \tev$. 

\begin{figure}[ht!]
\begin{center}
\vspace{1.0cm}
\psfig{file=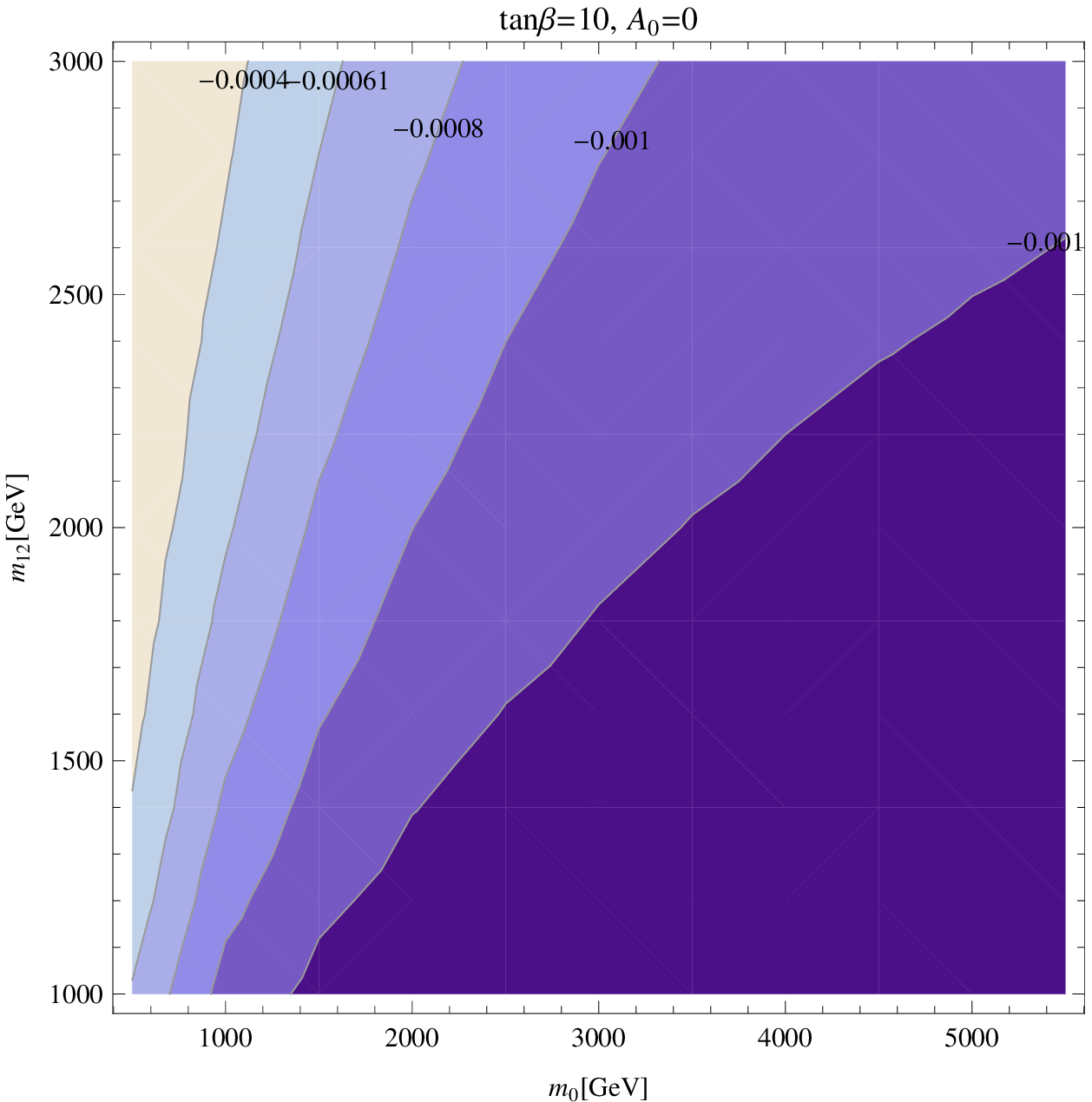  ,scale=0.57,angle=0,clip=}
\psfig{file=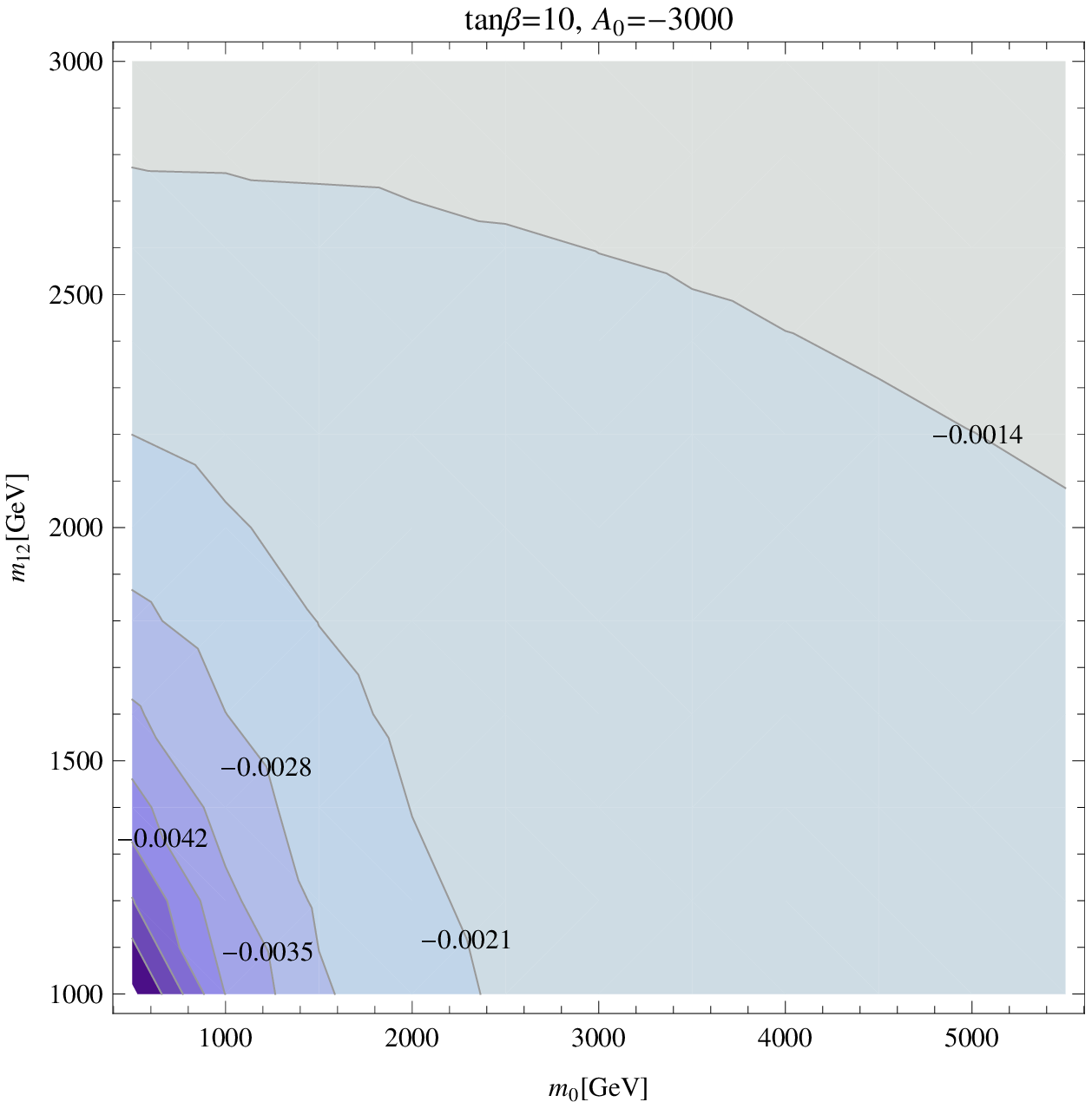  ,scale=0.57,angle=0,clip=}\\
\vspace{1.0cm}
\psfig{file=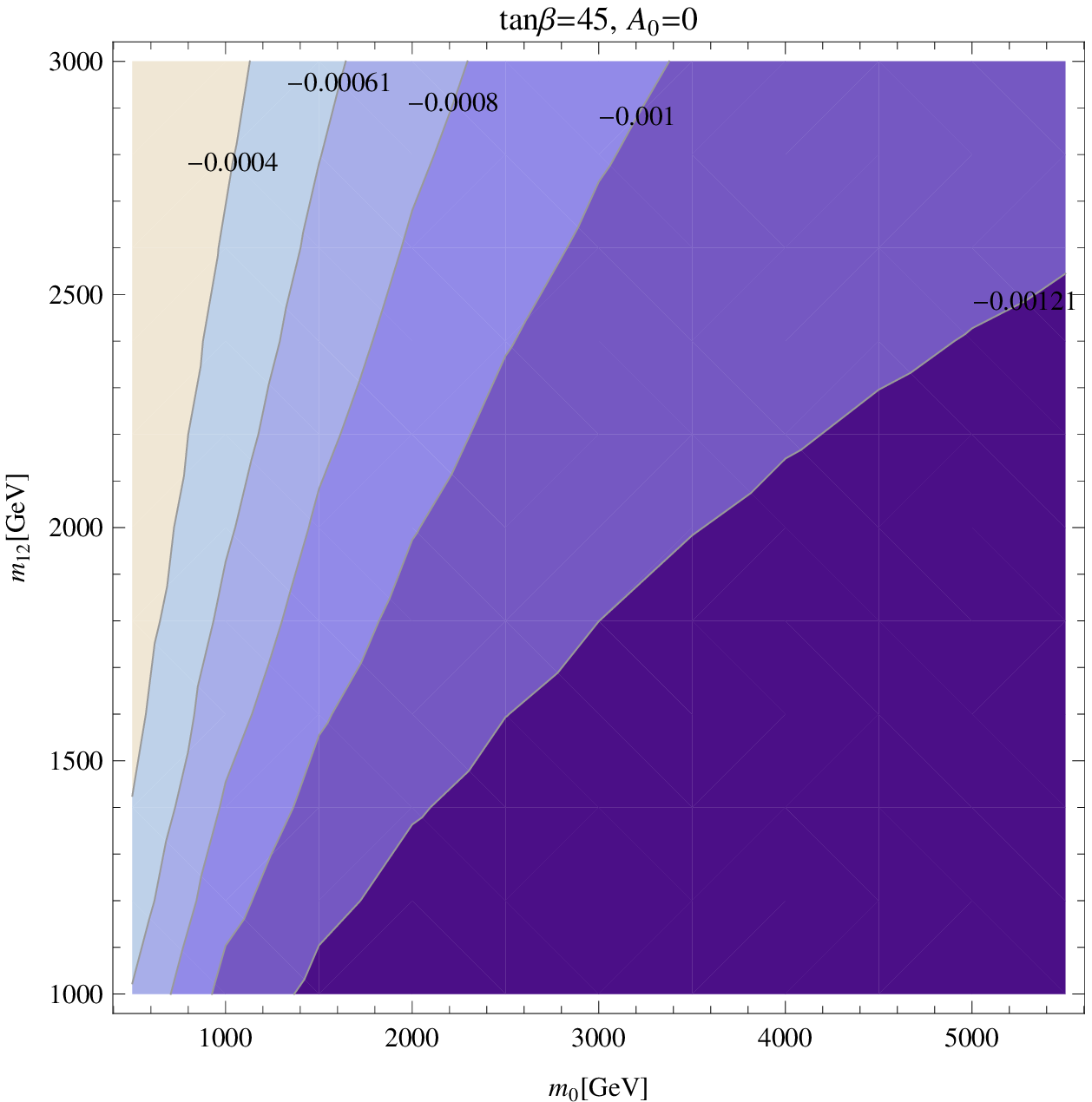 ,scale=0.56,angle=0,clip=}
\psfig{file=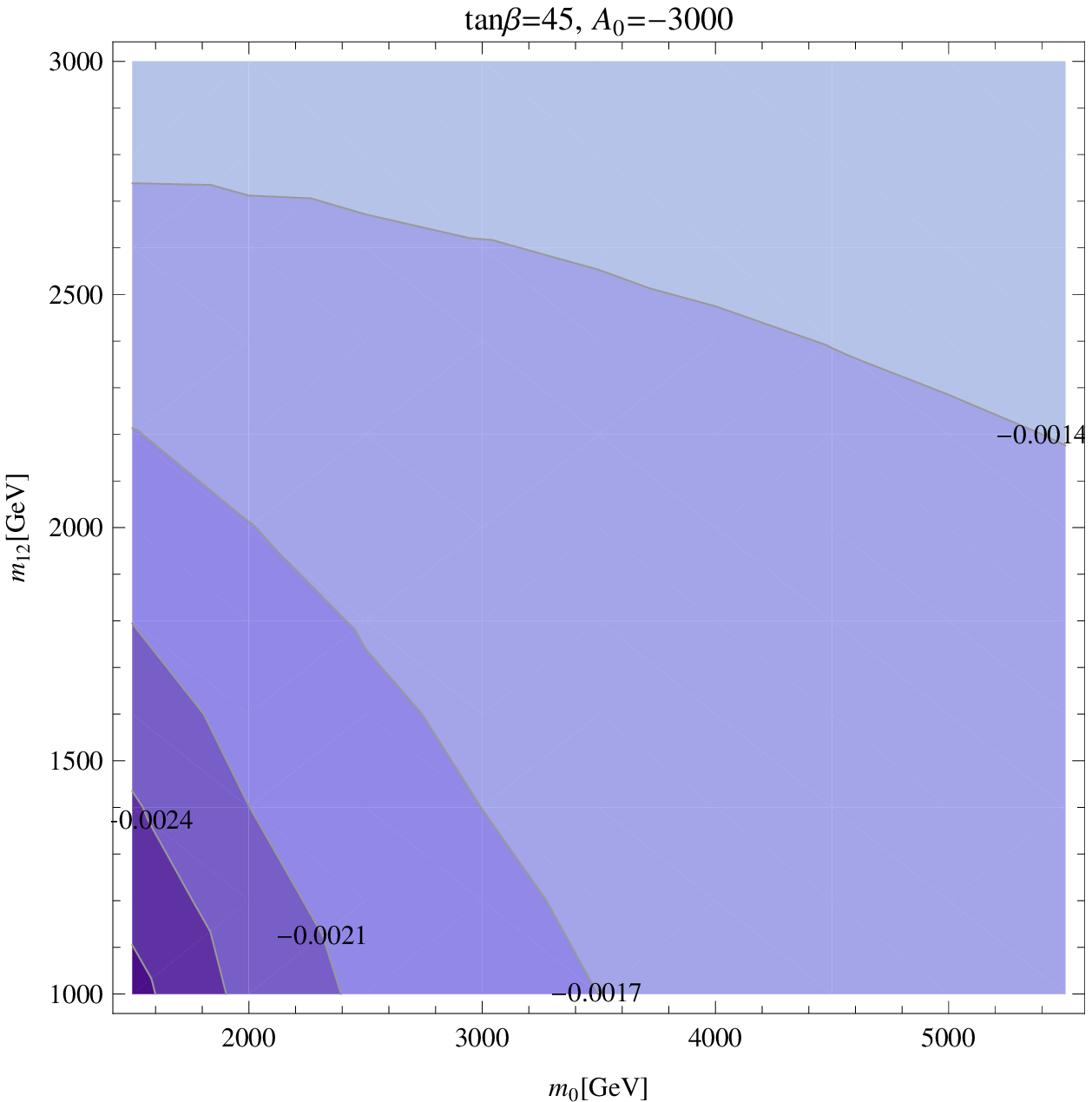   ,scale=0.56,angle=0,clip=}\\
\vspace{0.2cm}
\end{center}
\caption{Contours of $\delta^{LLL}_{12}$  in the
  $m_0$--$m_{1/2}$ plane for different values of $\tb$ and    
$A_0$ in the \CMSSMI. }  
\label{fig:DelLLL12}
\end{figure}

\begin{figure}[ht!]
\begin{center}
\vspace{3.0cm}
\psfig{file=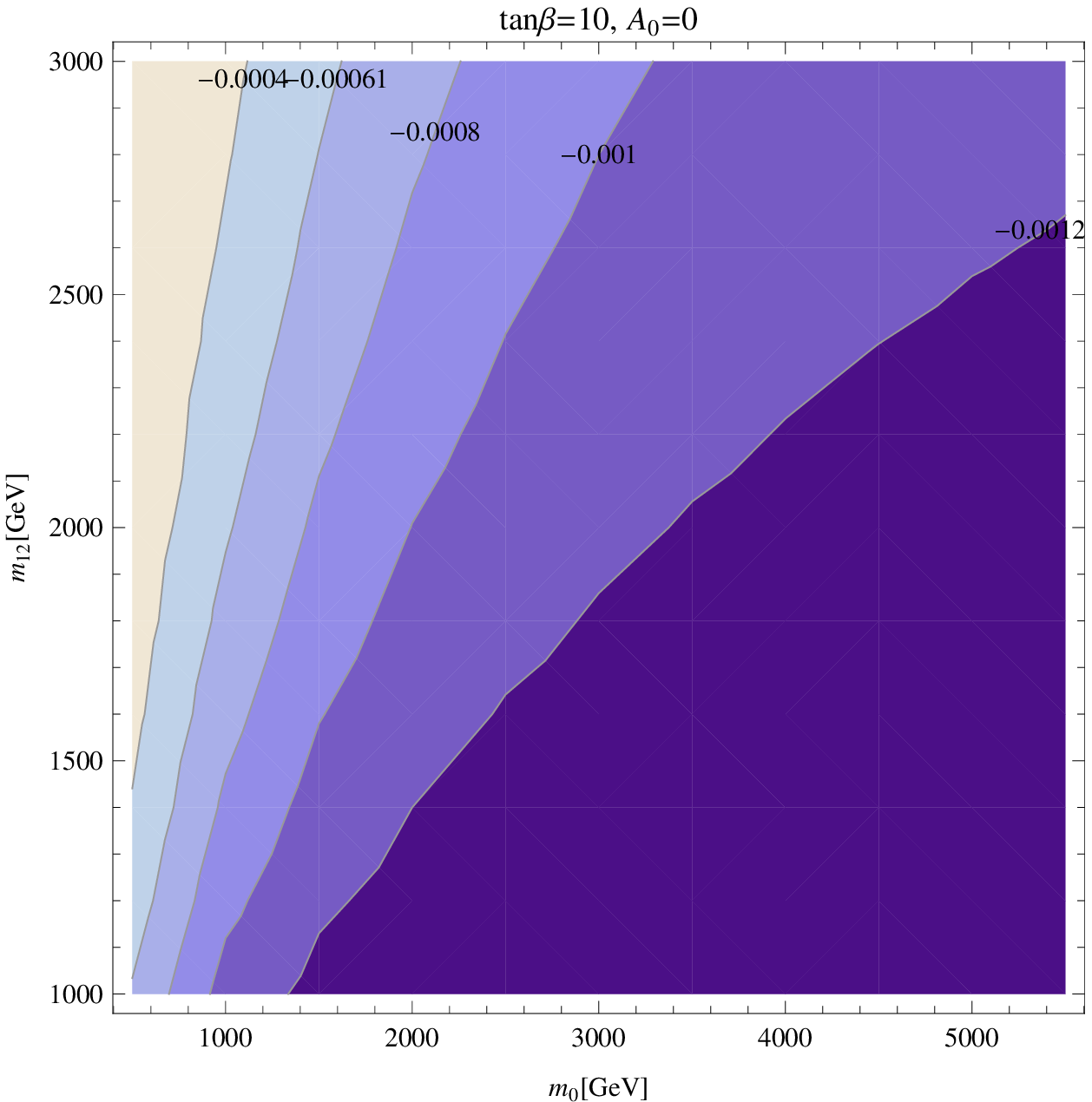  ,scale=0.57,angle=0,clip=}
\psfig{file=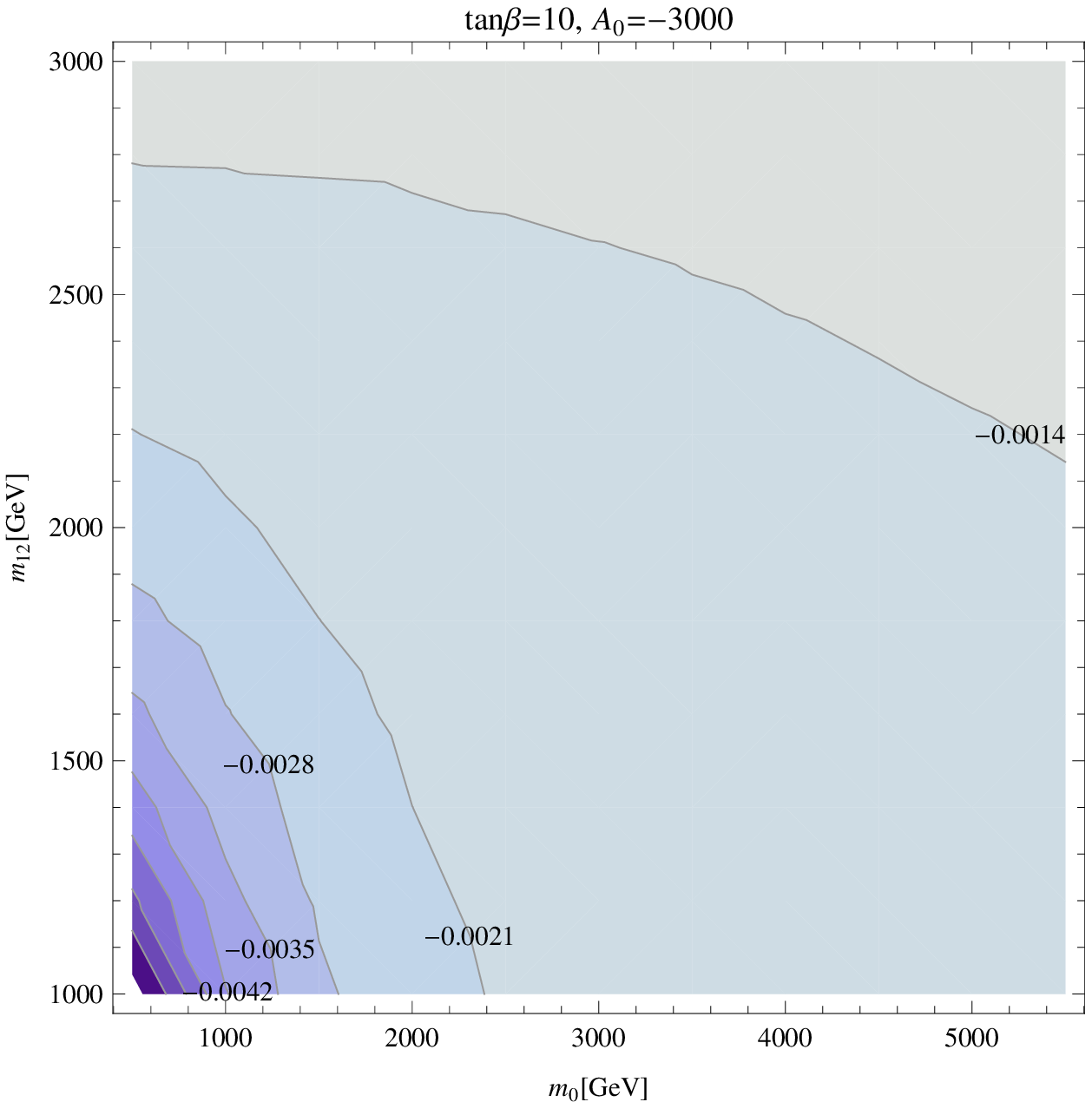  ,scale=0.57,angle=0,clip=}\\
\vspace{2.0cm}
\psfig{file=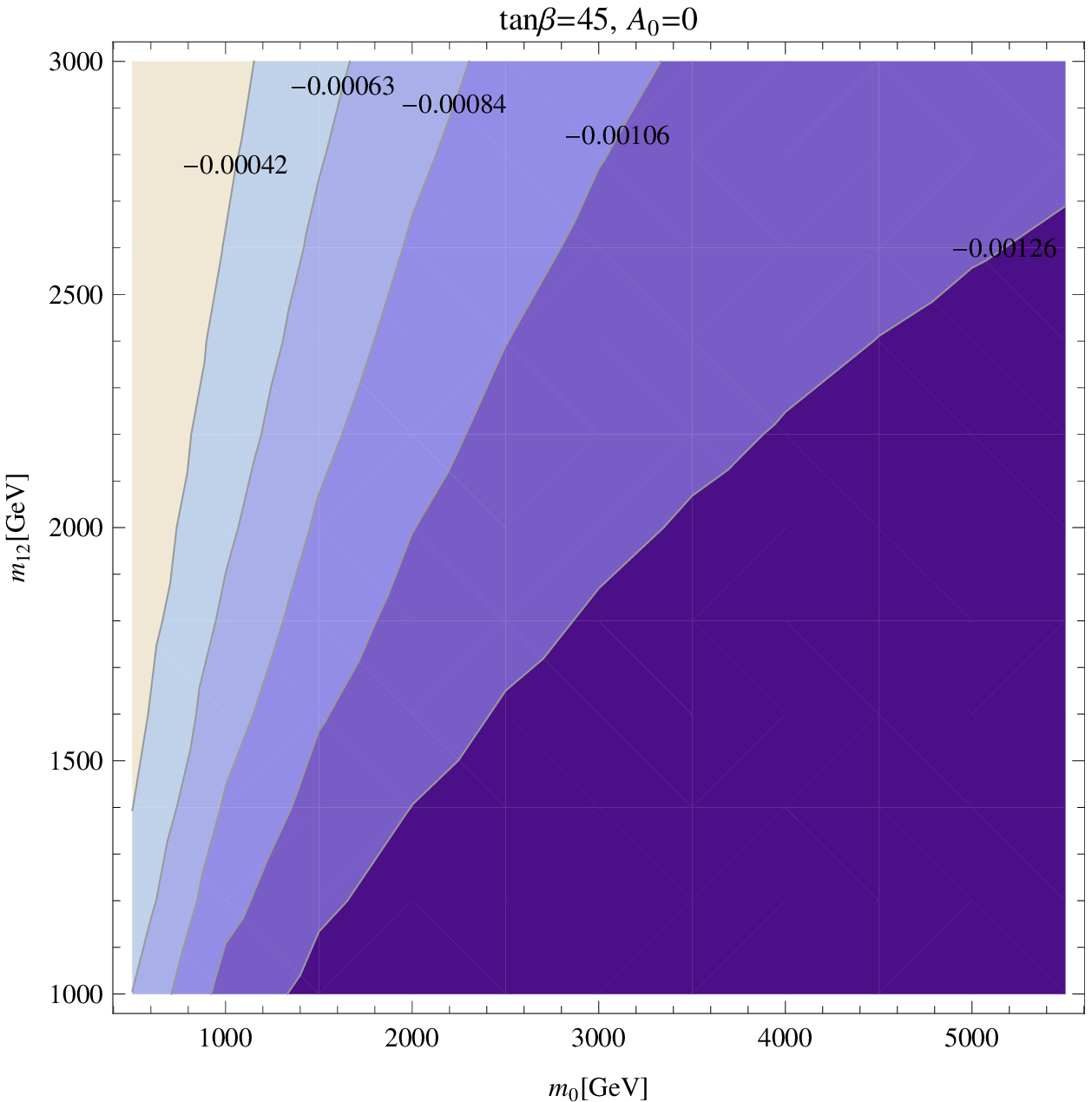 ,scale=0.56,angle=0,clip=}
\psfig{file=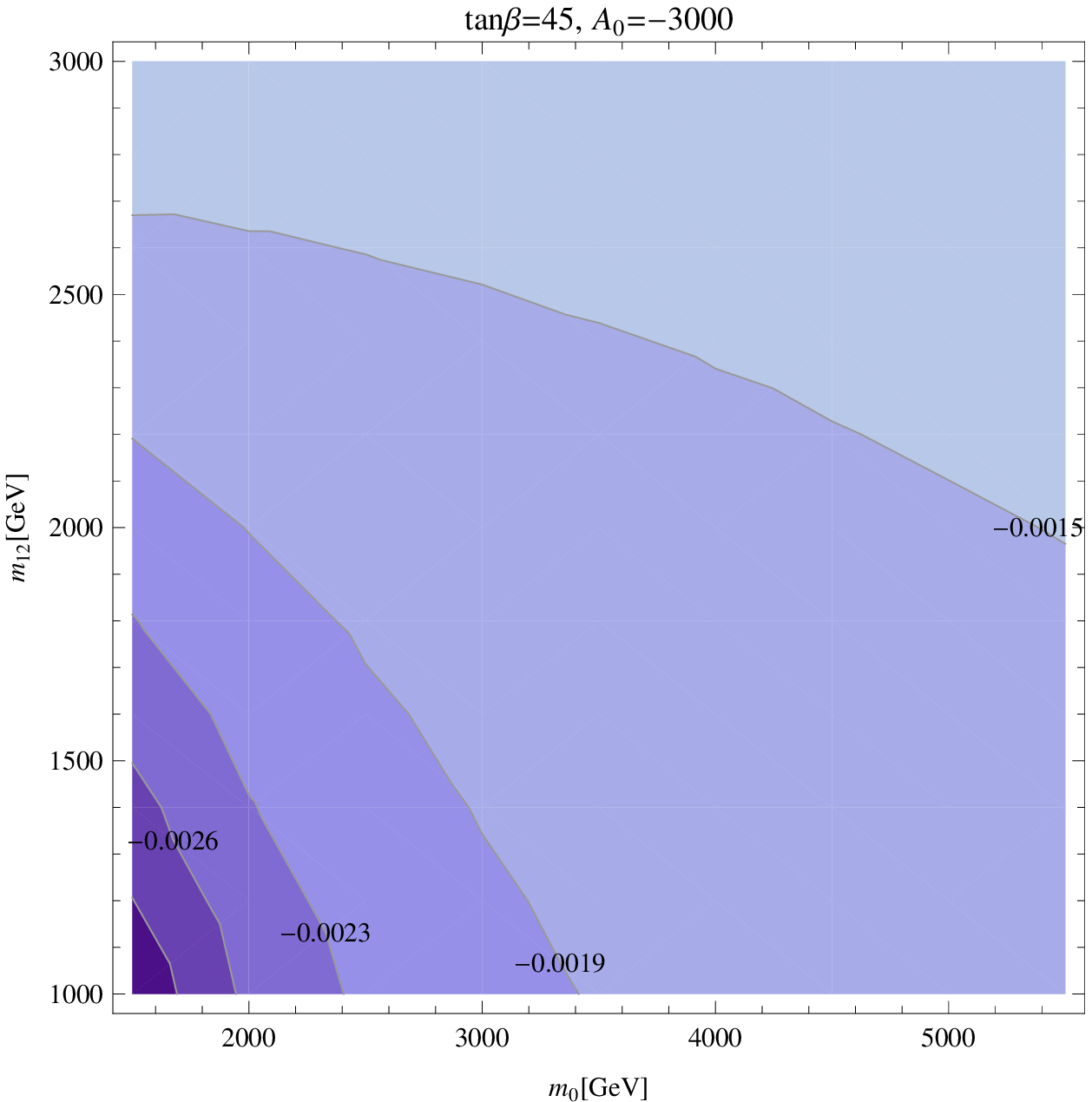   ,scale=0.56,angle=0,clip=}\\
\vspace{0.2cm}
\end{center}
\caption{Contours of $\delta^{LLL}_{13}$ in the
  $m_0$--$m_{1/2}$ plane for different values of $\tb$ and    
$A_0$ in the \CMSSMI. }  
\label{fig:DelLLL13}
\end{figure}

\begin{figure}[ht!]
\begin{center}
\vspace{3.0cm}
\psfig{file=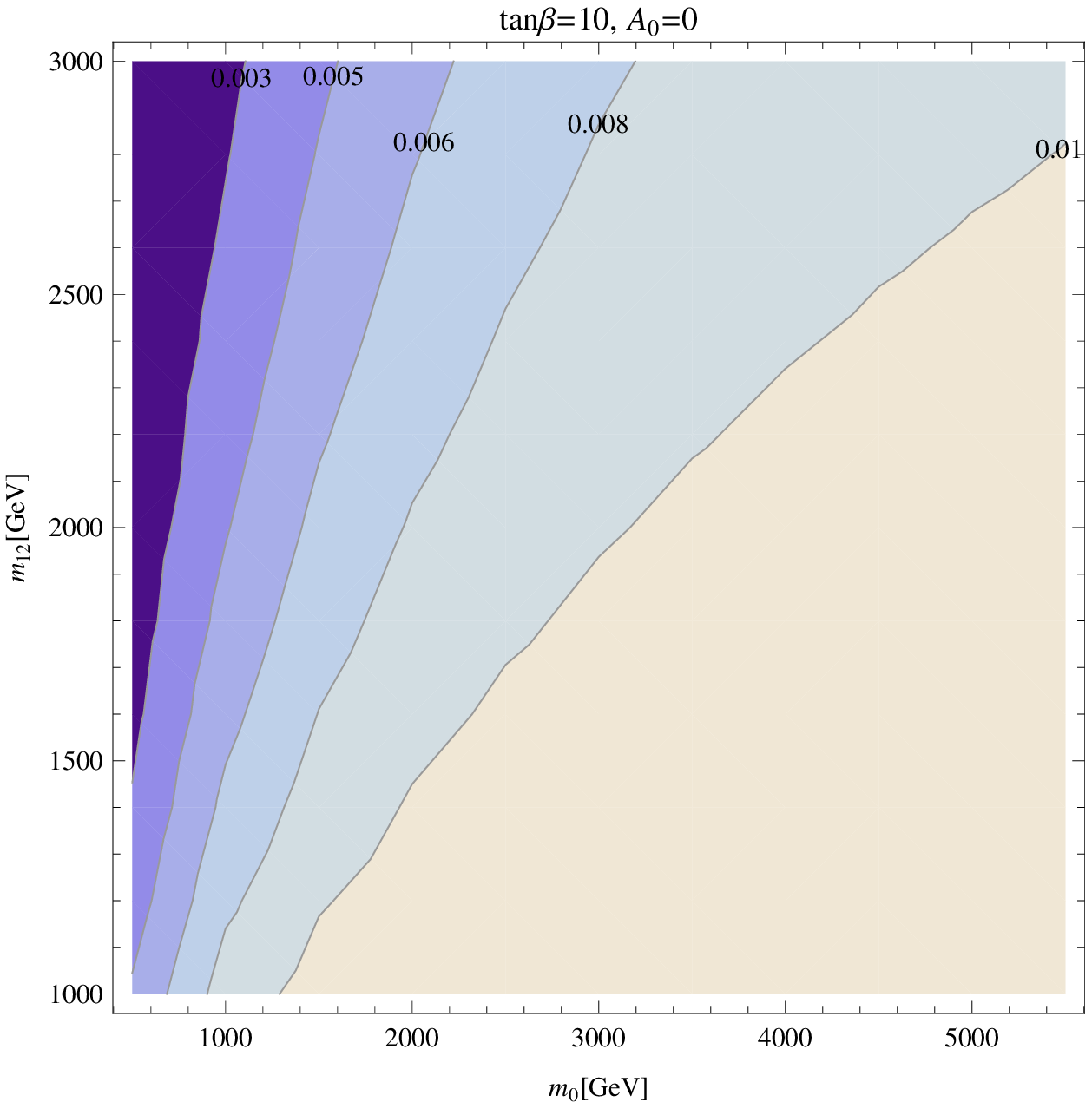  ,scale=0.57,angle=0,clip=}
\psfig{file=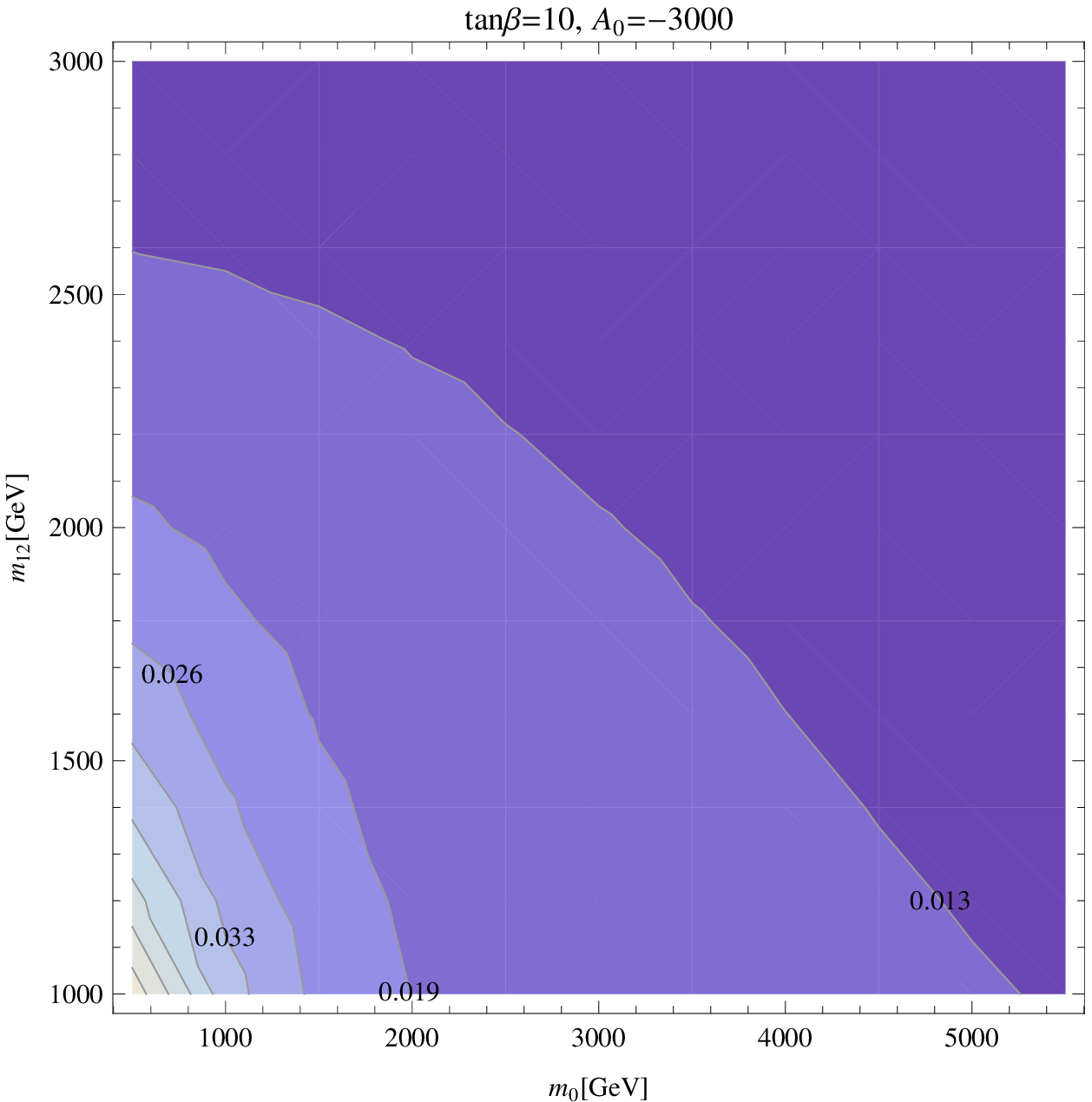  ,scale=0.57,angle=0,clip=}\\
\vspace{2.0cm}
\psfig{file=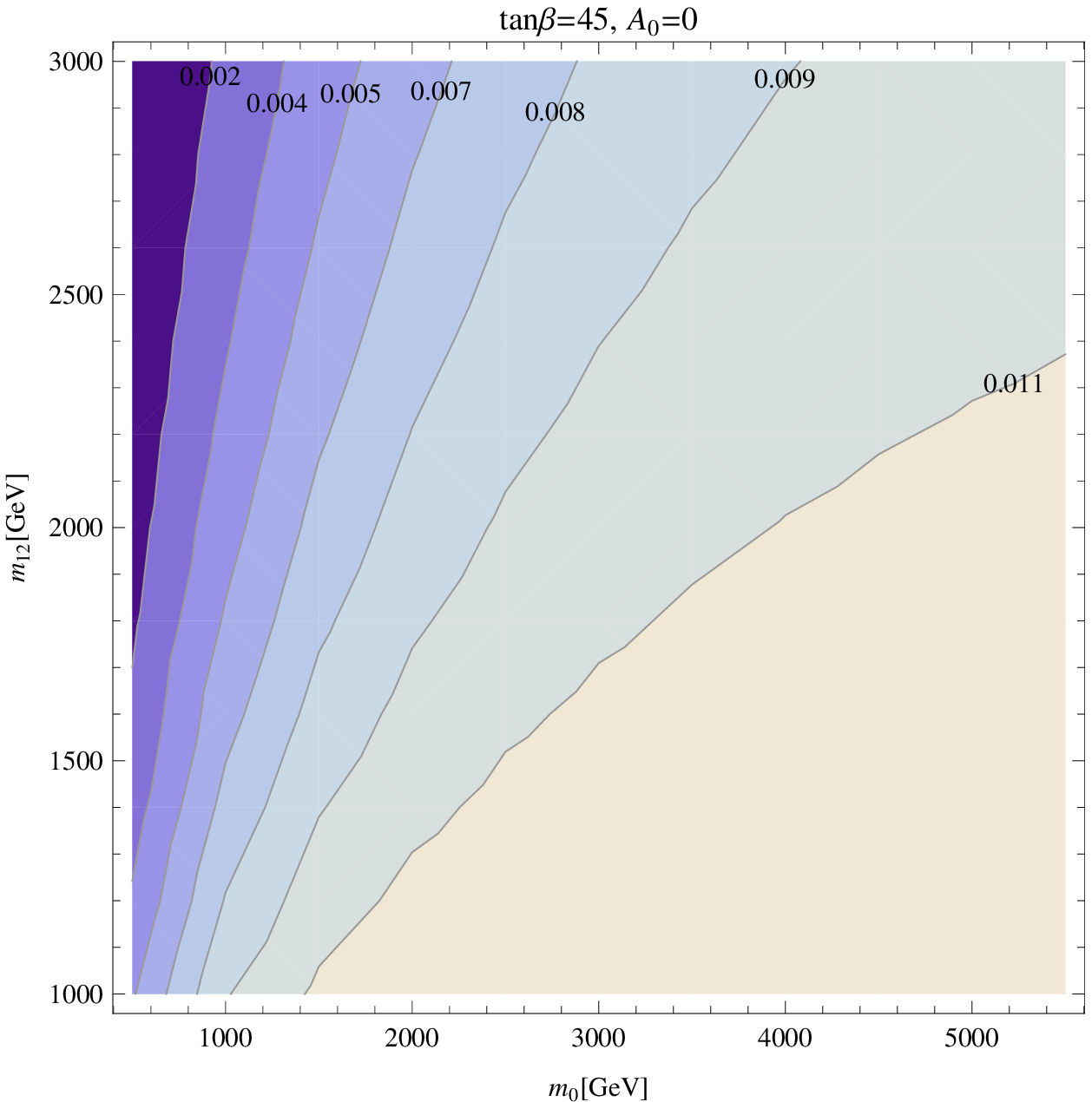 ,scale=0.56,angle=0,clip=}
\psfig{file=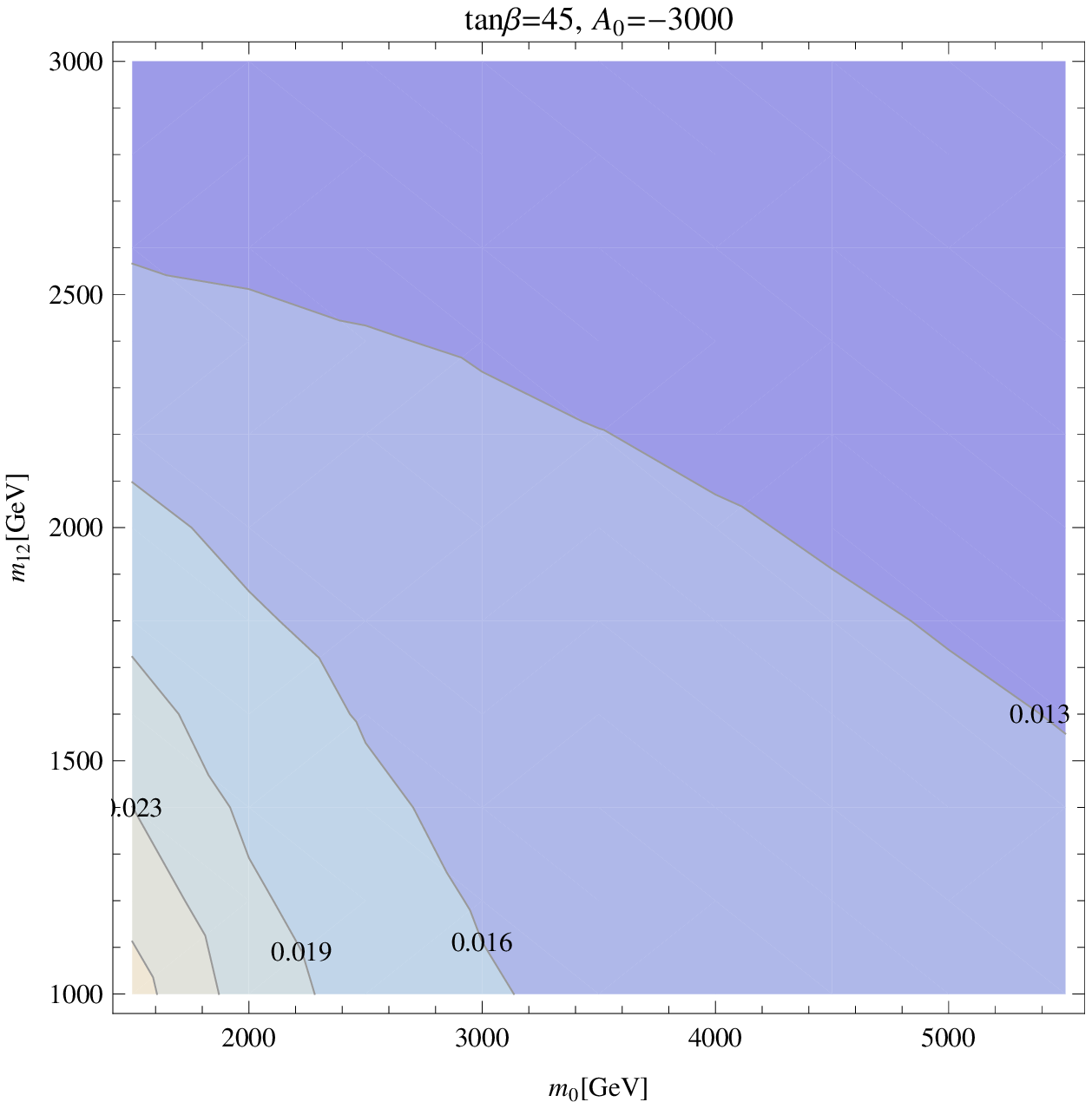   ,scale=0.56,angle=0,clip=}\\
\vspace{0.2cm}
\end{center}
\caption{Contours of $\delta^{LLL}_{23}$ in the
  $m_0$--$m_{1/2}$ plane for different values of $\tb$ and    
$A_0$ in the \CMSSMI.}  
\label{fig:DelLLL23}
\end{figure}

\begin{figure}[ht!]
\begin{center}
\vspace{3.0cm}
\psfig{file=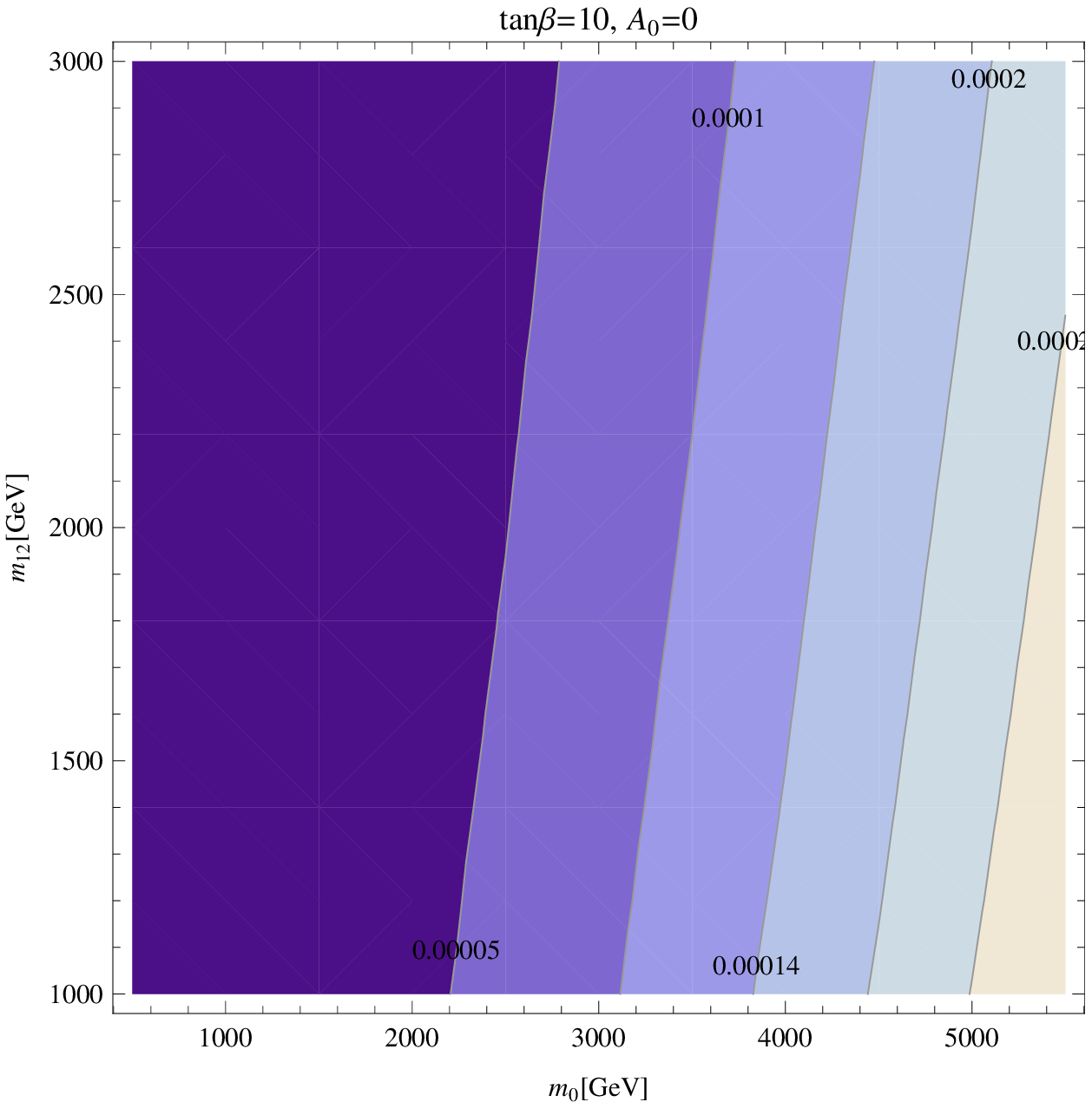  ,scale=0.57,angle=0,clip=}
\psfig{file=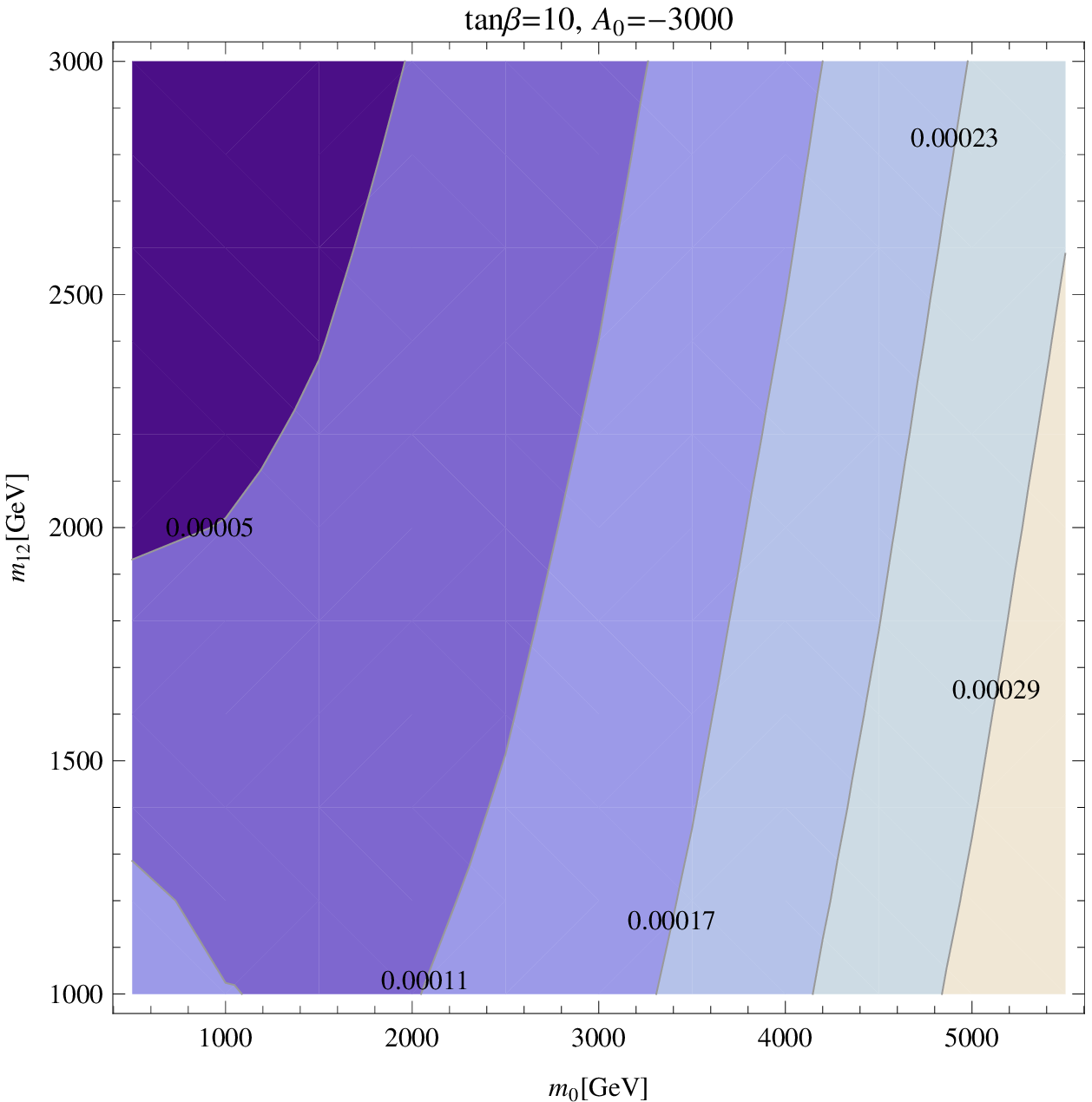  ,scale=0.57,angle=0,clip=}\\
\vspace{2.0cm}
\psfig{file=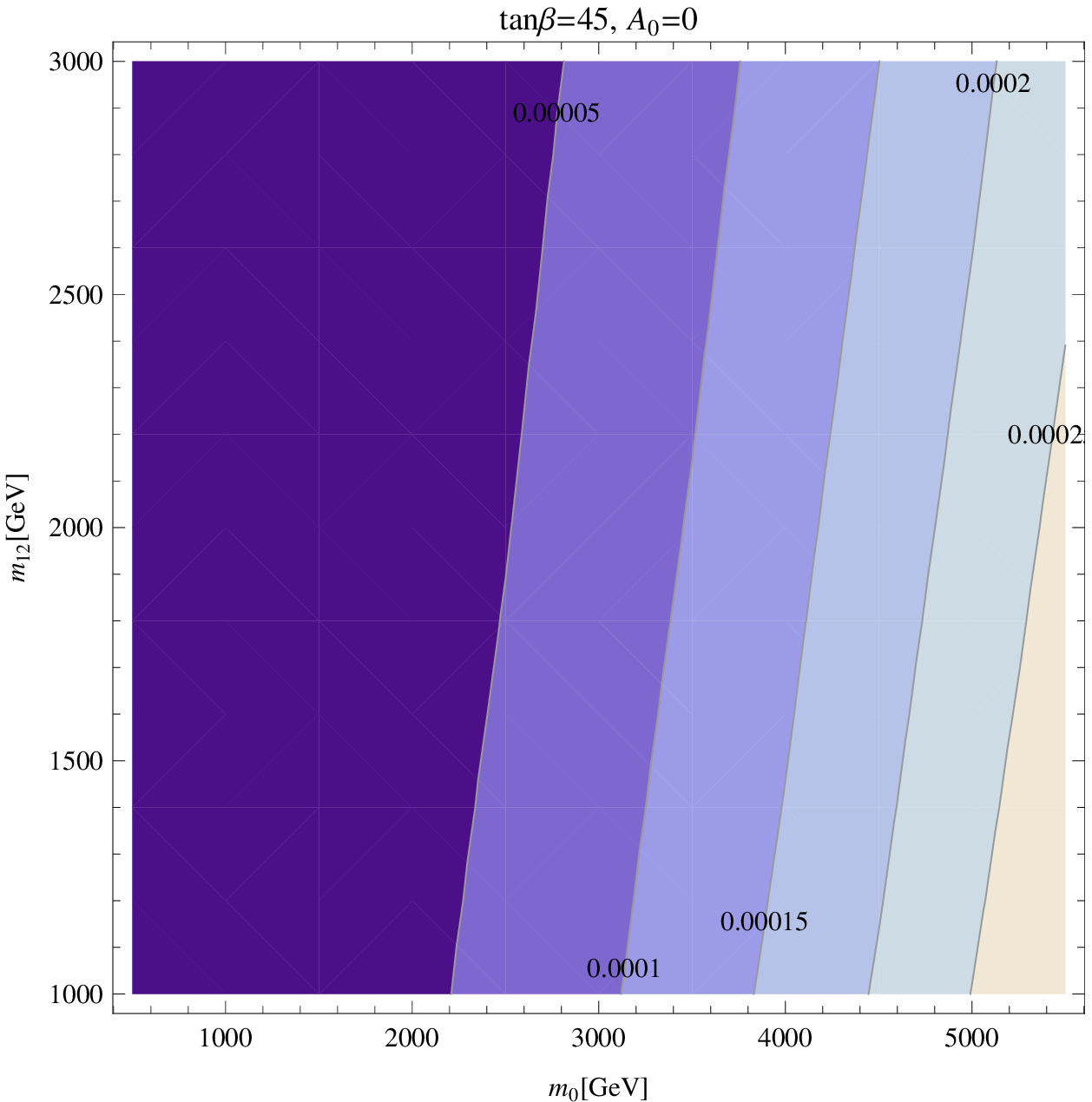 ,scale=0.56,angle=0,clip=}
\psfig{file=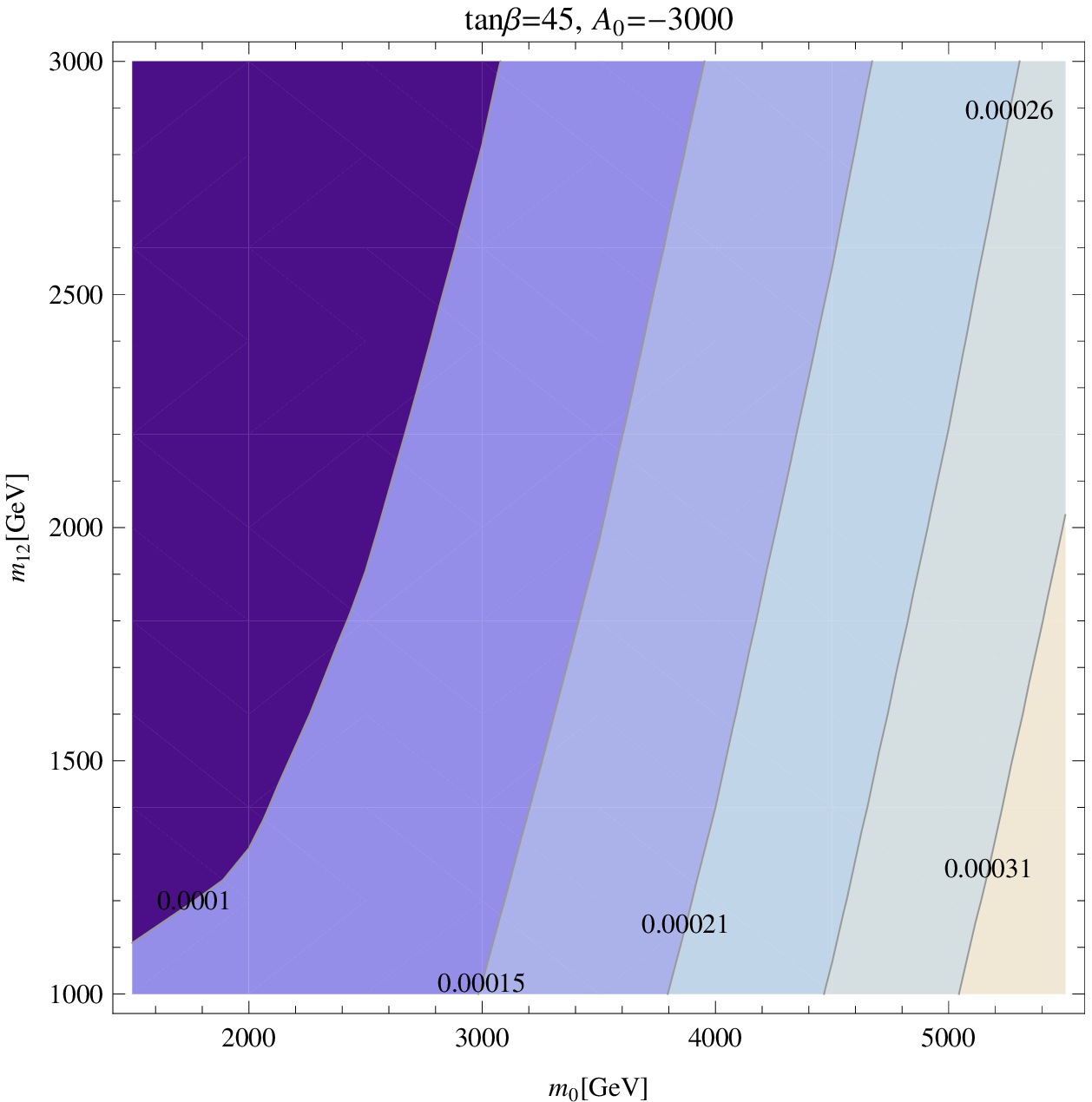   ,scale=0.56,angle=0,clip=}\\
\vspace{0.2cm}
\end{center}
\caption{Contours of \Drho\ in the
  $m_0$--$m_{1/2}$ plane for different values of $\tb$ and    
$A_0$ in the \CMSSMI.}  
\label{fig:SL-delrho}
\end{figure}

\begin{figure}[ht!]
\begin{center}
\vspace{3.0cm}
\psfig{file=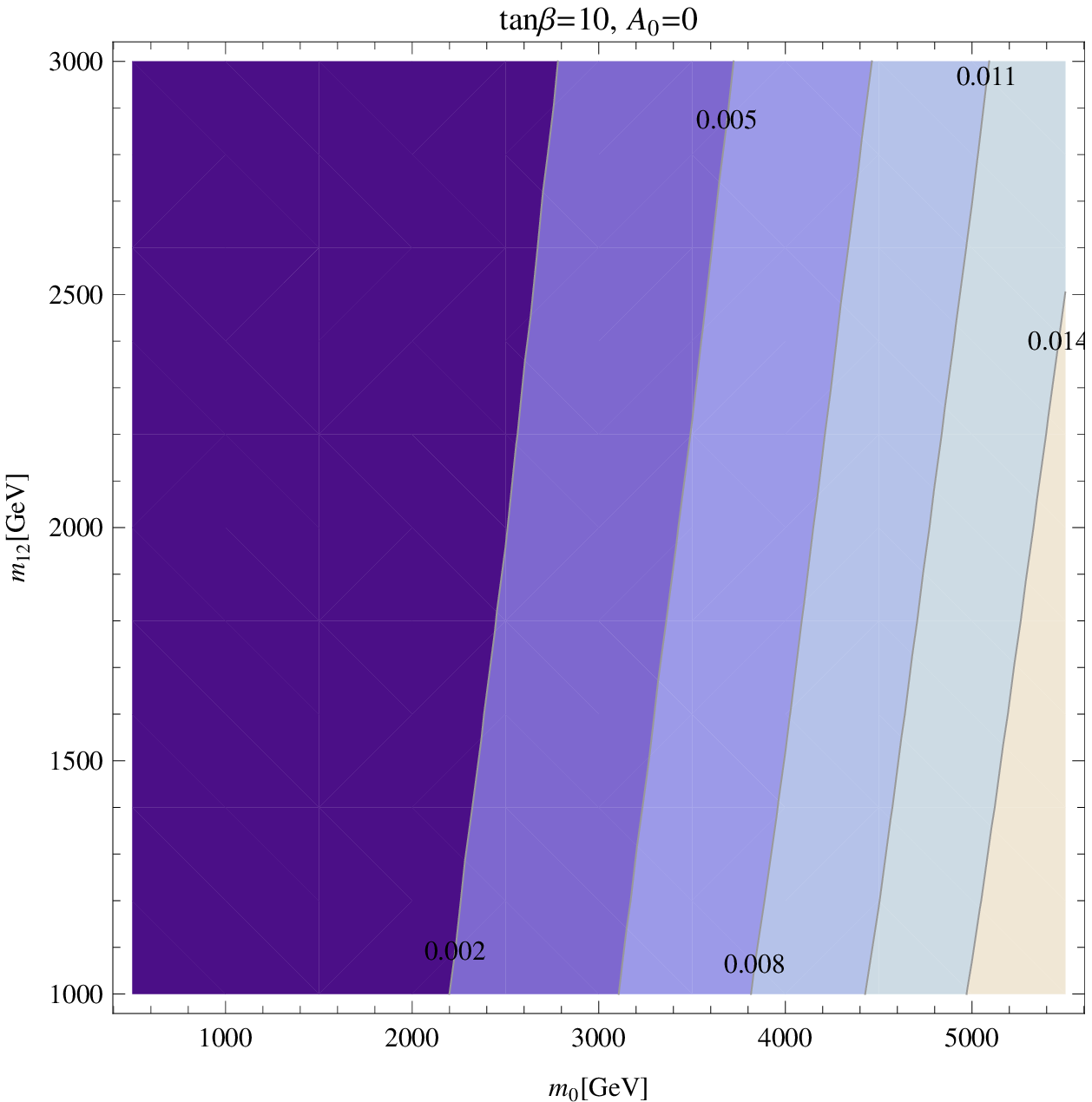  ,scale=0.57,angle=0,clip=}
\psfig{file=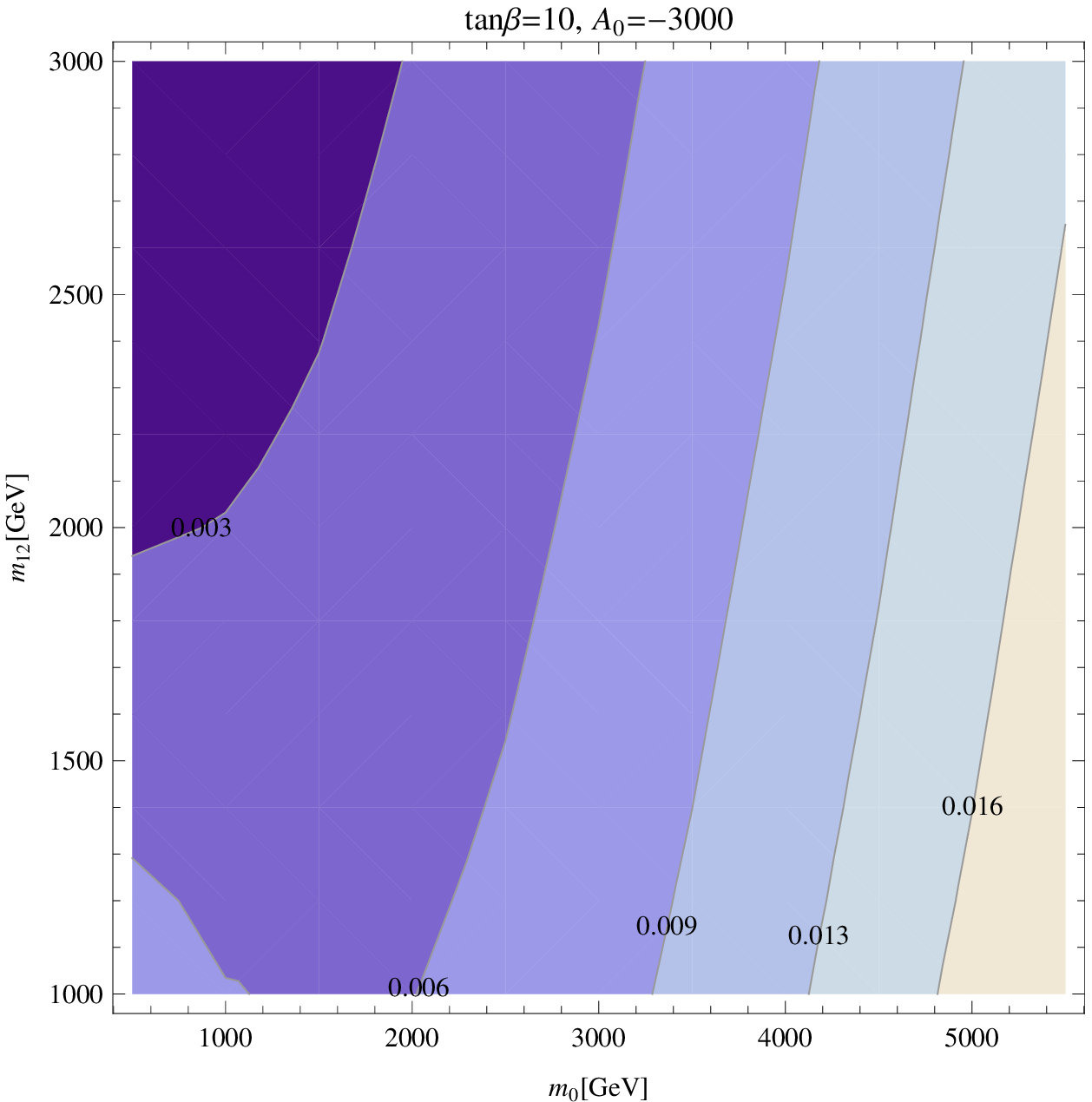  ,scale=0.57,angle=0,clip=}\\
\vspace{2.0cm}
\psfig{file=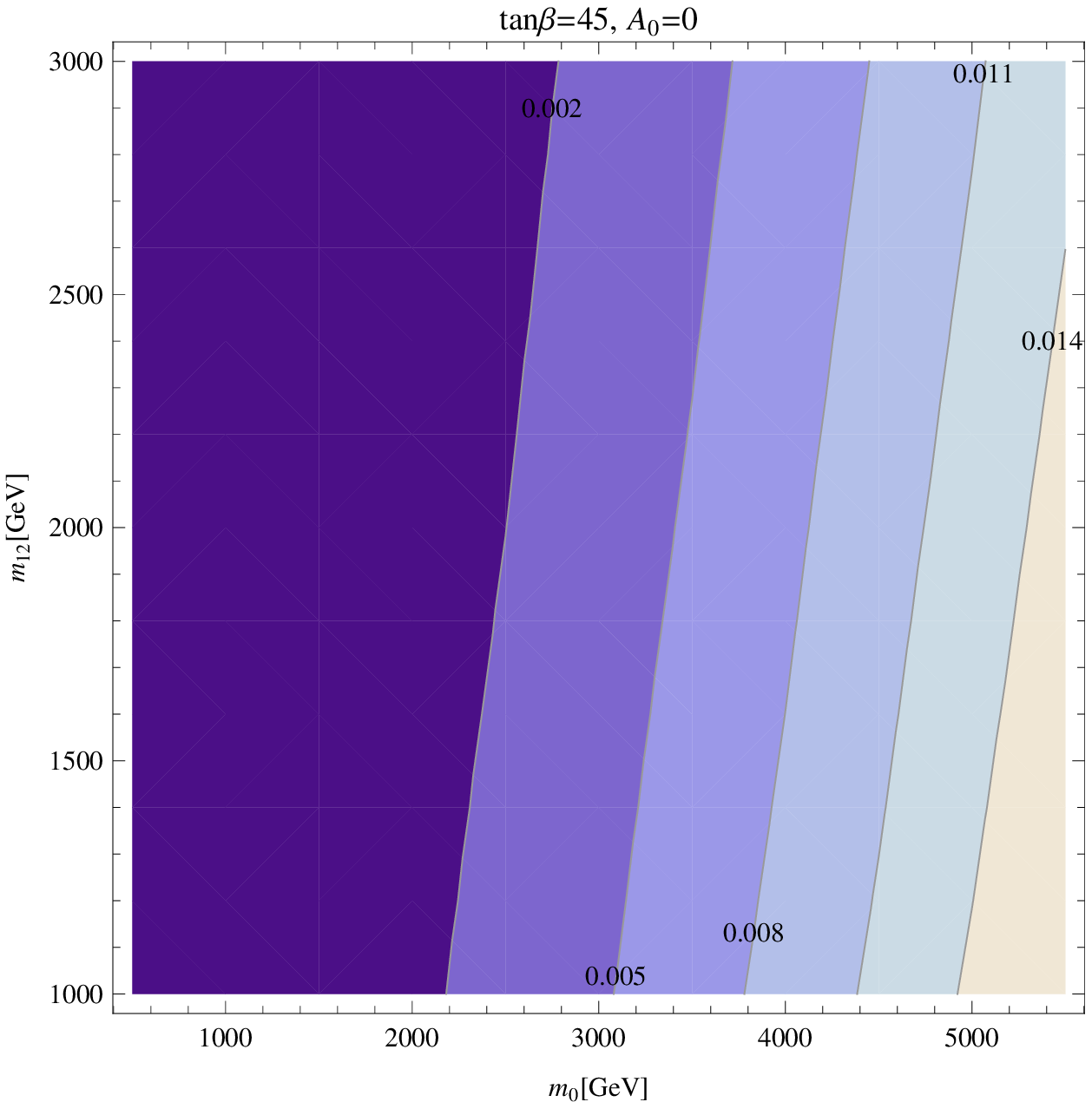 ,scale=0.56,angle=0,clip=}
\psfig{file=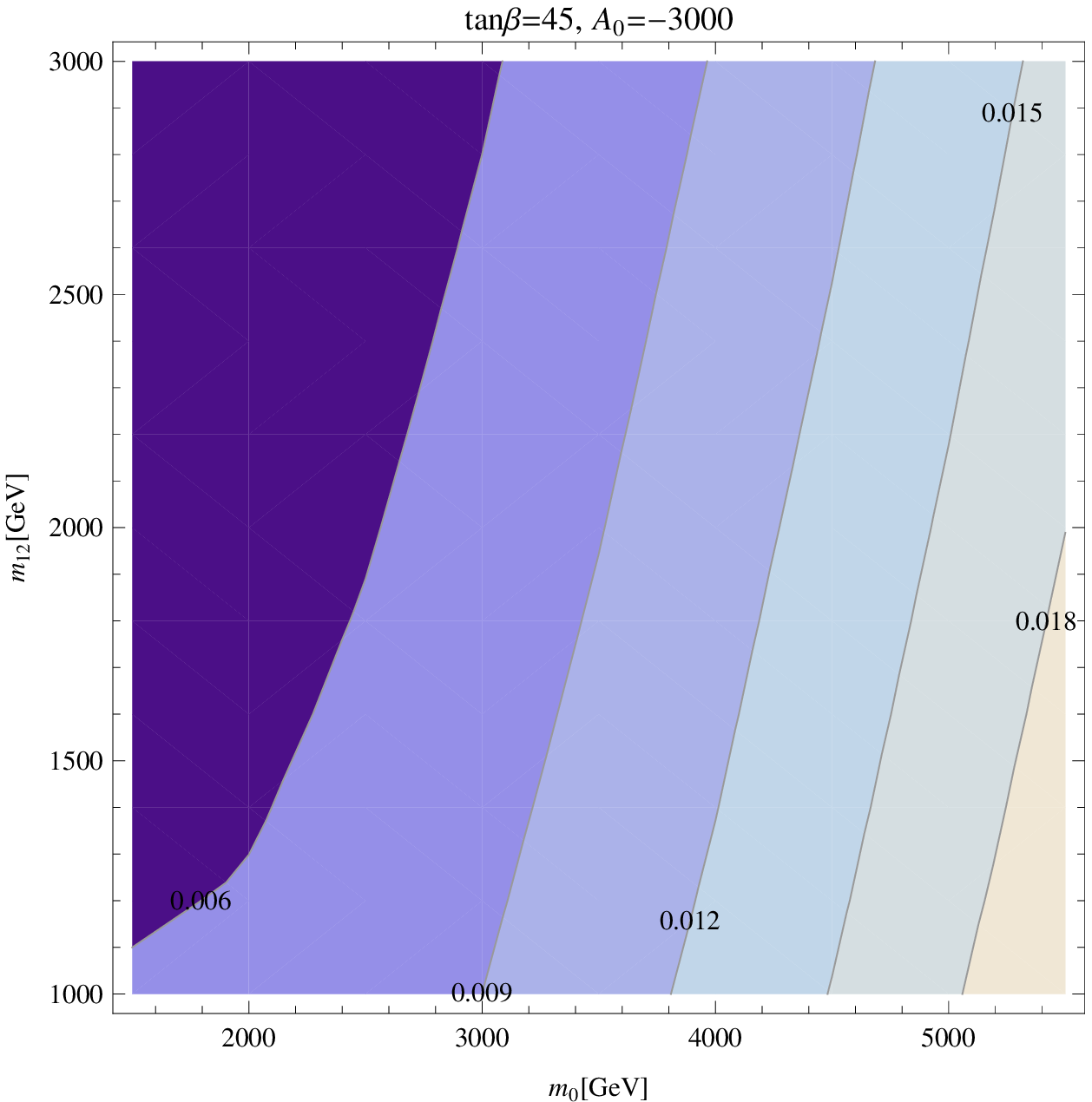   ,scale=0.56,angle=0,clip=}\\
\vspace{0.2cm}
\end{center}
\caption{Contours of \DMW\ in GeV in the
  $m_0$--$m_{1/2}$ plane for different values of $\tb$ and    
$A_0$ in the \CMSSMI.}  
\label{fig:SL-delMW}
\end{figure}

\begin{figure}[ht!]
\begin{center}
\vspace{3.0cm}
\psfig{file=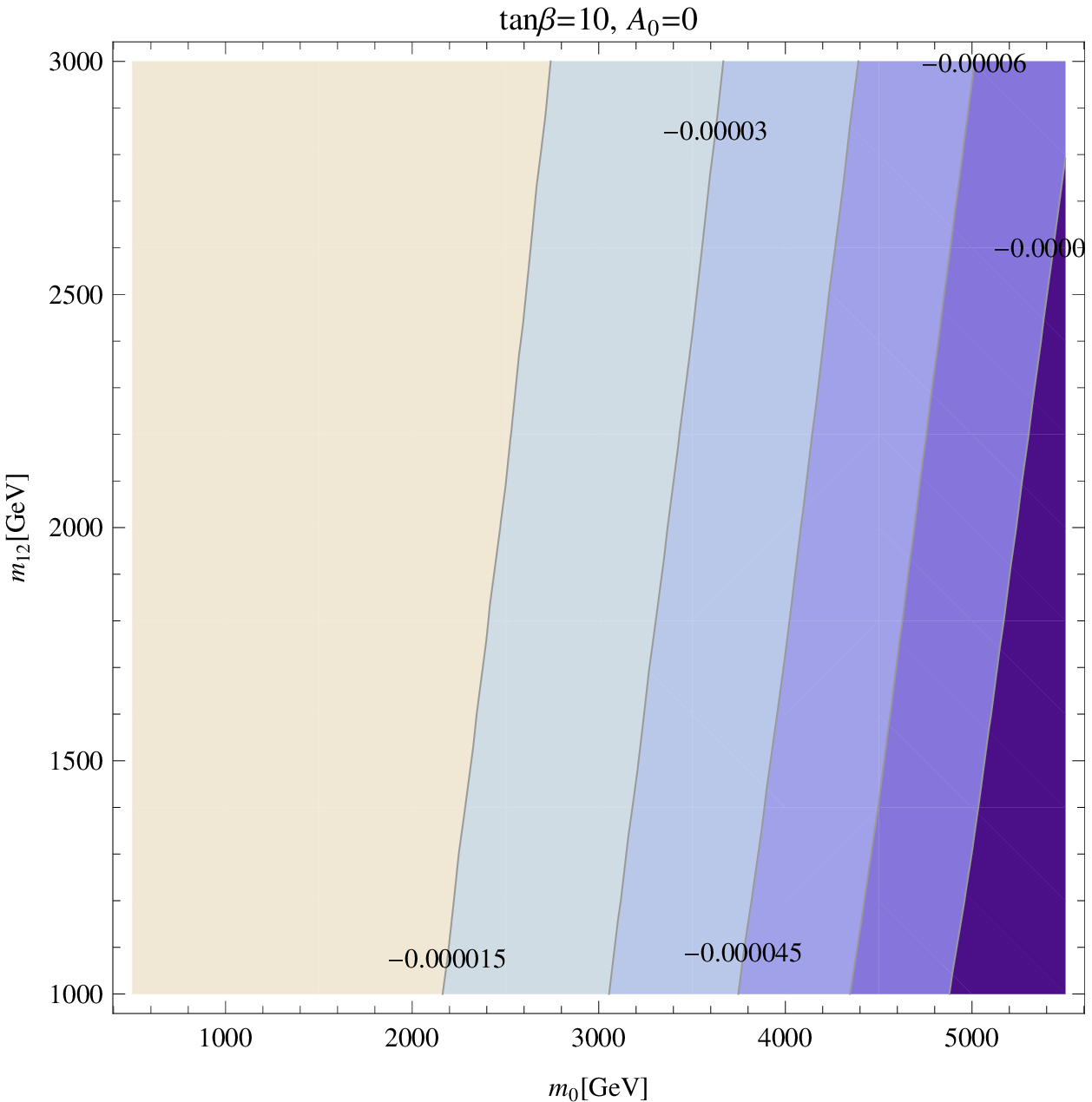  ,scale=0.57,angle=0,clip=}
\psfig{file=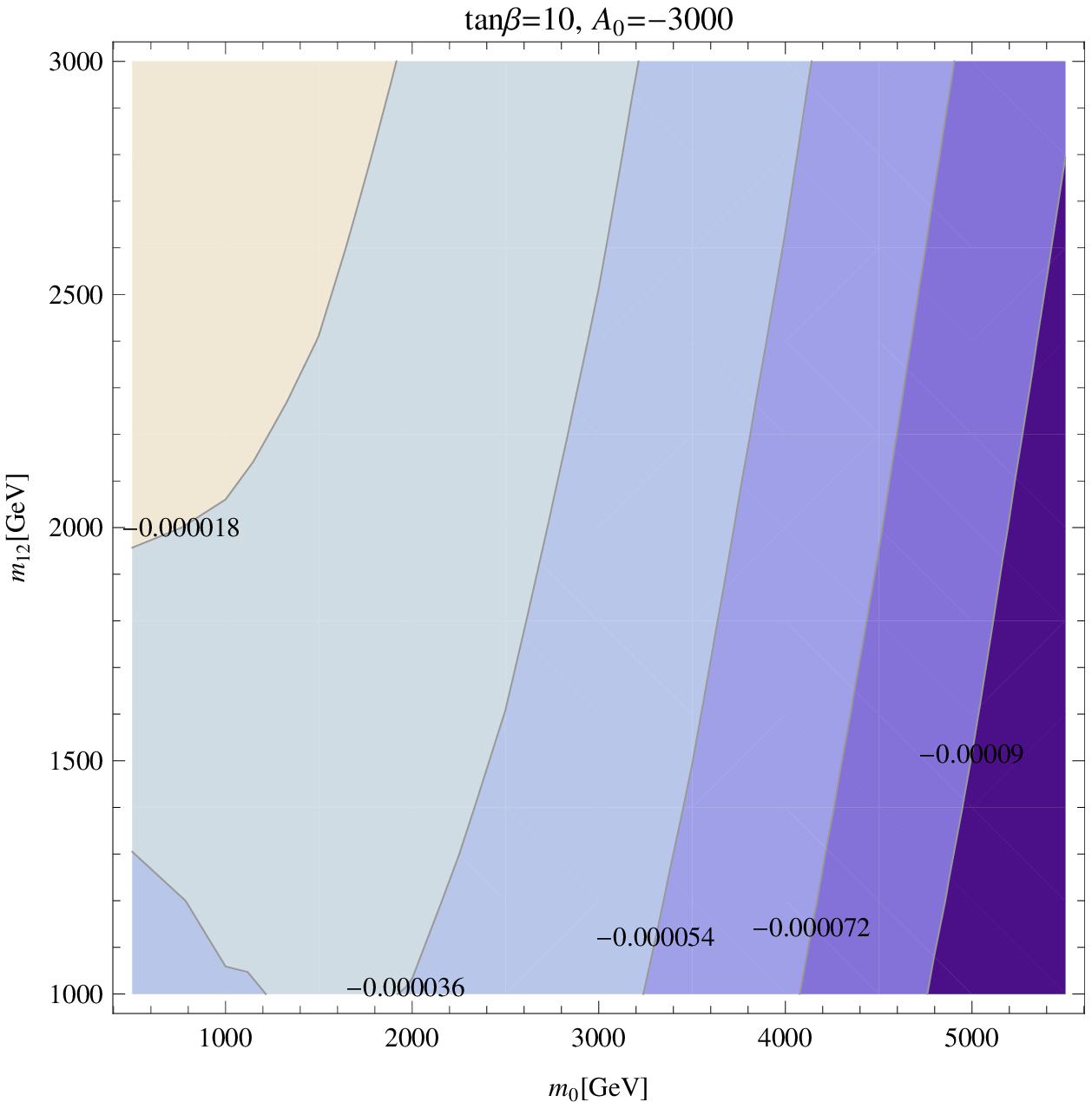  ,scale=0.57,angle=0,clip=}\\
\vspace{2.0cm}
\psfig{file=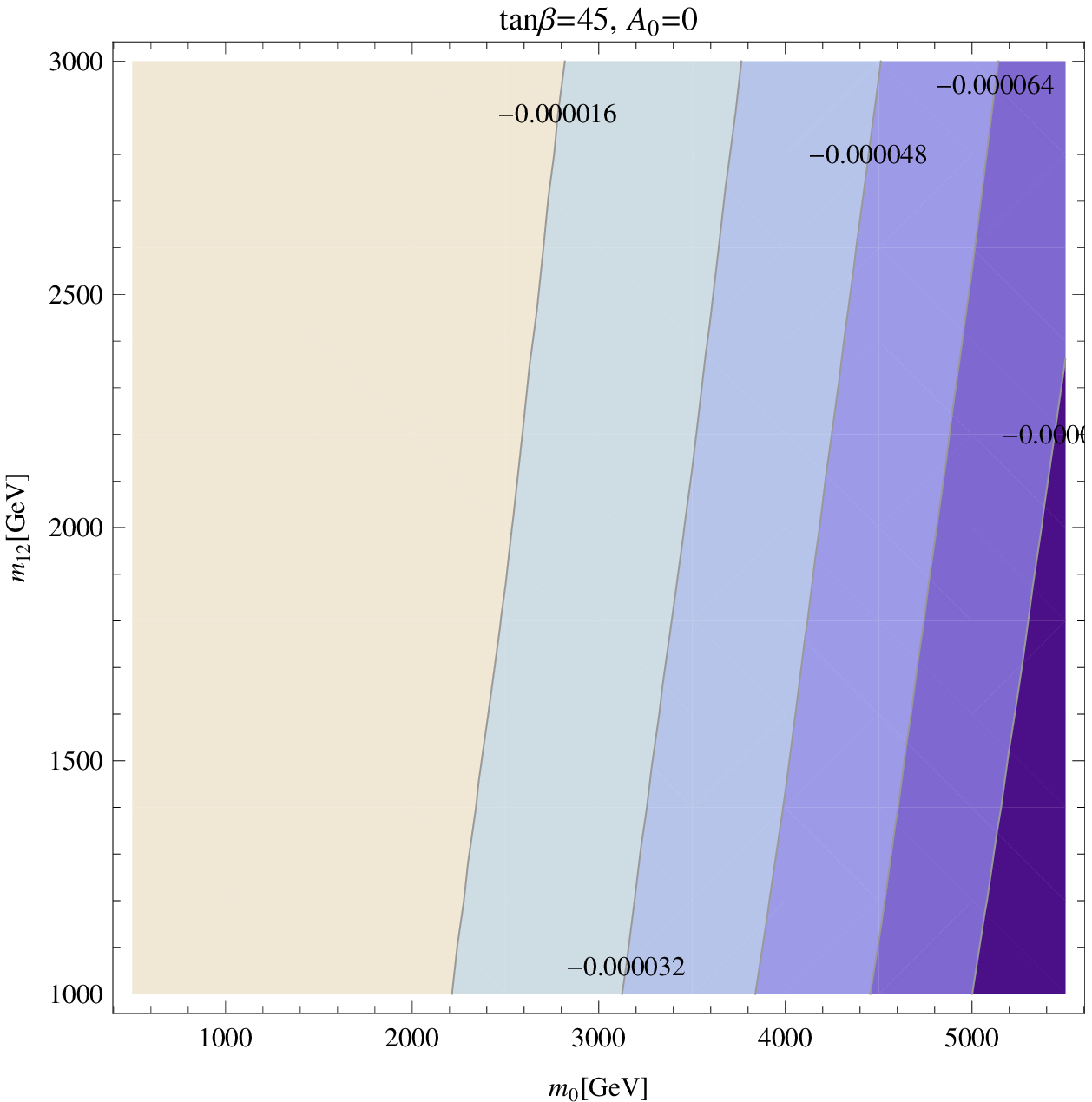 ,scale=0.56,angle=0,clip=}
\psfig{file=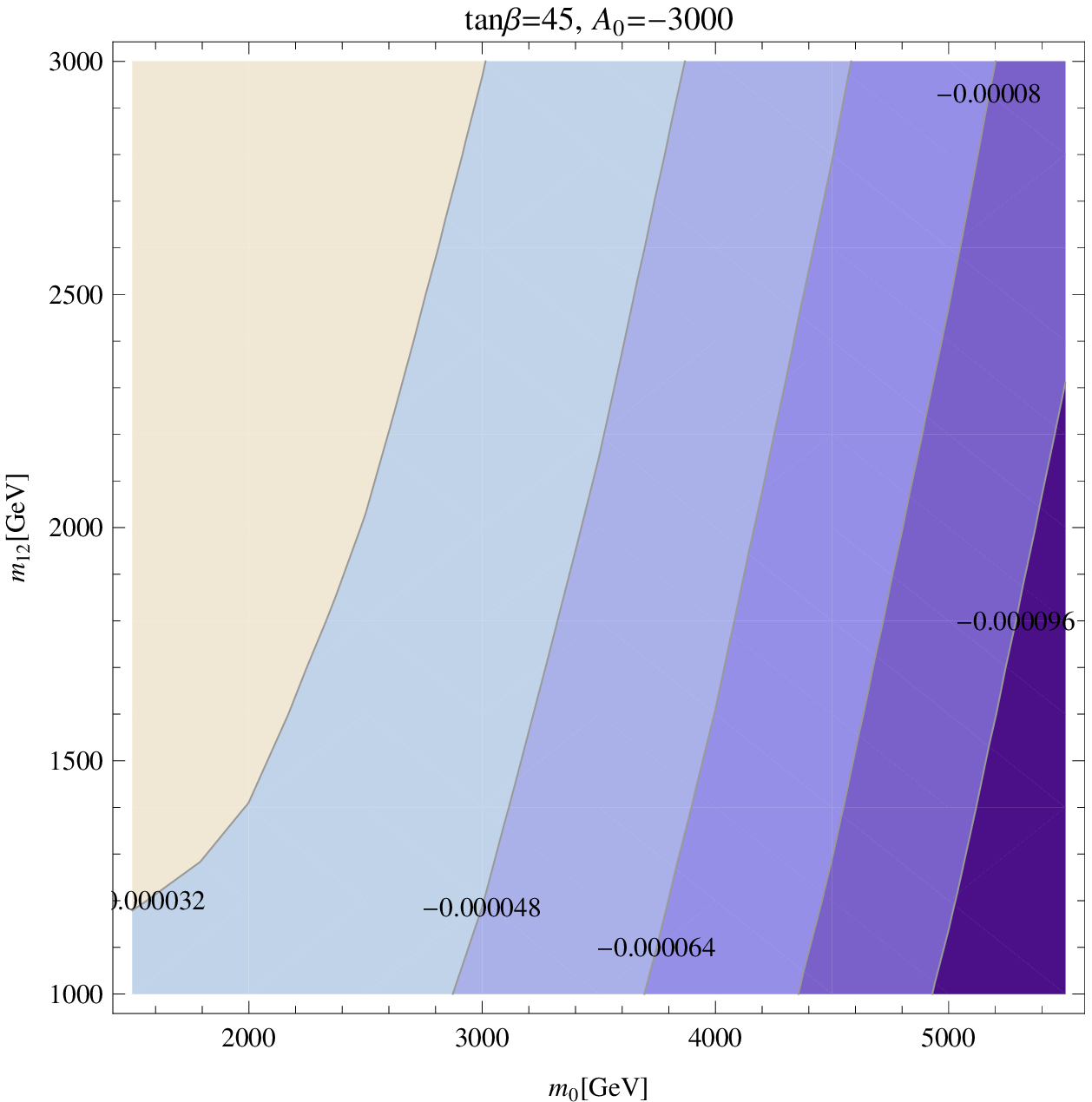   ,scale=0.56,angle=0,clip=}\\
\vspace{0.2cm}
\end{center}
\caption{Contours of \Dsweff\ in the
  $m_0$--$m_{1/2}$ plane for different values of $\tb$ and    
$A_0$ in the \CMSSMI. }  
\label{fig:SL-delSW2}
\end{figure}

\begin{figure}[ht!]
\begin{center}
\psfig{file=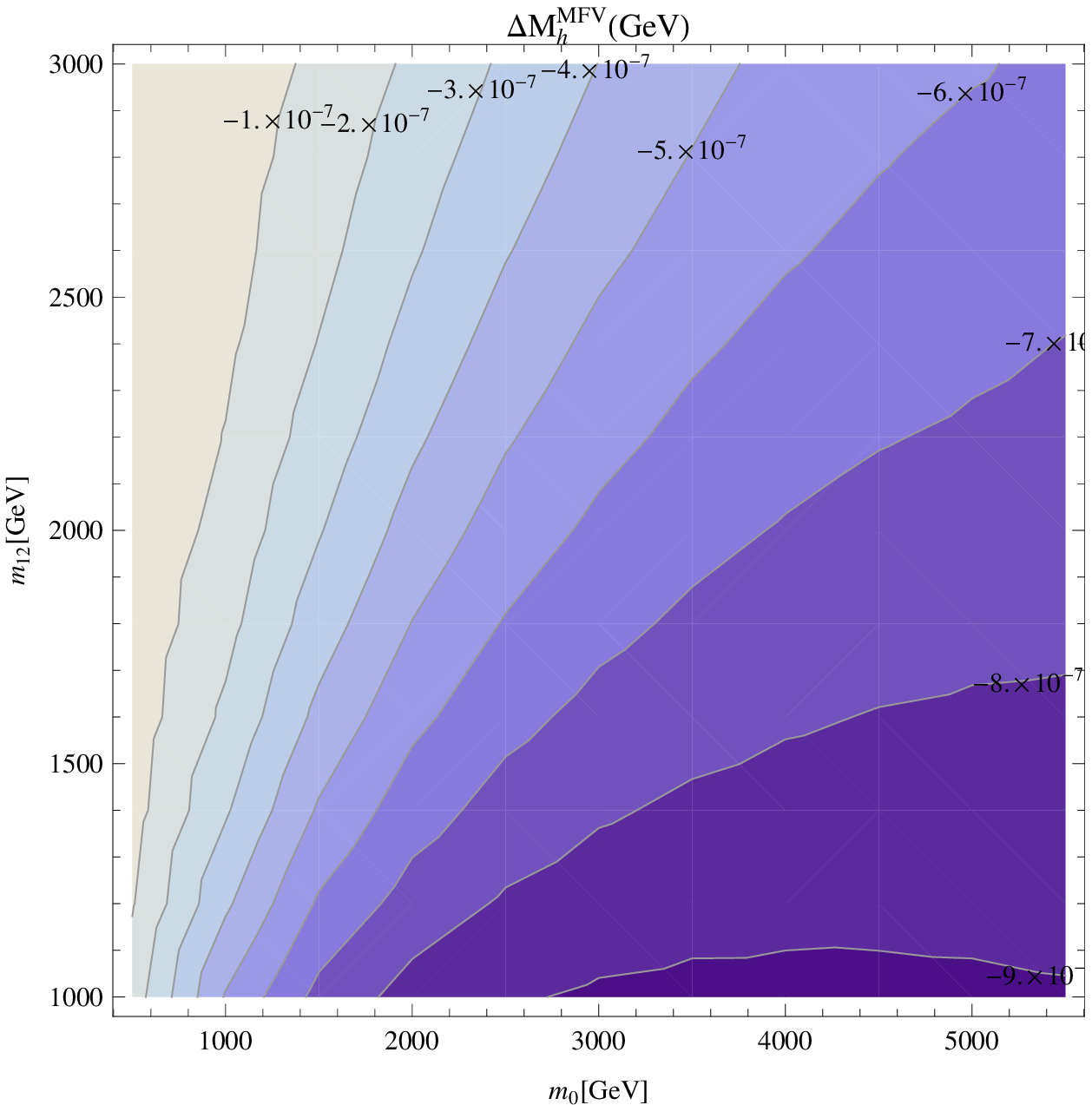 ,scale=0.56,angle=0,clip=}
\psfig{file=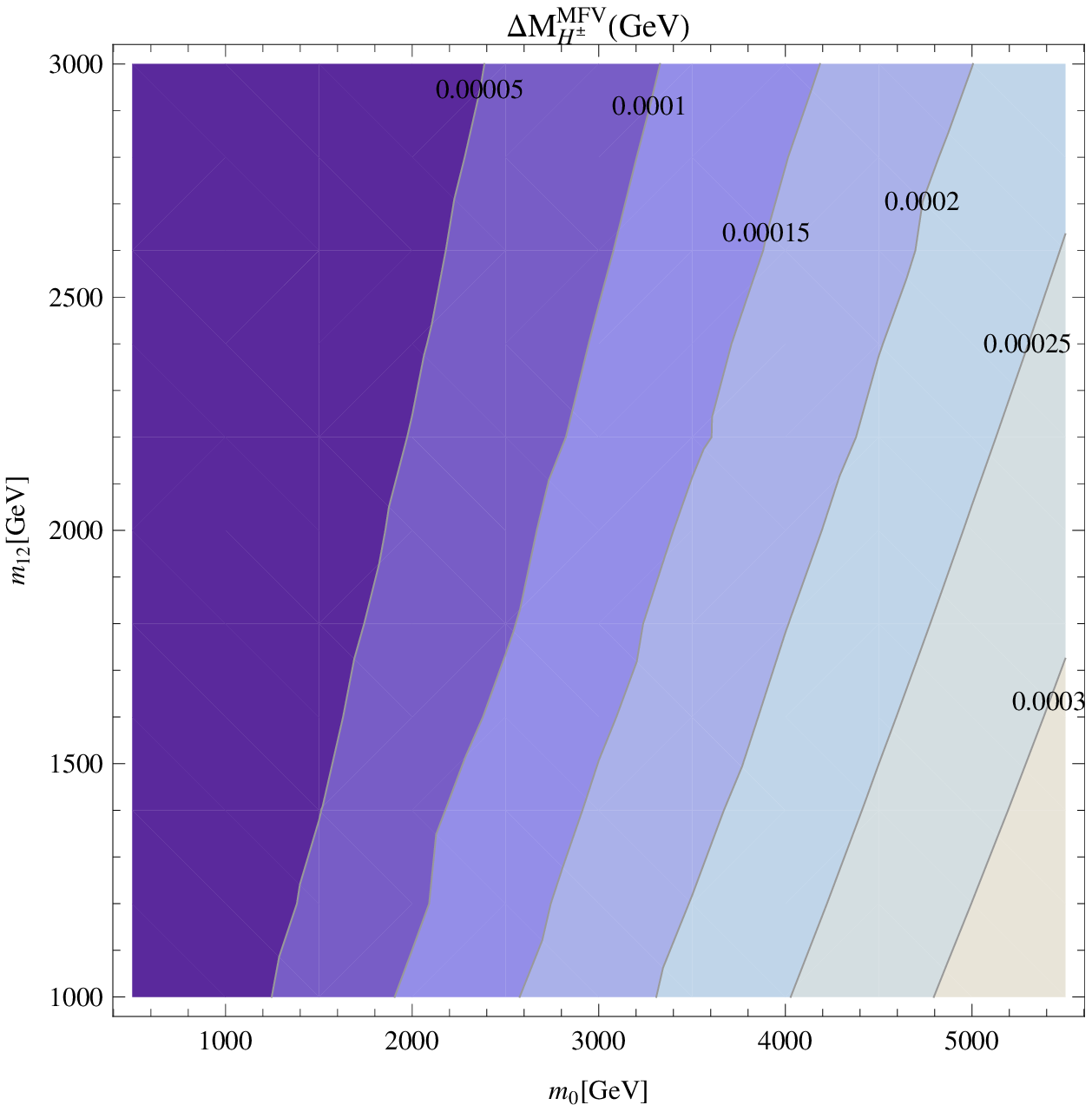 ,scale=0.56,angle=0,clip=}

\end{center}
\caption{Contours of \DMh\ (left) and \DMHp\ (right) 
in the $m_0$--$m_{1/2}$ plane for $\tb = 10$ and $A_0 = 0$ in
the \CMSSMI.}
\label{SL-MH} 
\end{figure}

Our numerical results in the \CMSSMI\ are shown in
\reffis{fig:DelLLL12} - \ref{SL-MH}. As in the CMSSM we present the
results in the $m_0$--$m_{1/2}$ plane for four combinations of 
$\tb = 10, 45$ (upper and lower row) and $A_0 = 0, -3000 \gev$ (left
and right column), again capturing the ``extreme'' cases.
We start presenting the three most relevant $\deFABij$. 
\reffis{fig:DelLLL12}-\ref{fig:DelLLL23} show $\del{LLL}{12}$, 
$\del{LLL}{13}$ and $\del{LLL}{23}$, respectively. As expected,
$\del{LLL}{23}$ turns out to be largest of \order{0.01}, while the
other two are about one order of magnitude smaller. The dependence on
$\tb$ is not very prominent, but going from $A_0 = 0$ to $-3000 \gev$
has a strong impact on the $\deFABij$. For small $A_0$ the size of the
$\deFABij$ is increasing with larger $m_0$ and $m_{1/2}$, for 
$A_0 = -3000 \gev$ the largest values are found for small $m_0$ and
$m_{1/2}$. Here one comment on flavor violating decays is in order.
The selected values of $Y_\nu$ result in a large prediction for, e.g.,
BR($\mu \to e \gamma$) that can eliminate some of the $m_0$--$m_{1/2}$
parameter plane, in particular combinations of low values of $m_0$ and
$m_{1/2}$. For our parameter settings these regions are small for 
$\tb = 10$ and reach up to roughly $m_0 + m_{1/2} \sim 2000 \gev$ for 
$A_0 = -3000 \gev$. For $\tb = 45$ they are larger and exclude roughly the
lower left half of the $m_0$--$m_{1/2}$ planes shown.

In \reffis{fig:SL-delrho}-\ref{fig:SL-delSW2} we show the results for
the EWPO. The same pattern and non-decoupling behavior for EWPO as
in the case of CMSSM (squark $\deFABij$) can be observed. However, the
corrections induced by slepton flavor violation 
are relatively small compared to squark case. For the most extreme
cases, i.e.\ the largest values of $m_0$, the corrections to $\MW$
turn out to be of the same order of the experimental uncertainty.
For those parts of the parameter space neglecting the effects of LFV
to the EWPO could turn out to be an insufficient approximation, in
particular in view of future improved experimental accuracies.

Finally, in \reffi{SL-MH} we present the corrections to the Higgs
boson masses induced by slepton flavor violation.
Here we only show \DMh\ (left) and \DMHp (right) for $\tb = 10$ and 
$A_0 = 0$. They turn out to be
negligibly small in both cases. Corrections to \DMH, which are not
shown, are even smaller. We have checked that these results hold also
for other combinations of $\tb$ and $A_0$. 
Consequently, within the Higgs sector the
approximation of neglecting the effects of the $\deFABij$ is fully
justified.

\newpage

%% file: Conclusions.tex
\section{Conclusions}
\label{sec:conclusions}

In this paper we have investigated the CMSSM and the \CMSSMI\ (i.e.\ the
CMSSM augmented by right-handed neutrinos to produce the observed
neutrino mass pattern via the seesaw type~I mechanism) under the
hypothesis of Minimal Flavor Violation (MFV, i.e.\ the only flavor
violating source is the CKM matrix and/or the PMNS matrix in the case of
the \CMSSMI). In many
phenomenological analyses of the CMSSM the effects of intergenerational
mixing in the squark and/or slepton sector is neglected. However, such
mixings are naturally induced, assuming no flavor violation at the GUT
scale, by the RGE running from the GUT to the EW scale exactly due to the
presence of the CKM and/or the PMNS matrix. 
The spectra of the CMSSM and \CMSSMI\ have been evaluated with the help
of the program {\tt SPheno} 
by taking the GUT scale input run down via the appropriate
RGEs to the EW scale.

We have evaluated the predictions for $B$-physics observables, MSSM
Higgs boson masses, electroweak precision observables in the CMSSM and
\CMSSMI. 
In order to analyze the effects of neglecting intergenerational mixing
these observables have been evaluated with the full spectrum at the EW
scale, as well as with the spectrum, but with all intergenerational
mixing set (artificially) to zero (as it has been done in many
phenomenological analyses). The difference in the various observables
indicates the size of the effects neglected in those analyses. In this
way it can be checked whether neglecting those mixing effects is a justified
approximation. 

Within the CMSSM we have taken a fixed grid of $A_0$ and $\tb$,
while scanning the $m_0$--$m_{1/2}$ plane. We found that the value of 
$\deFABij$ increases with the increase of the $A_0$ or $\tb$ values.
The Higgs boson masses receive corrections below current and future
experimental uncertainties, 
where the shifts in $\MHp$ were found largest at the level of 
\order{100 \mev}. Similarly for the $B$-physics observables the induced
effects are at least one order of magnitude smaller than the current
experimental uncertainty. For those two groups of observables the
approximation of neglecting intergenerational mixing explicitly is a
viable option. 

The picture changes for the electroweak precision observables. The
masses of the squarks grow with $m_0$, and thus do the mixing terms,
inducing a splitting between masses in an $SU(2)$ doublet, leading to
a non-decoupling effect. 
For $m_0 \gsim 3 \tev$ the effects induced in $\MW$ and $\sweff$ are
several times larger than the current experimental uncertainties and can
shift the CMSSM prediction outside the allowed experimental range. In
this way, taking the intergenerational mixing (correctly) into account
can set an {\em upper} bound on $m_0$ that is not present in recent
phenomenological analyses. 

Going to the \CMSSMI\ the numerical results depend on the
concrete model definition. We have chosen a set of parameter that 
reproduces correctly the observed neutrino data and simultaneously 
induces large LFV effects and induces {\em relatively} large corrections
to the calculated observables. Consequently, parts of the parameter
space are excluded by the experimental bounds on $\br(\mu \to e \ga)$.
Concerning the precision observables we find that
$B$-physics observables are not affected, we also find that the
additional effects induced by slepton flavor violation on Higgs boson
masses are negligible. Again the EWPO show the largest impact, where
for $\MW$ effects at the same level as the current experimental
accuracy can be observed for very large values of $m_0$.

\medskip
To summarize: artificially setting all flavor violating terms to zero in
the CMSSM and \CMSSMI\ is an acceptable approximation for $B$-physics
observables, Higgs boson masses. However, in the electroweak precision
observables the flavor violation in the MFV framework induced by the
presence of the CKM matrix 
in the RGE running from the GUT to the EW scale large effects can be
induced. Those effects can be substantially larger than the current
experimental accuracy in $\MW$ and $\sweff$. Taking those effects
correctly into account places new upper bounds on $m_0$ that are
neglected in recent phenomenological analyses.